\documentstyle[hip-artc,epsfig]{article}  
\volnumber{15}  \edyear{2002}  \frompage{1} \topage{80}                  
\recrevdate{27 July 2001}                                                

\def\taub{{\overline{\tau}}}
\def\etab{{\overline{\eta}}}
\def\bk{{\bf{k}}}
\def\dst{\displaystyle\strut}
\def\be{\begin{equation}}
\def\ee{\end{equation}}
\def\bea{\begin{eqnarray}}
\def\eea{\end{eqnarray}}
\def\l({\left(}
\def\r){\right)}
\def\ket#1{\vert#1\rangle}
\def\bra#1{\langle#1\vert}
\def\brak#1#2{\langle#1\vert #2\rangle}
\def\I#1{\int d^3#1}
\def\bx{{\bf{x}}}
\def\bp{{\bf{p}}}
\def\bpi{{\bf{\pi}}}
\def\bxi{{\bf{\xi}}}
\def\bq{{\bf{q}}}
\def\br{{\bf{r}}}
\def\axd{\hat{ a}^{\dag} (\bx)}
\def\apd{\hat a^{\dag} (\bp)}
\def\ax{\hat{ a}^{} (\bx)}
\def\ap{\hat  a^{} (\bp)}
\def\psxd{\hat \Psi^{\dag} (\bx)}
\def\psx{\hat \Psi^ (\bx)}
\def\psxde{\hat \Psi^{\dag} (\bx)}
\def\psxe{\hat \Psi^ (\bx)}
\def\pdpx#1{\hat \Psi^{\dag}(\mathbf{x}_{#1},\mathbf{\pi}_{#1})\ket{0}}
\def\ri{\right)}
\def\lef{\left(}
\def\om{\omega}
\renewcommand{\vec}[1]{{\bf #1}}
\def\dst{\displaystyle{\phantom{|}}}
\def\ov{\over\dst}
\def\rightmark{Particle Interferometry from 40 MeV to 40 TeV} 
\def\leftmark{T. Cs\"org\H {o}}
 
\title{Particle Interferometry from 40 MeV to 40 TeV} 
\authors{
{\twerm T. Cs\"org\H {o}
\index{Cs\"org\H {o}, T.}
}\\[2.812mm]
{\normalsize
MTA KFKI RMKI, H-1525 Budapest 114, POB 49, Hungary
\\[0.2ex]
}}
 
\abstract{Recent developments are summarized in the theory of
  Bose--Einstein and Fermi--Dirac correlations, with emphasis on the
  necessity of a simultaneous analysis of particle spectra and quantum
  statistical correlations for a detailed reconstruction of the
  space-time picture of particle emission.  The reviewed topics are as
  follows: basics and formalism of quantum-statistical correlations,
  model-independent analysis of short-range correlations, Coulomb
  wave-function corrections and the core/halo picture for $n$-particle
  Bose--Einstein correlations, the graph rules to calculate these
  correlations even with partial coherence in the core; particle
  interferometry in e$^+$e$^-$ collisions including the
  Andersson--Hofmann model; the invariant Buda--Lund particle
  interferometry; the Buda--Lund, the Bertsch--Pratt and
  Yano--Koonin--Podgoretskii parameterizations, the Buda--Lund hydro
  model and its applications to ($\pi$/K) + p and Pb + Pb collisions at
  CERN SPS, and to low energy heavy ion collisions; the binary source
  formalism and the related oscillations in the two-particle
  Bose--Einstein and Fermi--Dirac correlation functions; the
  experimental signs of expanding rings of fire and shells of fire in
  particle and heavy ion physics and their similarity to planetary
  nebulae in stellar astronomy; the signal of partial restoration of
  the axial $U_A(1)$ symmetry restoration in the two-pion
  Bose--Einstein correlation function; the back-to-back correlations
  of bosons with in-medium mass modifications; and the analytic
  solution of the pion-laser model.}  

\keyword{particle correlations, Bose--Einstein correlations,
        Fermi--Dirac correlations, quantum statistical correlations, 
        Coulomb final-state interactions,
        relativistic heavy ion collisions,
        low and intermediate energy heavy ion collisions,       
        hydrodynamical models, collective flow, 
        quark--gluon plasma, 
        hadron production by electron--positron collision} 
\PACS{27.75.Gz, 25.75.Ld, 25.75.-q, 25.70.-z, 
        24.10.Hz, 12.38.Mh, 13.65.+i\vspace*{6pt}}

\makeindex
\begin{document}
 
\maketitle
\setcounter{page}{1}
\rightline{\small\it ``Imagination is more important 
than knowledge."}\\[4pt]
\rightline{\small\sc A. Einstein}

\section{Introduction}

Although the concept of Bose--Einstein \cite{HBT,HBT56} or intensity
interferometry was discovered in particle and nuclear physics more
than 30 years ago \cite{gglp,gglp0}, some basic questions in the field
are still unanswered, namely, what the form of the Bose--Einstein
correlation functions is, and what this form means. However, even if
the ultimate understanding of the effect is still lacking, the level
of sophistication in the theoretical descriptions and the level of
sophistication in the experimental studies of Bose--Einstein
correlations and particle interferometry has increased drastically,
particularly in the field of heavy ion physics~\cite{QM}.

\subsection{W-mass determination and particle interferometry}
 
The study of Bose--Einstein correlations is interesting in its own
right, but it should be noted that consequences may spill over into
other fields of research, that are seemingly unrelated.  Such is the
topic of the W-mass determination at LEP2, a top priority research in
high energy physics.  It turned out that the non-perturbative
Bose--Einstein correlations between the pions from decaying
W$^+$W$^-$ pairs could be responsible for the presently largest
systematic errors in W-mass determination at
LEP2~\cite{lundww,lundww98}. Hence, the theoretical understanding and
the experimental control of Bose--Einstein correlations at LEP2 is
essential to make a precision measurement of the W mass, which in
turn may carry information via radiative corrections about the value
of the Higgs mass or signals of new physics beyond the Standard Model.

\subsection{Quark--gluon plasma and particle interferometry}

Heavy ion physics is the physics of colliding atomic nuclei.  At the
presently largest energies, the aim of heavy ion physics is to study
the sub-nuclear degrees of freedom by successfully creating and
identifying the quark--gluon plasma (QGP). This presently only
hypothetical phase of matter would consist of freely moving quarks and
gluons, over a volume which is macroscopical relative to the
characteristic 1 fm size of hadrons.

Theoretically proposed signals of the expected phase transition from
hot had\-ronic matter to QGP were tested till now by fixed target
experiments. At AGS, Brook\-haven, collisions were made with nuclei as
big as $^{197}$Au accelerated to\break 14.5~AGeV bombarding energy. At
CERN SPS, collisions were made with 60 and 200 AGeV beams of $^{16}$O
nuclei, 200~AGeV beams of $^{32}$S nuclei, 40 and 158 AGeV beams of
$^{208}$Pb nuclei~\cite{QM}.  The really heavy projectile runs were
made relatively recently, the data are being published and the
implications of the new measurements are explored theoretically, with
claims of a possible QGP production at CERN SPS Pb + Pb reactions,
however, without a clear-cut experimental proof of the identification
of the new phase~\cite{QM}.  Both at CERN and at BNL, new collider
experiments are planned and being constructed. The Relativistic Heavy
Ion Collider (RHIC) at Brookhaven will collide $100 + 100 $ AGeV
$^{197}$Au nuclei, which yields about 40 TeV total energy in the
center of mass frame.  RHIC started to deliver its first results in
2000.  The construction stage of the RHIC accelerator rings was
declared to be complete by the US Department of Energy on August 14,
1999, during a NATO Advanced Study Institute in Nijmegen, The
Netherlands, where the material of this review paper has been
presented.  The forthcoming Large Hadron Collider (LHC) at CERN is
scheduled to start in 2005. LHC will collide nuclei up to $^{208}$Pb
with 2.76 + 2.76 ATeV bombarding energy, yielding a total energy of
1150 TeV in the center of mass frame.  The status quo has been
summarized recently in Refs
\cite{harris-muller,muller-qm99,zajc,jacak-heinz,weiner_rep,uli_urs}.

At such large bombarding energies, the sub-nuclear structure of matter
is expected to determine the outcome of the experiments.  However, the
observed single particle spectra and two-particle correlations
indicated rather simple dependences on the transverse mass of the
produced particles~\cite{na44-slopes,na44-hbt}, that had a natural
explanation in terms of hydrodynamical parameterizations.  Although
hydrodynamical type of models are also able to fit the final hadronic
abundances, spectra and correlations,~\cite{muller-qm99} these models
are not able to describe the ignition part of the process, thus their
predictions are dependent on the assumed initial state. The hydro
models come in two classes: {\it i)} hydro {\it parameterizations},
that attempt to parameterize the flow, temperature and density
distributions on or around the freeze-out
hypersurface~\cite{nr,1d,3d,3d-qm,mpd95,schneder,uli_s,uli_l,nix} by
fitting the observed particle spectra and correlations, for
example~\cite{3d-qm,3d-s96,ster-cf98,ster-beier,ster-qm99}, but
without solving the time-dependent (relativistic) hydrodynamical
equations. The class {\it ii)} comes in the form of hydrodynamical
{\it solutions}, that assume an equation of state and an initial
condition, and follow the time evolution of the hydrodynamical system
until a freeze-out hypersurface.  These are better substantiated but
more difficult to fit calculations, than class {\it i)} type of
parameterizations.  The exact hydro solutions are obtained either in
analytical forms,
\cite{landau,milekhin,ZBG,bo-hydro,bjorken,hama-nava,raju_0,cspeter,sol,biro},
or from numerical solutions, see for example
Refs \cite{csernai,raju_1,hylander,rischke_plot}.  An even more
substantiated approach is hydrodynamical approach with continuous
emission of particles, which takes into account the small sizes of
heavy ion reactions as compared to the mean free path of the
particles~\cite{cem}. Such a continuous emission of hadrons during the
time evolution of the hot and dense hadronic matter is supported by
microscopic simulations~\cite{bravina-freez}.

In principle, the exact hydrodynamical solutions 
can be utilized in a time-reversed form: after fixing  
the parameters to describe the measured particle spectra 
and correlations at the time when the particles are produced,
the hydro code can be followed backwards in time, and one may learn
about the {\it initial condition}~\cite{gyulas-summ} in a given reaction:
was it a QGP or a conventional hadron gas initial state?

\subsection{Basics of quantum statistical correlations}

Essentially, intensity correlations appear due to the Bose--Einstein
or Fermi--Dirac symmetrization of the two-particle final states of
identical bosons or fermions, in short, due to quantum statistics.
        
The simplest derivation is as follows: suppose that a particle pair is
observed, one with momentum $k_1$ the other with momentum $k_2$. The
amplitude has to be symmetrized over the unobservable variables, in
particular over the points of emissions $x_1$ and $x_2$.  If Coulomb,
strong or other final-state interactions can be neglected, the
amplitude of such a final state is proportional to 
\be A_{12} \propto
\frac{1}{\sqrt{2}} \,\, [\, {\rm e}^{i k_1 x_1 + i k_2 x_2} \pm {\rm
  e}^{i k_1 x_2 + i k_2 x_1} \,], 
\ee 
where $+$ sign stands for bosons, $-$ for fermions.  If the particles
are emitted in an incoherent manner, the observable two-particle
spectrum is proportional to
\be 
N_2(k_1,k_2) \propto \int dx_1 \rho(x_1) \int dx_2
\rho(x_2) \,\, |A_{12}|^2 
\ee 
and the resulting two-particle intensity correlation function is 
\be 
C_2(k_1,k_2) =
\frac{N_2(k_1,k_2)}{N_1(k_1) N_2(k_2)} \, = 1 \pm |\tilde \rho(k_1 -
k_2) |^2 \label{e:rhoq} 
\ee 
that carries information about the Fourier-transformed space-time
distribution of the particle emission 
\be \tilde\rho(q) = \int dx \,\,
{\rm e}^{i q x} \,\, \rho(x)
\ee 
as a function of the relative momentum $q = k_1 - k_2$.
        
As compared to the idealized case when quantum-statistical
correlations are negligible (or neglected), Bose--Einstein or
Fermi--Dirac correlations modify the momentum distribution of the
hadron pairs in the final state by a weight factor $ \langle 1 \pm \rm
\cos[ (k_1 - k_2) \cdot (x_1 - x_2) ] \rangle$.

\subsection{Correlations between particle and heavy ion physics}
        
In case of pions, that are produced abundantly in relativistic heavy
ion experiments, Bose--Einstein symmetrization results in an
enhancement of correlations of pion pairs with small relative
momentum, and the correlation function carries information about the
space-time distribution of pion production points.  This in turn is
expected to be sensitive to the formation of a transient quark--gluon
plasma stage~\cite{bertsch}.

In particle physics, reshuffling or modification of the momentum of
pions in the fully hadronic decays of the W$^+$W$^-$ pairs happens due to
the Bose--Einstein symmetrization of the full final stage, that
includes symmetrization of pions with similar momentum from different
W-s.  As a consequence of this quantum interference of pions, a
systematic error as big as 100 MeV may be introduced to the W-mass
determination from reconstruction of the invariant masses of
$(\rm q\overline{\rm q})$ systems in 4-jet events \cite{lundww,lundww98}.  It
is very difficult to handle the quantum interference of pions from the
W$^+$ and W$^-$ jets with Monte-Carlo simulations, perturbative
calculations and other conventional methods of high energy physics.

Unexpectedly, a number of recent experimental results arose suggesting
that the Bose--Einstein correlations and the soft components of the
single-particle spectra in high energy collisions of elementary
particles show similar features to the same observables in high energy
heavy ion physics \cite{na22-hbt,na22,delphi-radii,l3-radii}.

These striking similarities of multi-dimensional Bose--Einstein
correlations and particle spectra in high energy particle and heavy
ion physics have no fully explored dynamical explanation yet.  This
review intends to give a brief introduction to various sub-fields of
particle interferometry, highlighting those phenomena that may have
applications or analogies in various different type of reactions. The
search for such analogies inspired a study of non-relativistic heavy
ion reactions in the 30 -- 80 AMeV energy domain and a search for new
exact analytic solutions of fireball hydrodynamics, reviewed briefly
for a comparison.

As some of the sections are mathematically more advanced, and other
sections deal directly with data analysis, I attempted to formulate
the various sections so that they be self-standing as much as
possible, and be of interest for both the experimentally and the
theoretically motivated readers.
%

\section{Formalism}
The basic properties of the Bose--Einstein $n$-particle correlation
functions (BECF-s) can be summarized as follows, using only the
generic aspects of their derivation.

The $n$-particle Bose--Einstein correlation function is defined as
\be 
        C_n(\bk_1, \cdots , \bk_n ) = {\dst N_n(\bk_1, \cdots , \bk_n) 
        \ov  N_1(\bk_1) \cdots N_1(\bk_n) }, 
\label{e:cndef}
\ee where $N_n(\bk_1, \cdots , \bk_n) $ is the $n$-particle inclusive
invariant momentum distribution, while \be N_n(\bk_1, \cdots , \bk_n )
= \frac{1}{\sigma} E_{\bk_1} \cdots E_{\bk_n}
\frac{d^{3n}\sigma}{d{\bk}_1 \cdots d{\bk}_n} \ee is the invariant
$n$-particle inclusive momentum distribution.  It is quite remarkable
that the complicated object of Eq.~(\ref{e:cndef}) carries quantum
mechanical information on the phase-space distribution of particle
production as well as on possible partial coherence of the source, can
be expressed in a relatively simple, straight-forward manner both in
the analytically solvable pion-laser model of
Refs \cite{pratt,zhang-clarif,cstjz,jzcst} as well as in the generic
boosted-current formalism of Gyulassy, Padula and
collaborators~\cite{gyu-ka,gyu-pa-ga,pa-gyu} as \be C_n(\bk_1, \cdots
, \bk_n ) = {\dst \sum_{\sigma^{(n)}} \prod_{i = 1}^n
  G(\bk_i,\bk_{\sigma_i}) \ov \prod_{i=1}^n G(\bk_i,\bk_i) },
\label{e:m2} \ee where $\sigma^{(n)}$ stands for the set of
permutations of indices $(1, 2, \cdots, n)$ and $\sigma_i $ denotes
the element replacing element $i$ in a given permutation from the set
of $\sigma^{(n)}$, and, regardless of the details of the two different
derivations, \be G(\bk_i,\bk_j) = \sqrt{E_{\bk_i} E_{\bk_j}} \langle
a^{\dagger}(\bk_i) a(\bk_j) \rangle \label{e:gdef} \ee stands for the
expectation value of $a^{\dagger}(\bk_i) a(\bk_j) $.  The operator
$a^{\dagger}({\bk})$ creates while operator $a({\bk})$ annihilates a
boson with momentum ${\bk}$.  The quantity $G(\bk_i,\bk_j)$
corresponds to the first order correlation function in the terminology
of quantum optics.  In the boosted-current formalism, the derivation
of Eq.~(\ref{e:m2}) is based on the assumptions that {\it i)} the
bosons are emitted from a semi-classical source, where currents are
strong enough so that the recoils due to radiation can be neglected,
{\it ii)} the source corresponds to an incoherent random ensemble of
such currents, as given in a boost-invariant formulation in
Ref.\ \cite{gyu-pa-ga}, and {\it iii)} that the particles propagate as
free plane waves after their production.  Possible correlated
production of pairs of particles is neglected here.  Note also the
recent clarification of the proper normalization of the two-particle
Bose--Einstein correlations~\cite{Miskowiec:1997ay}.

A formally similar result is obtained when particle production happens
in a correlated manner, generalizing the results of
Refs \cite{zhang-clarif,cstjz,jzcst,brood,zhang-n}.  Namely, the
$n$-particle {\it exclusive } invariant momentum distributions of the
pion-laser model read as 
\be N_n^{(n)} ( \bk_1, \cdots , \bk_n ) =
\sum_{\sigma^{(n)}} \prod_{i = 1}^n G_1(\bk_i, \bk_{\sigma_i}), 
\ee
with 
\be G_1(\bk_i \bk_j) = \sqrt{E_{\bk_i} E_{\bk_j}}
\mbox{Tr} \{\hat \rho_1 a^{\dagger}(\bk_i) a(\bk_j)\},
\label{e:bas1}
\ee
where $\hat \rho_1$ is the single-particle density matrix in the limit
when higher-order Bose--Einstein correlations are negligible.
Q.H. Zhang has shown~\cite{zhang-n}, that the $n$-particle
inclusive spectrum has a similar structure: 
\bea
        N_n
        ( \bk_1, \cdots , \bk_n ) & = & 
        \sum_{\sigma^{(n)}} \prod_{i = 1}^n G(\bk_i,\bk_{\sigma_i})\,,\\
         G(\bk_i,\bk_j) & = & \sum_{n=1}^\infty G_n(\bk_i,\bk_j) 
\label{e:bas}\,.
\eea
This result, valid only if the density of pions is below a critical
value~\cite{jzcst}, was obtained if the multiplicity distribution was
assumed to be a Poissonian one in the rare gas limit.  The formula of
Eq.~(\ref{e:bas}) has been generalized by Q.H. Zhang in
Ref.\ \cite{zhang-ngen} to the case when the multiplicity distribution
in the rare gas limit is arbitrary.

The functions $G_n(\bk_i,\bk_j)$ can be considered as representatives
of order $n$ symmetrization effects in exclusive events where the
multiplicity is fixed to $n$, see
Refs \cite{pratt,zhang-clarif,cstjz,jzcst} for more detailed
definitions.  The function $G(\bk_i,\bk_j)$ can be considered as the
expectation value of $a^{\dagger}(\bk_i) a(\bk_j) $ in an inclusive
sample of events, and this building block includes all the higher-order 
symmetrization effects.  In the relativistic Wigner function
formalism, in the plane wave approximation $G(\bk_1,\bk_2)$ can be
rewritten as 
\bea 
G(\bk_1,\bk_2) & \equiv & \tilde S(q_{12},K_{12}) \,
= \, \int d^4 x \,S(x,K_{12})\, \exp(i q_{12}\cdot x)\,,
                \label{e:gwig}\\
        K_{12} & = & 0.5 (k_1 + k_2)\,, \\
        q_{12} & = & k_1 - k_2\,,
\eea
where a four-vector notation is introduced, $k = (E_{\bk}, \bk)$,
and the energy of quanta with mass $m$ is given by $E_{\bk} =
\sqrt{m^2 + \bk^2}$, the mass-shell constraint.  Notation $a \cdot b$
stands for the inner product of four-vectors.  In the following, the
relative momentum four-vector shall be denoted also as $\Delta k = q =
(q_0, q_x, q_y, q_z) = (q_0, {\bq})$, the invariant relative
momentum is $Q = \sqrt{- q\cdot q}$.
       
The covariant Wigner transform of the source density matrix,
$S(x,{\bk})$ is a quantum-mechanical analogue of the classical
probability that a boson is produced at a given $(x,k)$ point in the
phase-space, where $ x = (t, {\br}) = (t,r_x,r_y,r_z)$.  The
quantity $S(x,K_{12})$ corresponds to the off-shell extrapolation of
$S(x,{\bk})$, as $K_{12}^0 \ne \sqrt{m^2 + {\bf K}_{12}^2}$.
Fortunately, Bose--Einstein correlations are non-vanishing at small
values of the relative momentum $q$, where $K_{12}^0 \simeq 
E_{{\bf K}_{12}}$.  Due to the mass-shell constraints, $G$ depends only on 6
independent momentum components.

For the two-particle Bose--Einstein correlation function,
Eqs~(\ref{e:m2},\ref{e:gdef},\ref{e:gwig}) yield the following
representation:
\be
        C_2({\bk}_1,{\bk}_2)  =  1 + 
                \frac{|\tilde S(q_{12},K_{12})|^2 }
                {\tilde S(0,{\bk}_1)\, \tilde S(0,{\bk}_2)}.
        \label{e:c2-wig}
\ee
Due to the unknown off-shell behavior of the Wigner functions, it is
rather difficult to evaluate this quantity from first principles, in a
general case.

When comparing model results to data, two kind of simplifying
approximations are frequently made:

{\it i) The on-shell approximation} can be used for
developing Bose--Einstein afterburners to Monte-Carlo event
generators, where only the on-shell part of the phase-space is
modeled. In this approximation, Eq.~(\ref{e:c2-wig}) is evaluated
with the on-shell mean momentum, $\tilde K = (\sqrt{m^2 + {\bf
    K}_{12}^2}, {\bf K}_{12})$.  This {on-shell approximation} was
used e.g.\ in Ref.\ \cite{pratt_csorgo} to sample $S(x,\tilde K )$ from
the single-particle phase-space distribution given by Monte-Carlo
event generators, and to calculate the corresponding Bose--Einstein
correlation functions in a numerically efficient manner.  The method
yields a straightforward technique for the inclusion of Coulomb and
strong final-state interactions as well, see e.g.\
Ref.\ \cite{pratt_csorgo}.
        
{\it ii) The smoothness approximation} can be used when
describing Bose--Einstein correlations from a theoretically
parameterized model, e.g.\ from a hydrodynamical calculation.  In this
case, the analytic continuation of $S(x,{\bk})$ to the off-shell
values of $K$ is providing a value for the off-shell Wigner function
$S(x,K_{12})$.  However, in the normalization of Eq.~(\ref{e:c2-wig}),
the product of two on-shell Wigner functions appear.  In the
smoothness approximation, one evaluates this product as a leading
order Taylor series in $q$ of the exact expression $\tilde S(0,{\bf K}
- {\bq}/2)S(0,{\bf K} + {\bq}/2)$.  The resulting formula, 
\be
C_2({\bk}_1,{\bk}_2) = 1 + \frac{|\tilde S(q_{12},K_{12})|^2 }
{|\tilde S(0,K_{12})|^2},
        \label{e:c2-wig-smooth}
\ee
relates the two-particle Bose--Einstein correlation function to the
Fourier-trans\-formed off-shell Wigner function $S(x,K)$. This provides
an efficient analytic or numeric method to calculate the BECF from
sources with known functional forms.  The correction terms to the
smoothness approximation of Eq.~(\ref{e:c2-wig-smooth}) are given in
Ref.\ \cite{uli_l}. These corrections are generally on the 5\% level
for thermal like momentum distributions.

\section{Model-Independent Analysis of Short-Range Correlations}

\def\bk{{\bf k}}
\def\vp{{\bf k}}
\def\vq{{\bf q}} 
\def\vk{{\bf k}}
\def\vK{{\bf K}}
\def\vx{{\bf x}}
\def\vy{{\bf y}}
\def\uk{{|{\bf k}|}}
\def\De{\Delta\eta}
\def\Des{\Delta\eta_*}
\def\Dk{\Delta k}
\def\Dt{\Delta\tau} 
\def\Dy{\Delta y}
\def\t0{\tau_0}
\def\tl{\tau_L}
\def\ch{\cosh}
\def\sh{\sinh}
\def\bea{\begin{eqnarray}}
\def\eea{\end{eqnarray}}
\def\I#1{\int d^3#1}
\def\bx{{\bf{x}}}
\def\bp{{\bf{p}}}
\def\bk{{\bf{k}}}
\def\bK{{\bf{K}}}
\def\bK{{\bf{K}}}
\def\bQ{{\bf{Q}}}
\def\bpi{{\bf{\pi}}}
\def\bxi{{\bf{\xi}}}
\def\bq{{\bf{q}}}
\def\br{{\bf{r}}}
\def\axd{\hat{ a}^{\dag} (\bx)}
\def\apd{\hat a^{\dag} (\bp)}
\def\ax{\hat{ a}^{} (\bx)}
\def\ap{\hat  a^{} (\bp)}
\def\ri{\right)}
\def\lef{\left(}
\def\dst{\displaystyle\phantom{|}}
\def\ov{\over\dst}
\def\om{\omega}
\def\eps{\epsilon}
\def\bak{{\bf K}}
\def\dek{{\bf \Delta k}}
\def\xb{{\overline{x}}}
\def\tb{{\overline{t}}}
\def\rb{{\overline{r}}}
\def\nb{{\overline{n}}}
\def\etab{{\overline{\eta}}}
\def\taub{{\overline{\tau}}}
%
%
\def\D{\Delta}
\def\De{\Delta\eta}
\def\t{{\tau}}
\def\ch{\cosh}
\def\sh{\sinh}
\def\ben{\begin{eqnarray}}
\def\enn{\end{eqnarray}}
\def\bea{\begin{eqnarray}}
\def\eea{\end{eqnarray}}
\def\be{\begin{equation}}
\def\ee{\end{equation}}
\def\ov{\over\displaystyle\strut}
\def\dst{\displaystyle\phantom{|}}
\def\l({\left(}
\def\r){\right)}
\def\o{{out}}
\def\s{{side}}
\def\e{{\eta}}
\def\bdk{{\bf \Delta k}}
\def\rl{R_l^2}
\def\ro{R_{o}^2}
\def\rs{R_{s}^2}
\def\rol{R_{ol}^2}
\def\rpa{R_{\parallel}^2 }
\def\rpe{R_{\perp}^2 }
\def\rta{R_{\tau}^2 }
\def\BL{{Buda}{--}{Lund}~} 

Can one {\it model-independently} characterize the shape of
two-particle correlation functions?  Let us attempt to answer this
question on the level of statistical analysis, without theoretical
assumptions on the thermal or non-thermal nature of the particle
emitting source. In this approach, the usual theoretical assumptions
are {\it not} made, neither on the presence or the negligibility of
Coulomb and other final-state interactions, nor on the presence or the
negligibility of a coherent component in the source, nor on the
presence or the negligibility of higher-order quantum statistical
symmetrization effects, nor on the presence or the negligibility of
dynamical effects (e.g.\ fractal structure of gluon-jets) on the
short-range part of the correlation functions.  The presentation
follows the lines of Ref.\ \cite{lagu}.  The reviewed method is {\it
  really} model-independent, and it can be applied not only to
Bose--Einstein correlation functions but to every experimentally
determined function, which features the properties {\it i)} and {\it
  ii)} listed below.

The following {\it experimental properties} are assumed:

{\it i) } The measured function tends to a constant for large values
of the relative momentum.

{\it ii)} The measured function has a non-trivial structure at a
certain value of its argument.

The location of the non-trivial structure in the correlation function
is assumed for simplicity to be close to $Q = 0$.

The properties {\it i)} and {\it ii)} are well satisfied by e.g.\ the
conventionally used two-particle Bose--Einstein correlation functions.
For a critical review on the non-ideal features of short-range
correlations, (e.g.\ non-Gaussian shapes in multi-dimensional
Bose--Einstein correlation studies), we recommend Ref.\ \cite{kittel}.

The core/halo intercept parameter $\lambda_*$ is defined as the {\it
  extrapolated} value of the two-particle correlation function at $Q =
0$, see Section~\ref{s:chalo} for greater details. It turns out that
$\lambda_*$ is an important physical observable, related to the degree
of partial restoration of $U_A(1)$ symmetry in hot and dense hadronic
matter~\cite{vck,ckv}, as reviewed in Section~\ref{s:ua1}.
\vfill\pagebreak

Various non-ideal effects due to detector resolution, binning,
particle mis-iden\-ti\-fi\-cation, resonance decays, details of the Coulomb
and strong final-state interactions etc.\ may influence this parameter
of the fit. One should also mention, that if all of these difficulties
are corrected for by the experiment, the extrapolated intercept
parameter $\lambda_*$ for like-sign charged bosons is (usually) not
larger, than unity as a consequence of quantum statistics for chaotic
sources, even with a possible admixture of a coherent component.
However, final-state interactions, fractal branching processes of
gluon jets, or the appearance of one-mode or two-mode squeezed
states~\cite{ac,acg} in the particle emitting source might provide
arbitrarily large values for the intercept parameter.

A really model-independent approach is to expand the measured
correlation functions in an abstract Hilbert space of functions.  It
is reasonable to formulate such an expansion so that already the first
term in the series be as close to the measured data points as
possible. This can be achieved if one identifies~\cite{lagu,edge-cst}
the approximate shape (e.g.\ the approximate Gaussian or the
exponential shape) of the correlation function with the abstract
measure $\mu(t)dt$ in the abstract Hilbert-space ${\cal H}$.  The
orthonormality of the basis functions $\phi_n(t)$ in ${\cal H}$ can be
utilized to guarantee the convergence of these kind of expansions, see
Refs \cite{lagu,edge-cst} for greater details.

\subsection{Laguerre expansion and exponential shapes
\label{s:lagu}}

If in a zeroth order approximation the correlation function has an
exponential shape, then it is an efficient method to apply the
Laguerre expansion, as a special case of the general formulation of
Refs \cite{lagu,edge-cst}:
\bea
C_2(Q)\! & =& \!{\cal N} \left\{ 
        1 + \lambda_L \exp(- Q R_L) 
        \left[ 1 + c_1 L_1(QR_L) + \frac{c_2}{2!} L_2(Q R_L) + ... \right]
        \right\}. 
        \nonumber \\
        && \label{e:laguerre}
\eea
In this and the next subsection, $Q$ stands symbolically for any,
experimentally chosen, one dimensional relative momentum variable.
The fit parameters are the scale parameters ${\cal N}$, $\lambda_L$,
$R_L$ and the expansion coefficients $c_1$, $c_2$, ... .  The
$n$th order Laguerre polynomials are defined as
\bea
        L_n(t) & = & \exp(t) \frac{d^n}{dt^n} t^n \exp(-t), 
\eea
they form a complete orthogonal basis for an exponential measure as
\bea
        \delta_{n,m} & \propto & 
        \int_0^{\infty} dt \, \exp(-t) L_n(t) L_m(t).
\eea
The first few Laguerre polynomials are explicitly given as
\bea
        L_0(t) & = & 1, \\
        L_1(t) & = & t  - 1,\\
        L_2(t) & = & t^2 - 4t + 2, \, ... \, .
\eea

As the Laguerre polynomials are non-vanishing at the origin, $C(Q = 0)
\ne 1 + \lambda_L$.  The physically significant core/halo intercept
parameter $\lambda_*$ can be obtained from the parameter $\lambda_L$
of the Laguerre expansion as
\bea
        \lambda_* & = & \lambda_L [1 - c_1 + c_2 - ... ] . 
\eea

\subsection{Edgeworth expansion and Gaussian shapes
\label{s:edge}}
If, in a zeroth-order approximation, the correlation function has a
Gaussian shape, then the general form given in Ref.\ \cite{edge} takes
the particular form of the Edgeworth
expansion~\cite{edge-cst,edge,edge0} as:
\bea
C(Q) & = &{\cal N} \left\{ 
        1 + \lambda_E \exp( - Q^2 R_E^2) 
                \right. \times \nonumber \\
                && 
        \left. 
        \left[ 1 + \frac{\kappa_3}{3!} H_3(\sqrt{2} Q R_E)
                +\frac{\kappa_4}{4!} H_4(\sqrt{2} Q R_E) + ... \right]
                \right\} .
                \label{e:edge}
\eea
The fit parameters are the scale parameters ${\cal N}$, $\lambda_E$,
$R_E$, and the expansion coefficients $\kappa_3$, $\kappa_4$,  ... 
 that coincide with the cumulants of rank 3, 4, ... of the
correlation function.  The Hermite polynomials are defined as
\bea
        H_n(t) & = & \exp( t^2/2) \left( - \frac{d}{dt} \right)^n
                \exp(-t^2/2), 
\eea
they form a complete orthogonal basis for a Gaussian measure as
\bea
        \delta_{n,m}  & \propto & 
        \int_{-\infty}^{\infty} dt 
        \, \exp(-t^2/2) H_n(t) H_m(t).
\eea
The first few Hermite polynomials are listed as
\bea
        H_1(t) & = & t, \\
        H_2(t) & = & t^2 -1, \\
        H_3(t) & = & t^3 - 3 t , \\
        H_4(t) & = & t^4 - 6 t^2 + 3,\, ... \,
\eea
The physically significant core/halo intercept parameter $\lambda_*$
can be obtained from the Edgeworth fit of Eq.~(\ref{e:edge}) as
\bea
\lambda_* & = & \lambda_E \left[ 1 + \frac{\kappa_4}{8} + ... \right].
\eea
This expansion technique was applied in the conference
contributions~\cite{edge-cst,edge} to the AFS minimum bias and 2-jet
events to characterize successfully the deviation of data from a
Gaussian shape.  It was also successfully applied to characterize the
non-Gaussian nature of the correlation function in two-dimensions in
case of the preliminary E802 data in Ref.\ \cite{edge-cst}, and it was
recently applied to characterize the non-Gaussian nature of the
three-dimensional two-pion BECF in e$^+ +$ e$^-$ reactions at
LEP1~\cite{l3-radii}.

\begin{figure}[h]
\vspace*{6pt}
\centerline{
\includegraphics[width=11cm,bb=35 220 530 615,clip]{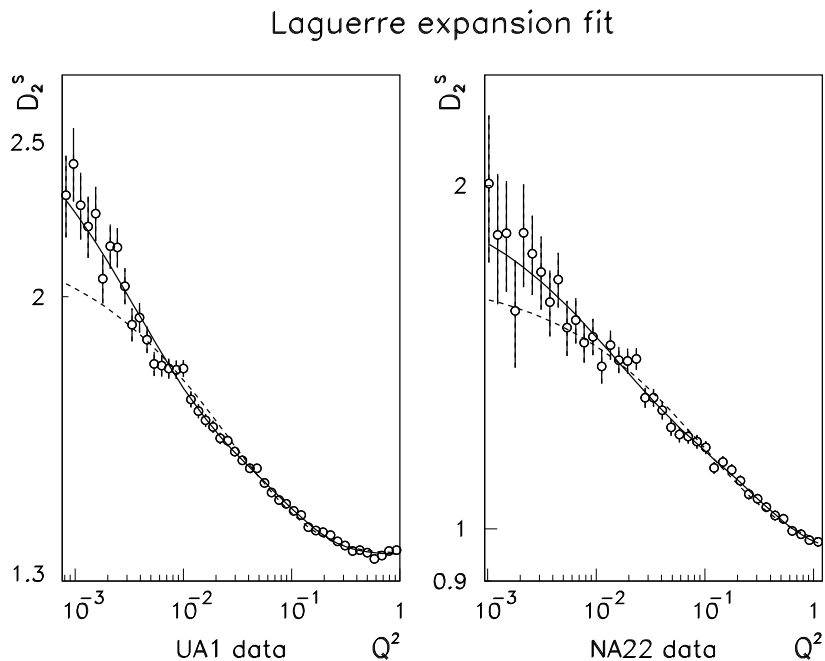}}
\vspace*{-12pt}
\caption{Laguerre expansion of NA22 and UA1 short-range correlations
  $D_2^s$ is shown by the solid line.  Dashed line stands for the best
  exponential fit, which clearly underestimates the strength of the
  measured points at low values of the squared invariant momentum
  difference $Q^2=-(k_1 - k_2)^2$. (Note the logarithmic horizontal
  and vertical scales.)}
\label{f:lagu}
\end{figure}

Figure \ref{f:lagu} indicates the ability of the Laguerre expansions to
characterize two well-known, non-Gaussian correlation
functions~\cite{lagu}: the second-order short-range correlation
function $D^s_2(Q)$ as determined by the UA1 and the NA22
experiments~\cite{ua1,na22-pwl}.  The convergence criteria of the
Laguerre and the Edgeworth expansions is given in Ref.\ \cite{lagu}.

\begin{table}[hbt]
\vspace*{-12pt}
\caption{Laguerre fits to UA1 and NA22 two-particle correlations}
\label{t:lagu}
\vspace*{-6pt}
\begin{center}
\begin{tabular}{ccc}
\hline\\[-10pt] 
\null & UA1 & NA22 \\ 
\hline\\[-10pt]
${\cal N}$    &  1.355 $\pm$ 0.003    & 0.95  $\pm$ 0.01  \\
$\lambda_{L}$ &  1.23  $\pm$ 0.07     & 1.37  $\pm$ 0.10  \\
$R_L$ [fm]    &  2.44  $\pm$ 0.12     & 1.35  $\pm$ 0.14  \\
$c_1$         &  0.52  $\pm$ 0.03     & 0.63  $\pm$ 0.06  \\
$c_2$         &  0.45  $\pm$ 0.04     & 0.44  $\pm$ 0.06  \\
\hline\\[-10pt]
$\chi^2/NDF$     & {41.2/41 = 1.01} 
                 & {20.0/34 = 0.59}  \\
\hline
\end{tabular}
\end{center}
\end{table}

From Table~\ref{t:lagu} the core/halo model intercept parameter is
obtained as $\lambda_* = 1.14 \pm 0.10$ (UA1) and $\lambda_* = 1.11
\pm 0.17$ (NA22). As both of these values are within errors equal to
unity, the maximum of the possible value of the intercept parameter
$\lambda_*$ in a fully chaotic source, we conclude that either there
are other than Bose--Einstein short-range correlations observed by
both collaborations, or the full halo of long lived resonances is
resolved in case of this measurement~\cite{chalo,dkiang,sk,bialas}.

If the two-particle BECF can be factorized as a product of (two or
more) functions of one variable each, then the Laguerre and the
Edgeworth expansions can be applied to the multiplicative factors ---
functions of one variable, each. This method was applied recently to
study the non-Gaussian features of multi-dimensional Bose--Einstein
correlation functions e.g.\ in Refs \cite{l3-radii,edge}.  The full,
non-factorized form of two-dimensional Edgeworth expansion and the
interpretation of its parameters is described in the handbook on
mathematical statistics by Kendall and Stuart~\cite{kendallstuart}.



\section{Coulomb Wave Corrections for Higher-Order Correlations}
\label{s:coulomb}

The short-range part of the two- and multi-particle correlation
function of charged particles is strongly effected by Coulomb interactions.
Even in the non-relativistic case, the $n$-body Coulomb scattering
problem is solvable exactly only for the $n = 2$ case, the full 3-body
Coulomb wave-function is unknown. However, when studying higher-order
Bose--Einstein correlations and e.g.\ searching for the onset of
(partial) coherence in the source, it is desired that the
Coulomb-induced correlations be removed from the data.

\begin{figure}[htb]
\vspace*{6pt}
\centerline{
\includegraphics[height=8cm,bb=20 150 550 650,clip]{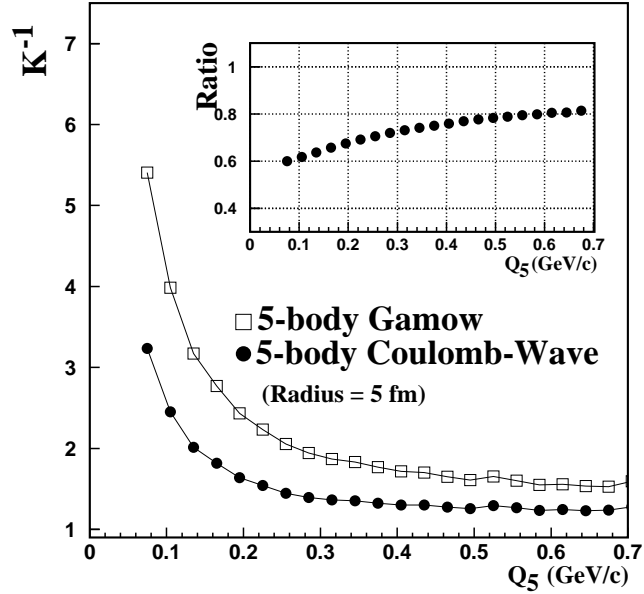}}
\vspace*{-12pt}
  \caption{Coulomb wave-function correction factor and generalized
    Gamow correction factor for 5-particle correlation functions, for
    a Gaussian source with $R_G = 5$~fm}
\label{f:Coulomb5_5}
\vspace*{-6pt}
\end{figure}

In any given frame, the boost-invariant decomposition of
Eq.~(\ref{e:gwig}) can be rewritten into the following, seemingly not
invariant form:
\bea
        G(\bk_1,\bk_2) & = & \int d^3 {\bf x}\,\, S_{{\bf K}_{12}}({\bf x})\,\,
                \exp(i {\bf q}_{12} {\bf x}),
                \label{e:gwignr}\\
        S_{{\bf K}_{12}}({\bf x}) 
        & = & \int dt\, \exp(- i  {\mbox {\boldmath $\beta$}}_{K_{12}} 
        {\bf q}_{12} t)\,\, S( {\bf x}, t, K_{12}), \label{e:Snr}\\
        {\mbox {\boldmath $\beta$}}_{K_{12}} & = & 
        ({\bf k}_1 + \bk_2) /(E_1 + E_2). \label{e:betak}
\eea
Note that the relative source function $S_{{\bf K}_{12}}({\bf x})$
reduces to a simple time integral over the source function $S(x,K)$ in
the frame where the mean momentum of the pair (hence the pair velocity
${\mbox {\boldmath $\beta$}}_{K_{12}}$) vanishes.

Based on a Poisson cluster picture, the effect of multi-particle
Coulomb final-state interactions on higher-order intensity
correlations is determined in general in Ref.\ \cite{alt-n}, with the
help of a scattering wave function which is a solution of the $n$-body
Coulomb Schr\"odinger equation in (a large part of) the asymptotic
region of the $n$-body configuration space.
        
If $n$ particles are emitted with similar momenta, so that their
$n$-particle Bose--Einstein correlation functions may be non-trivial,
Eqs~(\ref{e:gwignr}--\ref{e:betak}) form the basis for evaluation of
the Coulomb and strong final-state interaction effects on the
observables for any given cluster of particles, assuming that the
relative motion of the particles is non-relativistic within the
cluster, see Ref.\ \cite{alt-n}.  The Coulomb correction factor
$K^{-1}$ can be integrated for arbitrary large number of particles and
for any kind of model source, by replacing the plane wave
approximation with the approximate $n$-body Coulomb wave-function.  In
the limit of vanishing source sizes, the generalization of the Gamow
penetration factor was obtained to the correlation function of
arbitrary large number of particles~\cite{alt-n}.  In particular,
Coulomb effects on the $n$-particle Bose--Einstein correlation
functions of similarly charged particles were studied for Gaussian
effective sources, for $n = 3$ in Ref.\ \cite{alt-3} and for $n=4$ and
5 in Ref.\ \cite{alt-n}. For the typical $R=1$ fm effective source
sizes of the elementary particle reactions, the generalized $n$-body
Gamow penetration factor gave rather precise estimates of the Coulomb
correction (within 5\% from the Coulomb-wave correction). In
contrast, for typical effective source sizes observed in high energy
heavy ion reactions, Fig.\ \ref{f:Coulomb5_5} indicates that the new
Coulomb wave-function integration method allows for a removal of a
systematic error as big as 100\% from higher-order multi-particle
Bose--Einstein correlation functions.  See Ref.\ \cite{alt-n} for
greater details.


\section{Core/Halo Picture of Bose--Einstein Correlations}
\label{s:chalo}

%
%
%
%
\def\D{\Delta}
\def\l({\left(}
\def\r){\right)}
\def\ts{\tilde s_c}
\def\ave#1{\langle #1 \rangle}

The core/halo model~\cite{chalo,halo2,halo3,nhalo,pcoh} deals with the
consequences of a phenomenological situation, when the boson source
can be considered to be a superposition of a central core surrounded
by an extended halo.  In the forthcoming sections, final-state
interactions are neglected, we assume that the data are corrected for
final-state Coulomb (and possibly strong) interactions.

Bose--Einstein correlations are measured at small relative momenta of
particle pairs. In order to reliably separate the near-by tracks of
particle pairs in the region of the Bose enhancement, each experiment
imposes a cut-off $Q_{\rm min}$, the minimum value of the resolvable
relative momentum. The value of this cut-off may vary slightly from
experiment to experiment, but such a cut-off exists in each
measurement.

In the core/halo model, the following assumptions are made:

{\it Assumption 0:} The emission function does not have a
no-scale, power-law-like structure.  This possibility was discussed
and related to intermittency and effective power-law shapes of the
two-particle Bose--Einstein correlation functions in
Ref.~\cite{bialas}.

{\it Assumption 1:} The bosons are emitted either from a
{\it central} part or from the surrounding {\it halo}. Their emission
functions are indicated by $S_c(x,{\bf k})$ and $S_h(x,{\bf k})$,
respectively.  According to this assumption, the complete emission
function can be written as \bea S(x,{\bf k}) = S_c(x,{\bf k}) +
S_h(x,{\bf k}), \label{e:ax} \eea and $S(x,{\bf k})$ is normalized to
the mean multiplicity, $\int d^4x {\dst (d{\bf k}/E)} S(x,{\bf k}) \,
= \, \ave{n}$.

{\it Assumption 2:} The emission function that characterizes
the halo is assumed to change on a scale $R_h$ that is larger than
$R_{\rm max}\approx \hbar / Q_{\rm min}$, the maximum length-scale
resolvable~\cite{chalo} by the intensity interferometry microscope.
The smaller central core of size $R_c$ is assumed to be resolvable,
\bea 
R_h > R_{\rm max} > R_c\,.  
\eea 
This inequality is assumed to be satisfied by all characteristic
scales in the halo and in the central part, e.g.\ in case the side, out
or longitudinal components~\cite{bertsch,pratt-bp} of the correlation
function are not identical.

{\it Assumption 3:} The core fraction $f_c({\bf k}) =
N_c({\bf k})/N_1({\bf k})$ varies slowly on the relative momentum
scale given by the correlator of the core~\cite{nhalo}.

\begin{figure}[htb]
\vspace*{6pt}
\centerline{
\includegraphics[height=7cm]{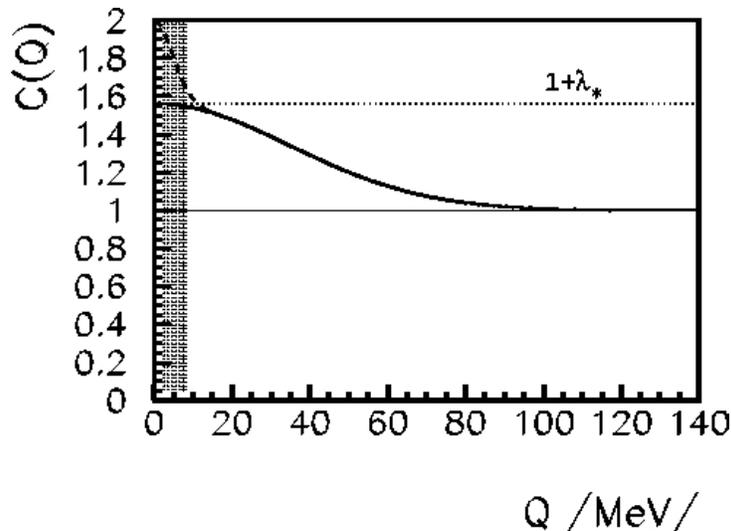}}
\vspace*{-12pt}
\caption{
  The shape of the BECF is illustrated for a source containing a core
  and a large halo. The contribution from the halo is restricted to
  the shaded area, while the shape of the BECF outside this interval
  is determined completely by the contribution of the core. If the
  resolution for a given experiment is restricted to $Q > 10 $ MeV,
  then an effective and momentum dependent intercept parameter,
  $\lambda_*(y,m_t)$ will be measured, which can be combined with the
  measured momentum distribution to determine the the momentum
  distribution of the particles emitted directly from the core.}
\label{f:halo_f1}
\end{figure}

The emission function of the core and the halo are normalized
as
\bea
\int d^4x {\dst d{\bf k}\ov E} S_c(x,{\bf k}) \, = \, \ave{n}_c
& \mbox{\rm and}&
\int d^4x {\dst d{\bf k}\ov E} S_h(x,{\bf k}) \, = \, \ave{n}_h.
\eea
One finds~\cite{chalo,nhalo} that
\bea
 N_1({\bf k}) \, =  \, N_c({\bf k}) + N_h({\bf k})
& \mbox{\rm and}&
\ave{n} \, = \, \ave{n}_c + \ave{n}_h.
\eea
\vfill\pagebreak\noindent
Note that in principle the core as well as the halo part of
the emission function could be decomposed into more detailed contributions,
e.g.\
\bea
\vspace*{-6pt}        
S_{h}(x,{\bf k}) & = &  \sum_{r=\omega,\eta,\eta^{\prime},K^0_S}
                S^{(r)}_{\rm halo}(x,{\bf k}).
\vspace*{-6pt}
\eea
In case of pions and NA44 acceptance, the $\omega$ mesons were shown
to contribute to the halo, Ref.\ \cite{sk}.  For the present
considerations, this separation is indifferent, as the halo is defined
with respect to $Q_{\rm min}$, the experimental two-track resolution.
For example, if $Q_{\rm min} = 10-15$ MeV, the decay products of the
$\omega$ resonances can be taken as parts of the halo~\cite{sk}. If
future experimental resolution decreases below $5$ MeV and the error
bars on the measurable part of the correlation function decrease {\it
  significantly} in the $Q < \hbar/\Gamma_{\omega} = 8$ MeV region,
the decay products of the $\omega$ resonances will contribute to the
resolvable core, see Refs \cite{chalo,sk} for greater details.

If Assumption 3 is also satisfied by some experimental data set, then
Eq.\ (\ref{e:c2-wig}) yields a particularly simple form of the
two-particle Bose--Einstein correlation function:

\noindent\vspace*{-12pt}\bea
C_2({\bf k}_1, {\bf k}_2) 
         & = & 1 +
      \lambda_*({\bf K}){\dst \mid 
                \tilde S_c({ \D k},{\bf K})\mid^2 
                \ov \tilde S_c( 0, {\bf k}_1) \tilde S_c(0,{\bf k}_2)}, \\
        & 
          \simeq & \, 1 +
      \lambda_*({\bf K}) {\dst \mid \tilde S_c( {\D k}, 
                {K}) \mid^2 \ov | \tilde S_c( 0, {K})|^2},
                           \label{e:lamq}
\eea
where mean and the relative momentum four-vectors are defined as
\be
        K  = 0.5 ( k_1 + k_2), \qquad \Delta k = k_1 - k_2,
\ee
with $K = (K^0,{\bf K})$ and $\Delta k = (\Delta k^0, 
        {\bf \Delta k})$, 
and the effective intercept parameter $\lambda_*({\bf K})$
is given as
\be 
        \lambda_*({\bf K})  =
                \left[N_c({\bf K}) / N_1({\bf K}) \right]^2.
\ee
As emphasized in Ref.\ \cite{chalo}, this {\it effective} intercept
parameter $\lambda_*$ shall in general depend on the mean momentum of
the observed boson pair, which within the errors of $Q_{\rm min}$
coincides with any of the on-shell four-momentum $k_1$ or $k_2$.  Note
that $\lambda_* \ne \lambda_{xct} = 1$, the latter being the
e\underline{x}a\underline{ct} intercept parameter at $Q = 0$~MeV.  The
core/halo model is summarized in Fig.~\ref{f:halo_f1}, see
Ref.\ \cite{chalo} for further details.  The core/halo model
correlation function is compared to the so-called
``model-independent", Gaussian approximation of
Refs \cite{uli_s,uli_l,uli_urs} and to the full correlation function
in Fig.~\ref{f:chalo_mig}, see appendix of Ref.\ \cite{3d} and that of
Ref.\ \cite{dkiang} for further details.

The measured two-particle BECF is determined for $\mid {\bf \D k} \mid
> Q_{\rm min}\approx 10$ MeV/c, and any structure within the $ \mid {\bf
  \D k} \mid < Q_{\rm min}$ region is not resolved.  However, the $(c,h)$
and $(h,h)$ type boson pairs create a narrow peak in the BECF exactly
in this $\D k $ region according to Eq.~(\ref{e:ax}), which cannot be
resolved according to {\it Assumption~2}.

The general form of the BECF of systems with large halo,
Eq.~(\ref{e:lamq}), coincides with the most frequently applied
phenomenological parameterizations of the BECF in high energy heavy
ion as well as in high energy particle reactions~\cite{bengt}.
Previously, this form has received a lot of criticism from the
theoretical side, claiming that it is in disagreement with quantum
statistics~\cite{weiner} or that the $\lambda $ parameter is just a
kind of fudge parameter, ``a measure of our ignorance".  In the
core/halo picture, Eq.~(\ref{e:lamq}) is derived with a standard
inclusion of quantum statistical effects.  Reactions including e$^+ +$
e$^-$ annihilations, lepton--hadron and hadron--hadron reactions,
nucleon--nucleus and nucleus--nucleus collisions are
phenomenologically well described~\cite{bengt} by a core/halo
picture.

\begin{figure}[htb]
\vspace*{6pt}
\centerline{
\includegraphics[height=7cm]{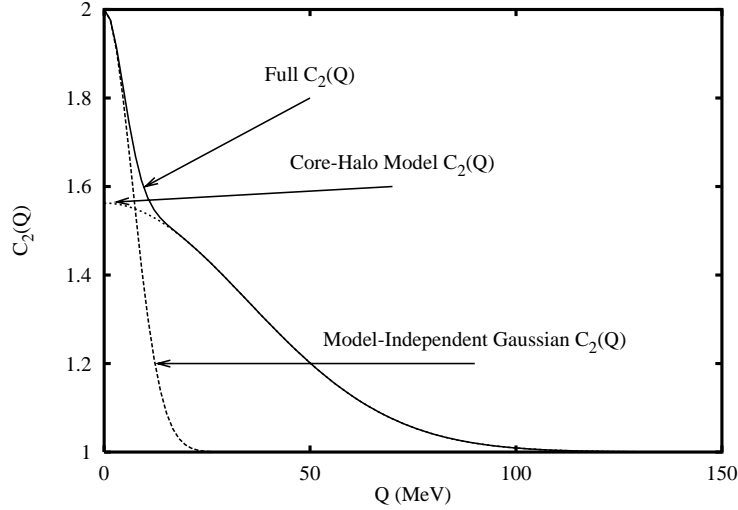}}
\vspace*{-12pt}
\caption{Comparison of the full correlation function (full line)
  to the core/halo model approximation (dashed line) and to the
  ``model-independent" Gaussian approximation (dotted line)}
\label{f:chalo_mig}
\end{figure}

\subsection{Partial coherence and  higher-order correlations }
        
In earlier studies of the core/halo model~\cite{chalo,nhalo} it was
assumed that $S_c(x,p)$ describes a fully incoherent (thermal) source.
In Ref.\ \cite{pcoh} an additional assumption was also made:

{\it Assumption 4}: A part of the core may emit bosons in a
coherent manner:
\be
        S_c(x,\bk)  =  S_c^p(x,\bk) + S_c^i(x,\bk), \label{e:pcs}
\ee
where upper index $p$ stands for coherent component (which leads to
partial coherence), upper index $i$ stands for incoherent component of
the source.

The invariant spectrum is given by 
\be N(\bk) = \int d^4x S(x,\bk) =
N_c(\bk) + N_h(\bk) 
\ee 
and the core contribution is a sum: 
\be
N_c(\bk) = \int d^4x S_c(x,\bk) = N_c^p(\bk) + N_c^i(\bk).  
\ee 
One can introduce the momentum dependent core fractions $f_c(\bk)$ and
partially coherent fractions $p(\bk)$ as 
\bea
f_c(\bk) & = & N_c(\bk)/N(\bk), \\
p_c(\bk) & = & N_c^p(\bk) / N_c(\bk).  
\eea 
Hence the halo and the incoherent fractions $f_h, p_i$ are 
\bea
f_h(\bk) & = & N_h(\bk)/N(\bk) = 1 - f_c(\bk), \\
f_i(\bk) & = & N_c^i(\bk) / N_c(\bk) = 1 - p_c(\bk).  
\eea

\subsection{Strength of the $n$-particle correlations}
We denote the $n$-particle correlation function of Eq.~(\ref{e:cndef}) as
\bea
        C_n(1,2,..., n) & = & C_n(\bk_1,\bk_2,...,\bk_n) \, = \,
        \frac{N_n(1,2,...,n) }{N_1(1) N_1(2) ... N_1(n)},
\eea
where a {\it symbolic notation} for $\bk_i$ is introduced, only the
index of $\bk$ is written out in the argument.  In the forthcoming, we
shall apply this notation consistently for the arguments of various
functions of the momenta, i.e. $f(\bk_i,\bk_j, ... , \bk_m)$ is
symbolically denoted by $f(i,j, ... ,m)$.

The strength of the $n$-particle correlation function (extrapolated
from a finite resolution measurement to zero relative momentum for
each pair) is denoted by $C_n(0)$, given~\cite{pcoh} by the following
simple formula,
\be
C_n(0) = 1 + \sum_{j=2}^n 
{\left( \null^{\dst n}_{\dst  j}\right)}
\alpha_j f_c^j 
\left[ (1-p_c)^j + j p_c (1-p_c)^{j-1} \right].
\label{e:lamnfp_sim}
\ee
Here, $\alpha_j$ indicates the number of permutations, that completely
mix exactly $j$ non-identical elements.  There are 
$\left( \null^{n}_{j} \right)$ different ways to choose $j$ different
elements from among $n$ different elements.  Since all the $n!$
permutations can be written as a sum over the fully mixing
permutations, the counting rule yields a recurrence relation for
$\alpha_j$, Refs \cite{nhalo,pcoh}:
\bea \alpha_n & = & n! -
\sum_{j=0}^{n -1} {\left( \null^{\dst n}_{\dst j}\right)} \alpha_j,
  \label{e:alp}\\
  \alpha_0 & = & 1.
\eea
The first few values of $\alpha_j$ are given as
\be
  \alpha_1  =  0, \quad
        \alpha_2  =  1, \quad
        \alpha_3  =  2, \quad
        \alpha_4  =  9, \quad
        \alpha_5  =  44, \quad
        \alpha_6  =  265,
\ee
the first few intercept parameters, $\lambda_{*,n} = C_n(0) - 1$, are
given as
\bea 
        \lambda_{*,2} & = & f_c^2
                [ (1 - p_c)^2 + 2 p_c (1-p_c)],
                \label{e:l2} \\
        \lambda_{*,3} & = & 3 f_c^2 
                [ (1 - p_c)^2 + 2 p_c (1-p_c)]  \nonumber
        \\
        \null & \null & \qquad
                + 2 f_c^3 
                [ (1 - p_c)^3 + 3 p_c (1-p_c)^2],
                \\
        \lambda_{*,4} & = & 6 f_c^2 
                [ (1 - p_c)^2 + 2 p_c (1-p_c)]  \nonumber
        \\
        \null & \null & \qquad
                + 8 f_c^3
                [ (1 - p_c)^3 + 3 p_c (1-p_c)^2] \nonumber
        \\
        \null & \null & \qquad
                + 9 f_c^4
                [ (1 - p_c)^4 + 4 p_c (1-p_c)^3],
                \\
        \lambda_{*,5} & = & 
                10 f_c^2 
                [ (1 - p_c)^2 + 2 p_c (1-p_c)] \nonumber
        \\
        \null & \null & \qquad
                + 20 f_c^3
                [ (1 - p_c)^3 + 3 p_c (1-p_c)^2] \nonumber
        \\
        \null & \null & \qquad
                + 45 f_c^4 
                [ (1 - p_c)^4 + 4 p_c (1-p_c)^3]  \nonumber
        \\
        \null & \null & \qquad
                + 44 f_c^5 
                [ (1 - p_c)^5 + 5 p_c (1-p_c)^4].
        \label{e:lamfp)}
\eea
        
In general, terms proportional to $f_c^j$ in the incoherent case shall
pick up an additional factor $ [ (1 - p_c)^j + j p_c (1 - p_c)^{j-1}
]$ in case the core has a coherent component~\cite{nhalo,pcoh}.  This
extra factor means that either all $j$ particles must come from the
incoherent part of the core, or one of them must come from the
coherent, the remaining $j-1$ particles from the incoherent part.  If
two or more particles come from the coherent component of the core,
the contribution to intensity correlations vanishes as the intensity
correlator for two coherent particles is zero~\cite{suzuki}.
                
If the coherent component is present, one can introduce the normalized
{\underline i}ncoherent and {\underline p}artially coherent core
fractions as
\bea
        \tilde s_c^i(j,k)
                & = & {\tilde S^i_c(j,k) \over \tilde S^i_c(j,j)},\\
        \tilde s_c^p(j,k)
                & = & {\tilde S^p_c(j,k) \over \tilde S^p_c(j,j)}.
\eea
In the partially coherent core/halo picture, one obtains the following
closed form for the order $n$ Bose--Einstein correlation
functions~\cite{pcoh}:
\bea
        C_n(1,...,n)\! & = & \! 1 + \sum_{j = 2}^n
        \sum_{m_1 ... m_j = 1}^{\null \,\,\, n \,\,\, _{\prime}}
        \sum_{\rho^{(j)}} 
        \left\{
        \prod_{k=1}^{j}
        f_c(m_k) [1 - p_c(m_k)] \,\, \tilde s^i_c(m_k,m_{\rho_k})
        \right. \nonumber
        \\
        \null & \null & \null \hspace{-3.0cm} 
        \left. +
         \sum_{l = 1}^j f_c(m_l) p_c(m_l)  \,\, \tilde s^p_c(m_l,m_{\rho_l}) \!
        \prod_{k=1, k \ne l}^j \!\!
        f_c(m_k) [1 - p_c(m_k)] \, \tilde s^i_c(m_k,m_{\rho_k})
        \right\}.
        \label{e:fpmix}
\eea
Here, $\rho^{(j)}$ stands for the set of permutations that completely
mix exactly $j$ elements, $\rho_i$ stands for the permuted value of
index $i$ in one of these permutations. By definition, $\rho_i \ne i$
for all $i = 1, 2, ..., j$. The notation $\Sigma^{\prime}$ indicates
summation for different values of indexes, $m_i\ne m_l$ for all $i,l$
pairs.  The expression Eq.~(\ref{e:fpmix}) contains two (momentum
dependent) phases in the Fourier-transformed, normalized source
distributions: one denoted by $\phi^i(\bk_m,\bk_n)$ in the
Fourier-transformed normalized {\underline{i}ncoherent} core emission
function, $\tilde s_c^i(\bk_m,\bk_n)$ and another independent phase
denoted by $\phi^c(\bk_m,\bk_n)$ is present in the the
Fourier-transformed normalized {\underline{c}oherent} core emission
function, $\tilde s_c^p(\bk_m,\bk_n)$. One can write
\bea
        \tilde s_c^i(\bk_m,\bk_n)
        & = & |\tilde s_c^i(\bk_m,\bk_n)| \exp[i \phi^i(\bk_m,\bk_n)],\\
        \tilde s_c^p(\bk_m,\bk_n)
        & = & |\tilde s_c^p(\bk_m,\bk_n)| \exp[i \phi^p(\bk_m,\bk_n)].
\eea
The shape of both the coherent and the incoherent components is
arbitrary, but corresponds to the space-time distribution of particle
production.  If the variances of the core are finite, the emission
functions can be parameterized by Gaussians, for the sake of
simplicity~\cite{sk}.  If the core distributions have power-law-like
tails, like in case of the Lorentzian distribution~\cite{3d}, then the
Fourier-transformed emission functions correspond to exponentials or
to power-law structures.  For completeness, we list these
possibilities below:
\bea
        |\tilde s_c^i(\bk_m,\bk_n)|^2
        & = & \exp( - R_{i}^2 Q_{mn}^2) \qquad\!  \mbox{\rm or} \\ 
        |\tilde s_c^i(\bk_m,\bk_n)|^2
        & = & \exp( - R_{i} Q_{mn})  \qquad  \mbox{\rm or} \\ 
        |\tilde s_c^i(\bk_m,\bk_n)|^2
        & = & a_i  (R_{i} Q_{mn})^{b_i}  \qquad\ \  \mbox{\rm etc ... , } \\
\nonumber \\[-6pt] 
        |\tilde s_c^p(\bk_m,\bk_n)|^2
        & = & \exp( - R_{p}^2 Q_{ij}^2) \qquad\ \ \!  \mbox{\rm or} \\ 
        |\tilde s_c^p(\bk_m,\bk_n)|^2
        & = & \exp( - R_p Q_{mn})  \qquad  \mbox{\rm or} \\ 
        |\tilde s_c^p(\bk_m,\bk_n)|^2
        & = & a_p  (R_p Q_{mn})^{b_p}  \qquad\ \ \!  \mbox{\rm etc ...  } .
\eea
In the above equations, subscripts $i$ and $p$ index the parameters
belonging to the incoherent or to the partially coherent components of
the core, and $Q_{mn}$ stands for certain experimentally defined
relative momentum component determined from $\bk_m$ and $\bk_n$.

A straightforward counting yields that in the limiting case when all
momenta are equal, the simple formula of Eq.~(\ref{e:lamnfp_sim})
follows from the shape of the $n$-particle Bose--Einstein correlation
functions of Eq.~(\ref{e:fpmix}), as $\tilde s^i_c(i,i) = \tilde
s^p_c(i,i) = 1$. 

\subsection{Graph rules}
\begin{figure}[h]
\vspace*{6pt}
\centerline{
\includegraphics[height=7cm]{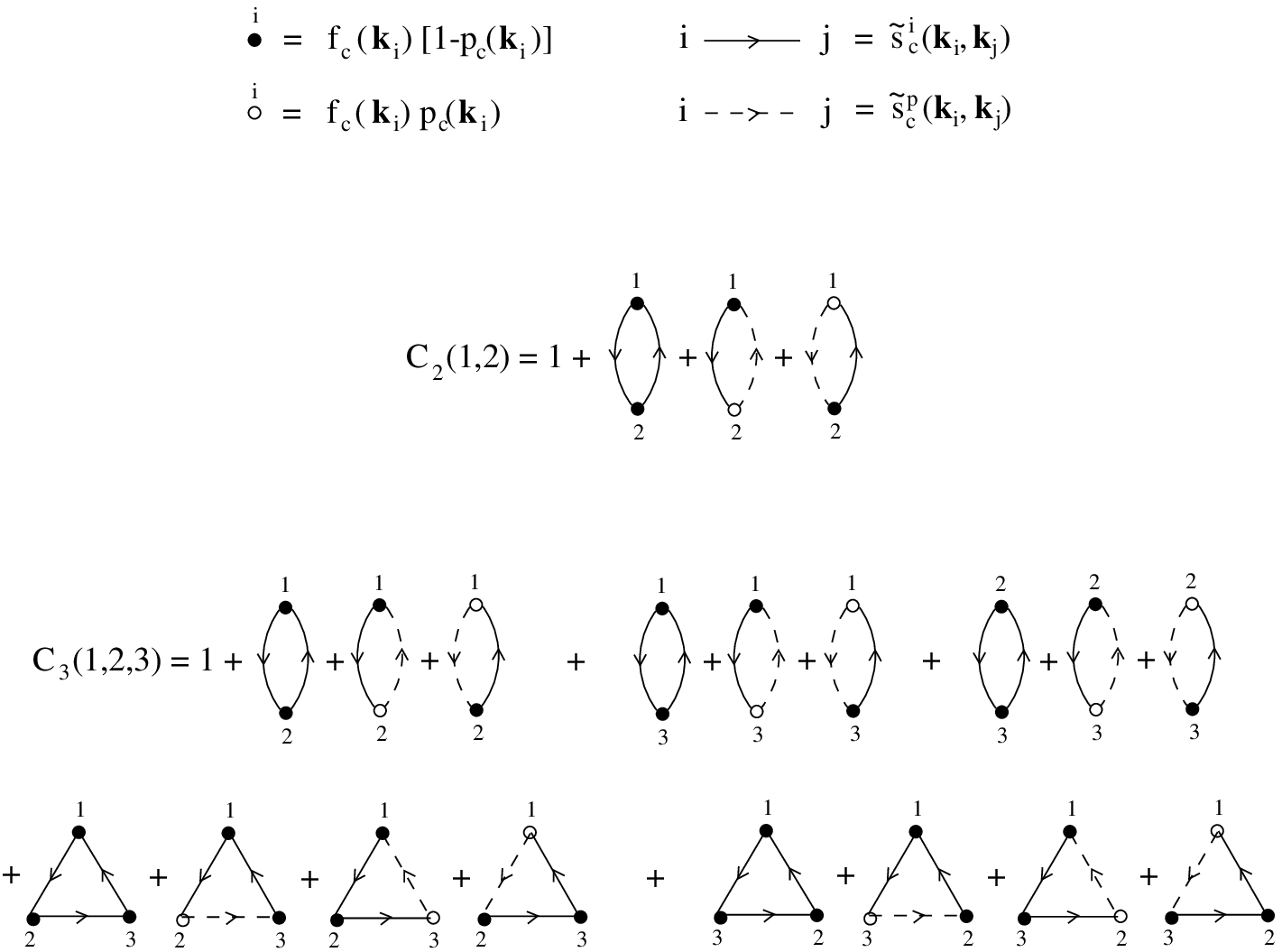}}
\vspace*{-12pt}
  \caption{Graphs determining the second and the third order correlation
  function for partially coherent core/halo sources}
  \label{f:graph_23}
\end{figure}

\begin{figure}[ht]
\vspace*{6pt}
\centerline{
\includegraphics[height=16.5cm]{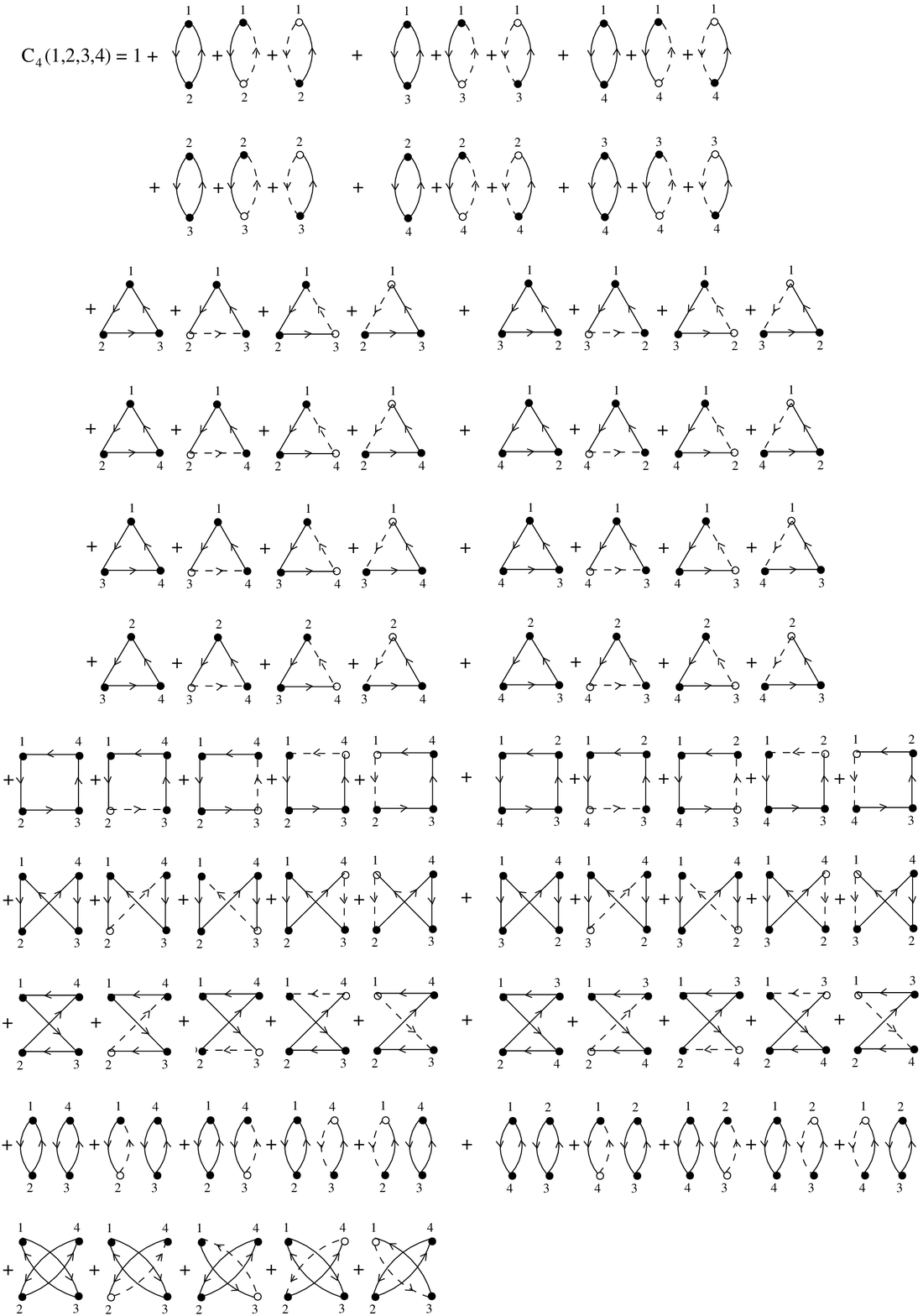}}
\vspace*{-6pt}
  \caption{Graphs determining the fourth order correlation
  function for partially coherent core/halo sources}
  \label{f:graph_4}
\end{figure}

Graph rules were derived for the evaluation of the $n$-particle
correlation function $C_n(\bk_1, ... ,\bk_n)$ in Ref.\ \cite{pcoh}.
Graphs contributing to the $n$ = 2 and 3 case are shown in
Fig.\ \ref{f:graph_23}, the case of $n = 4$ is shown in Fig.\ 
\ref{f:graph_4}.
        
Circles can be either open or full. Each circle carries one label
(e.g.\ $j$) standing for a particle with momentum $\bk_j$.  Full
circles represent the incoherent core component by a factor $f_c(j) [1
- p_c(j)]$, whereas open circles correspond to the coherent component
of the core, a factor of $f_c(j) p_c(j)$.

For the $n$-particle correlation function, all possible $j$-tuples of
particles have to be found.  Such $j$-tuples can be chosen in ${\left(
    \null^{n}_{j}\right)}$ different manner.  In a
$j$-tuple, either each circle is filled, or the circle with index $k$
is open and the other $j-1$ circle is filled, which gives $j+1$
different possibilities.  All the permutations that fully mix either
$j = 2$ or 3, ..., or $n$ different elements have to be taken into
account for each choice of filling the circles.  The number of
different fully mixing permutations that permute the elements $i_1,
..., i_j$ is given by $\alpha_j$ and can be determined from the
recurrence of Eq.~(\ref{e:alp}).

Lines, that connect a pair of circles (or vertexes) $(i,j)$ stand for
factors that depend both on $\bk_i$ and $\bk_j$.  Full lines represent
incoherent--incoherent particle pairs, and corresponds to a factor of
$\tilde s_c^i(i,j)$.  Dashed lines correspond to incoherent-coherent
pairs, and carry a factor of $\tilde s_c^p(i,j)$.  The lines are
oriented, they point from circle $i$ to circle $j$, corresponding to
the given permutation, that replaces element $j$ by element $i$.
Dashed lines start from an open circle and point to a full circle.

All graphs contribute to the order $n$ correlation function, that are
in agreement with the above rules. The result corresponds to the fully
mixing permutations of all possible $j$-tuples $(j = 2, ... , n)$ chosen
in all possible manner from elements $(1,2, ... , n)$.

Each graph adds one term to the correlation function, given by the
product of all the factors represented by the circles and lines of the
graph.  Note that the directions of the arrows matter.  The
correlation function $C(1, ... ,n)$ is given by 1 plus the sum of all
the graphs.

Note that for the $n$-particle {\it cumulant} correlation function,
$n$ circles, representing the $n$ particles, should be connected in
all possible manner corresponding only to the fully mixing
permutations of elements $(1, ..., n)$.  Disconnected graphs do not
contribute to the cumulant correlation functions, as they correspond
to permutations, that either do not mix all of the $n$ elements or can
be built up from two or more independent permutations of certain
sub-samples of elements $(1, 2, ... , n)$.

\subsection{Application to three-particle correlation data}

\begin{figure}[htb]
\vspace*{6pt}
\centerline{\epsfig{file=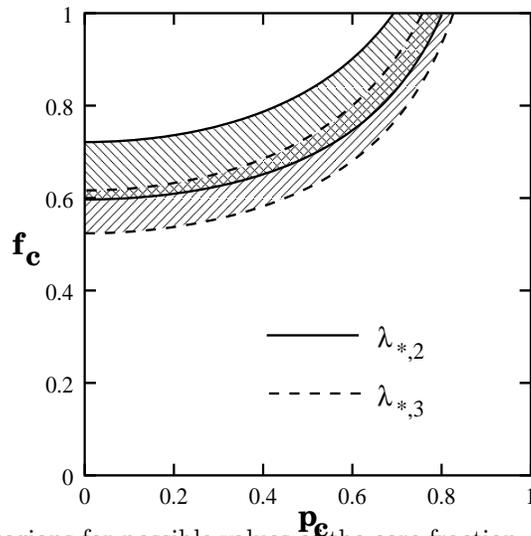,height=7cm,width=7.cm,angle=270}}
\vspace*{-12pt}
\caption{Allowed regions for possible values of the core fraction $f_c$ and
  the partially coherent fraction $p_c$ are evaluated on the two
  standard deviation level from the intercept parameter of the second
  and the third order BE correlation functions, $\lambda_{2,*}$ and
  $\lambda_{3,*}$}
\label{f:fc_pc}
\end{figure}

In the CERN SPS S + Pb reactions, the strength of the two- and
three-particle correlation functions was determined experimentally by
the NA44 Collaboration as $\lambda_{*,2} = 0.44 \pm 0.04$ in Ref.\ 
\cite{na44-hbt} and by $\lambda_{*,3} = 1.35 \pm 0.12 \pm 0.09$, Ref.\ 
\cite{na44-3pi}.  Note that the value of $\lambda_{*,3}$ was
determined with the help of the Coulomb 3-particle wave-function
integration method of Ref.\ \cite{alt-3}, reviewed in
Section~\ref{s:coulomb}, because the estimate\break\vfill\pagebreak\noindent 
based only on the 3-body
Gamow penetration factor introduced unacceptably large systematic
errors to the three-particle Bose--Einstein correlation function.

The two experimental values, $\lambda_{*,2}$ and $\lambda_{*,3}$ can
be fitted with the two theoretical parameters $f_c$ and $p_c$, as done
in Ref.\ \cite{pcoh}.  Figure \ref{f:fc_pc} illustrates the 2 $\sigma$
contour plots in the $(f_c,p_c)$ plane, obtained using the published
value of $\lambda_{*,2} = 0.44 \pm 0.04$ and the preliminary value of
$\lambda_{*,3} = 1.35 \pm 0.12$. A range of $(f_c,p_c)$ values is
found to describe simultaneously the strength of the two-particle and
the three-particle correlation functions within two standard
deviations from these values.  Thus neither the fully chaotic, nor the
partially coherent source picture can be excluded at this level of
precision.

\begin{table}[htb]
\vspace*{-12pt}
\caption{Strength of higher-order correlation functions for various 
core fractions and partially coherent fractions allowed by NA44 2- 
and 3-particle correlation data}
\label{t:pcnhalo}
\vspace*{-6pt}
\begin{center}
\begin{tabular}{llllll}
\hline
\multicolumn{1}{c}{$f_c$}  &  \multicolumn{1}{c}{$p_c$}  & 
\multicolumn{1}{c}{$\lambda_{*,2}$} & 
\multicolumn{1}{c}{$\lambda_{*,3}$} & \multicolumn{1}{c}{$\lambda_{*,4}$} &
\multicolumn{1}{c}{$ \lambda_{*,5}$}\\[2pt]
\hline\\[-10pt]
0.60  &  0.00  &  0.36  &  1.51  &  5.05  &  17.17      \\
0.70  &  0.50  &  0.37  &  1.45  &  4.25  &  11.87      \\
1.00  &  0.75  &  0.44  &  1.63  &  4.33  &  10.47      \\
\hline
\end{tabular}
\end{center}
\end{table}

Cramer and Kadija pointed out, that for higher values of $n$ the
difference between a partially coherent source and between the fully
incoherent particle source with an unresolvable component (halo or
mis-identified particles) will become larger and larger~\cite{crak}.
Indeed, similar values can be obtained for the strength of the second
and third order correlation function, if the source is assumed to be
fully incoherent $(f_c = 0.6, p_c=0)$ or if the source has no halo but
a partially coherent component $(f_c = 1, p_c=0.75)$, but the strength
of the 5th order correlation function is almost a factor of 2 larger
in the former case, as can be seen from Table~\ref{t:pcnhalo}.
Precision measurements of 4th and 5th order correlations are
necessary to determine the value of the degree of partial coherence in
the pion source.


\section{Particle Interferometry in e$^+ +$ e$^-$ Reactions}

The hadronic production in e$^+$e$^-$ annihilations is usually
considered to be a basically coherent process and therefore no
Bose--Einstein effect was expected, whereas  hadronic reactions
should be of a more chaotic nature giving rise to a sizable
effect. It was even argued that the strong ordering in rapidity,
preventing neighboring $\pi^-\pi^-$ or $\pi^+ \pi^+$ pairs,
would drastically reduce the effect~\cite{veneciano}. Therefore
it was a surprise when G. Goldhaber at the Lisbon Conference in 1981
\cite{goll} presented data which showed that correlations between
identical particles in e$^+ $e$^-$ annihilations were very similar
in size and shape to those seen in hadronic reactions, see the review paper 
Ref.\ \cite{bengt} for further details.

\subsection{The Andersson--Hofmann model }
\begin{figure}[htb]
\vspace{6pt}
\centerline{\epsfig{file=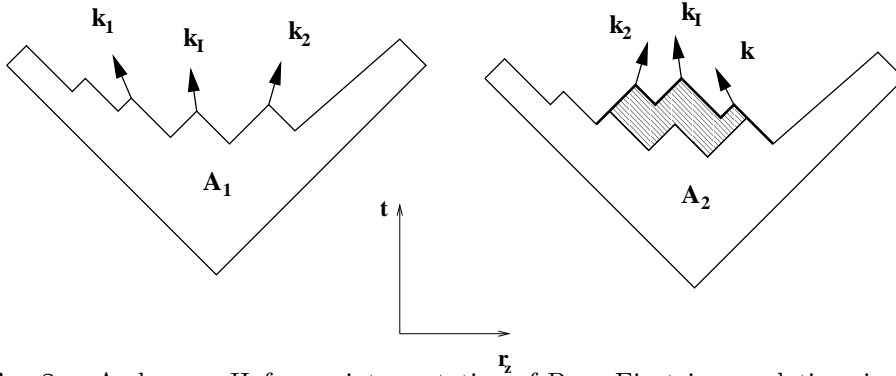,height=12cm,angle=270}}
\vspace{-12pt}
\caption{\label{f:andersson} 
  Andersson--Hofmann interpretation of Bose--Einstein correlations in
  the Lund string model.  $A_{1,2}$ denotes the space-time area of a
  color field enclosed by the quark loop in e$^+ $e$^-$ annihilation.
  Two particles 1 and 2 are separated by the intermediate system $I$.
  When the particles 1 and 2 are identical, the configuration in the
  left side is indistinguishable from that of the right side, and
  their amplitudes for production must be added. The probability of
  production will depend on the difference in area $\Delta A = A_1
  -A_2$, shown as the hatched area.}
\end{figure}

The Bose--Einstein correlation effect, {\it a priori} unexpected for a
coherent process, has been given an explanation within the Lund string
model by B. Andersson and W. Hofmann~\cite{andersson}. The space-time
structure of an e$^+ $e$^-$ annihilation is shown for the Lund string
model \cite{lundm} in Fig.\ \ref{f:andersson}.  The probability for a
particular final state is given by the expression
\begin{equation}
{\rm Prob.} \sim {\rm phase space} \cdot \exp(-bA),\label{e:pb24}
\end{equation}
where $A$ is the space-time area spanned by the string before it
breaks and $b$ is a parameter. The classical string action is given by
$S= \kappa A$, where $\kappa $ is the string tension. It is natural to
interpret the result in Eq.\ (\ref{e:pb24}) as resulting from an
imaginary part of the action such that
\begin{equation}
S= (\kappa + i b/2) A,
\end{equation}
and an amplitude $M$ given by
\begin{equation}
M \sim \exp(iS),
\end{equation}
which implies
\begin{equation}
{\rm Prob.} \sim \mid M \mid^2 \sim \exp(-bA).
\end{equation}

Final states with two identical particles are indistinguishable and
can be obtained in different ways. Suppose that the two particles
indicated as 1 and 2 in Fig.~\ref{f:andersson} are identical, then the
hadron state in the left panel can be considered as being the same as
that in the right panel (where 1 and 2 are interchanged).  The
amplitude should, for bosons, be the sum of two terms
\begin{equation}
M \sim \exp[i(\kappa + ib/2)A_1] +\exp[i(\kappa +i b/2) A_2]\,,
\end{equation}
where $A_1$ and $A_2$ are the two string areas, giving a
probability proportional to
\begin{equation}
\mid M \mid^2 \sim [\exp(-bA_1) + \exp(-bA_2)]\cdot
\left[1 +{\cos(\kappa\Delta A)\over \cosh(b\Delta A/2)}\right]
\end{equation}
with $\Delta A \equiv A_1 - A_2$.  The magnitudes of $\kappa $ and $b$
are known from phenomenological studies.  The energy per unit length
of the string is given by $\kappa \approx 1$ GeV/fm, and $b$ describes
the breaking of the string at a constant rate per unit area,
$b/\kappa^2 \approx 0.7$~GeV$^{-2}$ \cite{lundm}. The difference in
space-time area $\Delta A$ is marked as the hatched area in
Fig.~\ref{f:andersson}. It can be expressed by the $(t,r_z)$
components $(E,k)$ of the four-momenta of the two identical particles
1 and 2, and the intermediate system $I$:
\begin{equation}
\Delta A = [E_2 k_1 - E_1 k_2 + E_I(k_1 - k_2) - k_I(E_1-E_2)]/\kappa^2\,.
\end{equation}
To take into account also the component transverse to the string a
small additional term is needed. The change in area $\Delta A$ is
Lorentz invariant to boosts along the string direction and is
furthermore approximately proportional to $Q = \sqrt{ - (k_1 -
  k_2)^2}$.

\begin{figure}[htb]
\vspace{-12pt}
\centerline{\epsfig{file=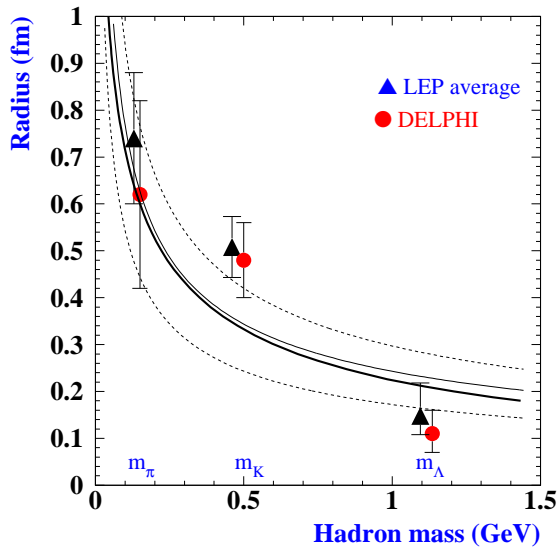,height=8cm,width=8cm,angle=0}}
\vspace{-12pt}
\caption{Mass dependence of the length of homogeneity in
  e$^+ $e$^-$ annihilation at~LEP}
\label{f:alexander}
\vspace{0.3cm}
\end{figure}
 
The interference pattern between the amplitudes will be dominated by
the phase change of $\Delta \Phi = \kappa \Delta A$.  It leads to a
Bose--Einstein correlation which, as a function of the four-momentum
transfer, reproduces the data well but shows a steeper dependence at
small $Q$ than a Gaussian function. A comparison to TPC data confirmed
the existence of such a steeper than Gaussian dependence on $Q$,
although the statistics at the small $Q$-values did not allow a firm
conclusion~\cite{bengt,TPC}.

Recently, the interest for multi-dimensional analysis of
Bose--Einstein correlations increased also in the particle physics
community, see Ref.\ \cite{kittel} for a critical review of the present
status.

I would like to highlight three interesting features: {\it i)} The
effect seems to depend on the transverse momentum of the produced pion
pairs, decreasing effective radii were observed for increasing
transverse mass~\cite{delphi-radii,l3-radii}.  This effect is also
seen in the LUBOEI algorithm of JETSET, although no intrinsic
momentum-dependent scale is plugged into the algorithm~\cite{luboei}.
{\it ii)} The three-dimensional Bose--Einstein correlations of L3
indicate a non-Gaussian structure~\cite{l3-radii}.  {\it iii)} The
effective source sizes of heavier particles (K, $\Lambda$) were
measured recently~\cite{alexander-opal}, based on spin statistics
developed by Alexander and Lipkin~\cite{alexander-lipkin}.  The
measured source sizes show a clear decrease with increasing particle
masses.  The latter effect was explained by Alexander, Cohen and
Levin~\cite{alexander-coh-lev} by arguments based on the Heisenberg
uncertainty relation, and independently with the help of virial
theorem applied for a QCD motivated confining potential.  See
Fig.\ \ref{f:alexander}, reproduced from Ref.\ \cite{alexander-ismd}.
Note that a similar decrease was predicted in Ref.\ \cite{csorgo-zim},
which would depend not on the mass, but on the transverse mass of the
particles, if the particle production happens so that the position of
the emission is very strongly correlated with the momentum of the
emitted particle~\cite{csorgo-zim}.  So, it would be timely to check
whether the effect depends on the particle mass, or on the transverse
mass.  Although the side radius components indicate such a decrease in
case of pions, similar measurements for kaons and $\Lambda$-s would be
indispensable to clarify the origin of the observed behavior.

The question arises: can the effects {\it i)--iii)} be explained in a
unified framework, that characterizes the hadronization process in
e$^+$e$^-$ annihilation? An explanation of the rather small effective
size of the source of the $\Lambda$-s seems to be a challenge for the
Lund string model.

The three-dimensional analysis of the NA22 data on h + p reactions
indicated a strong decrease of all the characteristic radii with
increasing values of transverse momenta of the pair in the NA22
experiment~\cite{na22-hbt}.  A decrease of the effective source sizes
with increasing values of the transverse mass for a given kind of
particle is seen in heavy ion collisions, similarly to effect {\it i)}
in particle physics.  The property {\it iii)}, the decrease of the
effective source size with the increase of the mass of the particle is
seen in heavy ion physics and is explained in terms of hydrodynamical
expansion, similarly to the explanation of effect {\it i)}, see
Figs~\ref{f:xna49} and \ref{f:xna44} in Section~\ref{s:bl-pbpb}.  Can
one give a unified explanation of these similarities between results
of particle interferometry in e$^+ +$ e$^-$, h + p and heavy ion
physics? We do not yet know the answer to this question.


\section{Invariant Buda--Lund Particle Interferometry } 
\label{s:bl}

\def\vp{{\bf k}}
\def\vq{{\bf q}}
\def\vk{{\bf k}}
\def\vK{{\bf K}}
\def\vx{{\bf x}}
\def\vy{{\bf y}}
\def\uk{{|{\bf k}|}}
\def\bc{\begin{center}}
\def\ec{\end{center}}
\def\De{\Delta\eta}
\def\Det{\Delta\eta_T}
\def\Des{\Delta\eta_*}
\def\Dk{\Delta k}
\def\Dt{\Delta\tau}
\def\Dy{\Delta y}
\def\t0{\tau_0}
\def\tl{\tau_L}
\def\ch{\cosh}
\def\sh{\sinh}
\def\bx{{\bf{x}}}
\def\bp{{\bf{p}}}
\def\bk{{\bf{k}}}
\def\bK{{\bf{K}}}
\def\bK{{\bf{K}}}
\def\bQ{{\bf{Q}}}
\def\bpi{{\bf{\pi}}}
\def\bxi{{\bf{\xi}}}
\def\bq{{\bf{q}}}
\def\br{{\bf{r}}}
\def\bak{{\bf K}}
\def\dek{{\bf \Delta k}}
\def\xb{{\overline{x}}}
\def\tb{{\overline{t}}}
\def\rb{{\overline{r}}}
\def\nb{{\overline{n}}}
\def\etab{{\overline{\eta}}}
\def\taub{{\overline{\tau}}}

%
%
\def\D{\Delta}
\def\o{{out}}
\def\s{{side}}
\def\e{{\eta}}
\def\bdk{{\bf \Delta k}}
\def\rl{R_l^2}
\def\ro{R_{o}^2}
\def\rs{R_{s}^2}
\def\rol{R_{ol}^2}
\def\rpa{R_{\parallel}^2 }
\def\rpe{R_{\perp}^2 }
\def\rta{R_{\tau}^2 }
\def\BL{{Buda}{--}{Lund}~} 

The $n$-particle Bose--Einstein correlation function of
Eq.~(\ref{e:cndef}) is defined as the ratio of the $n$-particle
invariant momentum momentum distribution divided by an $n$-fold
product of the single-particle invariant momentum distributions.
Hence these correlation functions are boost-invariant.

The invariant Buda--Lund parameterization (or BL in short) deals with
a boost-invariant, multi-dimensional characterization of the building
blocks $\langle a_{{\bf k}_1}^{\dagger} a_{{\bf k}_2} \rangle $ of
arbitrary high order Bose--Einstein correlation functions, based on
Eqs~(\ref{e:gdef},\ref{e:gwig}).  The BL parameterization was
developed by the Buda--Lund Collaboration in Refs \cite{3d,mpd95}.

\begin{figure}[htb]
\vspace{6pt}
\centerline{\epsfig{file=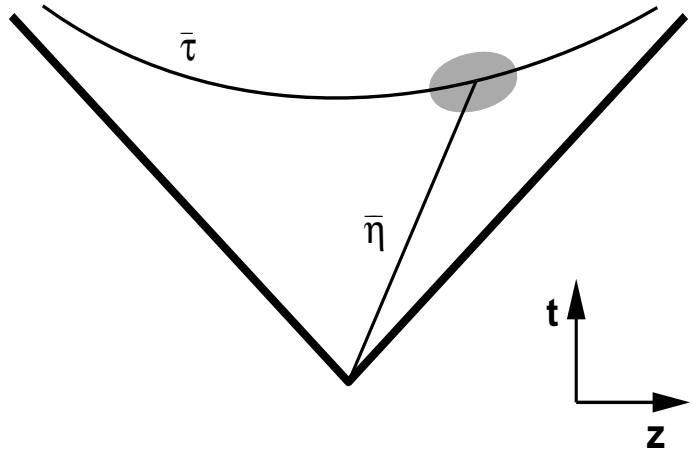,height=6cm,width=9cm,angle=0}}
\vspace{-12pt}
\caption{Space-time picture of particle emission for a given fixed mean
  momentum of the pair.  The mean value of the proper-time and the
  space-time rapidity distributions is denoted by $\taub$ and $\etab$.
  As the rapidity of the produced particles changes from the target
  rapidity to the projectile rapidity the $[\taub(y),\etab(y)]$
  variables scan the surface of mean particle production in the
  $(t,r_z)$ plane.}
\label{f:bl}
\vspace{0.3cm}
\end{figure}

The essential part of the BL is an invariant decomposition of the
relative momentum $q$ in the $\exp(i q \cdot \Delta x)$ factor into a
temporal, a longitudinal and two transverse relative momentum
components.  This decomposition is obtained with the help of a
time-like vector in the coordinate space, that characterizes the
center of particle emission in space-time, see Fig.~\ref{f:bl}.

Although the BL parameterization was introduced in Ref.\ \cite{3d} for
high energy heavy ion reactions, it can be used for other physical
situations as well, where a dominant direction of an approximate
boost-invariant expansion of the particle emitting source can be
identified and taken as the longitudinal direction $r_z$. For example,
such a direction is the thrust axis of single jets or of back-to-back
two-jet events in case of high energy particle physics.  For
longitudinally almost boost-invariant systems, it is advantageous to
introduce the boost invariant variable $\tau$ and the space-time
rapidity $\eta$, 
\bea
\tau &  = & \sqrt{t^2 - r_z^2}, \\
\eta & = & 0.5 \log\left[(t+r_z)/(t-r_z)\right] .  
\eea 
Similarly, in momentum space one introduces the transverse mass $m_t$
and the rapidity $y$ as
\bea
m_t & = & \sqrt{E^2 - p_z^2}, \\
y & = & 0.5 \log\left[(E+p_z)/(E-p_z)\right] .  
\eea 
The source of particles is characterized in the boost invariant
variables $\tau$, $m_t$ and $\eta - y$.  For systems that are only
approximately boost-invariant, the emission function may also depend
on the deviation from mid-rapidity, $y_0$.  The scale on which the
approximate boost-invariance breaks down is denoted by $\Delta \eta$,
a parameter that is related to the width of the rapidity distribution.

The correlation function is defined with the help of the
Wigner function formalism, Eq.~(\ref{e:gwig}), the intercept parameter
$\lambda_*$ is introduced in the core/halo picture of
Eq.~(\ref{e:lamq}).  The case of $ n = 2$ particles and a chaotic core
with $p_c = 0$ was discussed in Ref.\ \cite{3d}.  In the following, we
evaluate the building block for arbitrary high order Bose--Einstein
correlation functions.  We assume for simplicity that the core is
fully incoherent, $p_c(j) = 0$ in Eq.~(\ref{e:fpmix}).  A further
simplification is obtained if we assume that the emission function of
Eqs~(\ref{e:gwig},\ref{e:lamq}) factorizes as a product of an {\it
  effective} proper-time distribution, a space-time rapidity
distribution and a transverse coordinate distribution~\cite{lcms,3d}:
\bea S_c(x,K) d^4 x & = & H_*(\tau) G_*(\eta) I_*(r_x,r_y) \, d\tau \,
\taub d\eta dr_x dr_y .
\label{e:fact} 
\eea 
The subscript $*$ stands for a dependence on the mean momentum $K$,
the mid-rapidity $y_0$ and the scale of violation of boost-invariance
$\Delta \eta$, using the symbolic notation $f_* \equiv f[K, y_0,
\Delta\eta]$.  The function $H_*(\tau) $ stands for such an effective
proper-time distribution (that includes, by definition, an extra
factor $\tau$ from the Jacobian $d^4x = d\tau\, \tau\, d\eta, dr_x
dr_y$, in order to relate the two-particle Bose--Einstein correlation
function to a Fourier transformation of a distribution function in
$\tau$).  The effective space-time rapidity distribution is denoted by
$G_*(\eta) $, while the effective transverse distribution is denoted
by $I_*(r_x,r_y)$.  In Eq.~(\ref{e:fact}), the mean value of the
proper-time $\taub$ is factored out, to keep the distribution
functions dimensionless.  Such a pattern of particle production is
visualized in Fig.\ \ref{f:bl}.
        
In case of hydrodynamical models, as well as in case of a decaying
Lund strings~\cite{lcms,mpd95}, production of particles with a given
momentum rapidity $y$ is limited to a narrow region in space-time
around $\etab$ and $\taub$.  If the sizes of the effective source are
sufficiently small (if the Bose--Einstein correlation function is
sufficiently broad), the $\exp( i q \cdot \Delta x)$ factor of the
Fourier transformation is decomposed in the shaded region in
Fig.~\ref{f:bl} as
\bea
        \exp[i (q^0 \Delta t -  q_z \Delta r_z) ]  
        & \simeq  &
        \exp[i ( Q_= \Delta \tau - Q_{\parallel} \taub \Delta\eta) ], 
        \label{e:blexp} \\
        \exp[ - i ( q_x \Delta r_x + i q_y \Delta r_y) ]  
         & \equiv & 
        \exp[ - i ( Q_:\Delta r_:  +  Q_{..} \Delta r_{..} ) ].
                \label{e:bltexp}
\eea
The invariant {\it temporal}, {\it parallel}, {\it side}ward, {\it
  out}ward (and {\it perp}endicular) relative momentum components are
defined, respectively, as
\bea
        Q_= & = &   
                q_0 \cosh[\etab] - q_z \sinh[\etab],
                 \label{e:q=} \\
        Q_{\parallel} & = & 
                q_z \cosh[\etab] - q_0 \sinh[\etab],
                \label{e:qpar} \\
        Q_{..} & = &  (q_x K_y - q_y K_x)/\sqrt{K_x^2 + K_y^2}, \\ 
        Q_{:} & = & (q_x K_x + q_y K_y)/\sqrt{K_x^2 + K_y^2}, \\ 
        Q_{\perp} & = & \sqrt{ q_{x}^2 + q_{y}^2} = \sqrt{Q_{:}^2+Q_{..}^2}.
\enn
The time-like normal-vector $\nb$ indicates an invariant direction of
the source in coordinate space~\cite{3d}.  It is parameterized as
$\nb^{\mu} = (\cosh[\etab],0,0,\sinh[\etab])$, where $\etab$ is a mean
space-time rapidity \cite{3d,3d-cf98,mpd95}.  The parameter $\etab$
is one of the fitted parameters in the BL type of decomposition of the
relative momenta.  The above equations are invariant, they can be
evaluated in any frame. To simplify the presentation, in the following
we evaluate $q$ and $\etab$ in the LCMS.  The acronym LCMS stands for
the Longitudinal Center of Mass System, where the mean momentum of a
particle pair has vanishing longitudinal component, $K_z = 0.5
(k_{1,z} + k_{2,z}) = 0.$ In this frame, introduced in
Ref.\ \cite{lcms}, ${\bf K}$ is orthogonal to the beam axis, and the
time-like information on the duration of the particle emission couples
to the out direction.  The rapidity of the LCMS frame can be easily
found from the measurement of the momentum vectors of the particles.
As $\etab$ is from now on a space-time rapidity measured in the LCMS
frame, it is invariant to longitudinal boosts: $\etab^\prime = (\etab
- y) - (0-y) = \etab$.

The symbolic notation for the {\it side} direction is two dots side by
side as in $Q_{..}$.  The remaining transverse direction, the {\it
  out} direction was indexed as in $Q_{:}$, in an attempt to help to
distinguish the zeroth component of the relative momentum $Q_0$ from
the out component of the relative momentum $Q_{:} \equiv Q_o =
Q_{\rm out}$, $Q_0 \ne Q_o$.  Hence $ K_{:} = |{\bf K}_\perp|$ and $K_{..}
= 0$.  The geometrical idea behind this notation is explained in
details in Ref.\ \cite{3d-cf98}.
        
The perpendicular (or transverse) component of the relative momentum
is denoted by $Q_{\perp}$.  By definition, $Q_{..}$, $Q_{:}$ and
$Q_{\perp}$ are invariants to longitudinal boosts, and $Q^2 = - q\cdot
q = Q_{..}^2 + Q_:^2 + Q_{||}^2 - Q_=^2$.
 
With the help of the {\it small source size} (or large relative
momentum) expansion of Eq.~(\ref{e:blexp}), the amplitude $\tilde
s_c(1,2) = \tilde s_c^i(1,2)$ that determines the arbitrary order
Bose--Einstein correlation functions in Eq.~(\ref{e:fpmix}) can be
written as follows:
\be
        \tilde s^i_c(1,2) = \frac{ 
        \tilde H_*(Q_=) \tilde G_*(Q_{\parallel})  \tilde I_*(Q_{:},\, Q_{..})}
        {\tilde H_*(0) \tilde G_*(0) \tilde I_*(0,0) }. 
        \label{e:blamp}
\ee
This expression and Eq.~(\ref{e:fpmix}) yield a general, invariant,
multi-\-dimensional Buda--Lund parameterization of order $n$
Bose--Einstein correlation functions, valid for all $n$.  The
Fourier-transformed distributions are defined as
\bea
        \tilde H_*(Q_=)  & = & 
                \int_0^\infty d\tau \exp(i Q_= \tau) H_*(\tau),
                        \label{e:htild} \\
        \tilde G_*(Q_\parallel)  & = & \int_{-\infty}^{\infty} d\eta 
                \exp(- i Q_\parallel \taub \eta ) G_*(\eta),
                        \label{e:gtild}\\
        \tilde I_*(Q_:,Q_{..}) 
                         & = & \int_{-\infty}^{\infty} dr_{:}
                        \int_{-\infty}^{\infty}  dr_{..} 
                        \exp(- i Q_: r_: - i Q_{..} r_{..}) 
                        I_*(r_:,r_{..}) . \label{e:itild}
\eea

As a particular case of Eqs~(\ref{e:blamp},\ref{e:fpmix}) for $n = 2$
and $p_c(j) = 0$, the two-particle BECF can be written into a
factorized Buda--Lund form as
\bea
        C({\bf k}_1, {\bf k}_2) & = & 1 + \lambda_*(K) 
                {\dst |\tilde H_*(Q_=)  |^2 \ov |\tilde H_*(0) |^2} \,
                {\dst |\tilde G_*(Q_{\parallel})  |^2 \ov |\tilde G_*(0) |^2}\,
                {\dst |\tilde I_*(Q_{:},\, Q_{..})|^2 \ov 
                |\tilde I_*(0,0) |^2} .
        \label{e:blf}
\eea
\vfill\pagebreak\noindent
Thus, the BL results are rather generic. For example, BL
parameterization may in particular limiting cases yield the {\it
  power-law}, the {\it exponential}, the {\it double-Gaussian}, the
{\it Gaussian}, or the less familiar {\it oscillating} forms of
Eq.~(\ref{e:blosc}), see also Ref.\ \cite{3d-cf98}.  The {\it
  Edgeworth}, the {\it Laguerre} or other similarly constructed
low-momentum expansions~\cite{lagu} can be applied to any of the
factors of one variable in Eq.~(\ref{e:blf}) to characterize these
unknown shapes in a really model-independent manner, relying only on
the convergence properties of expansions in terms of complete
orthonormal sets of functions~\cite{lagu}.

In a Gaussian approximation and assuming that $R_{:} = R_{..} =
R_{\perp}$, the Buda--Lund form of the Bose--Einstein correlation
function reads as follows:
\be
        {C_2({\bf k}_1,{\bf k}_2) = }
        {1}
        {+} \lambda_* \,
        { \exp\left( - R^2_= Q^2_=
                     - R^2_{\parallel} Q^2_{\parallel}
                     - R_{\perp}^2 Q_{\perp}^2 \right)} ,
        \label{e:gaussblperp}
\ee
where the 5 fit parameters are $\lambda_*$, $R_=$, $R_\parallel$,
$R_\perp$ and the value of $\etab$ that enters the definitions of
$Q_=$ and $Q_\parallel$ in Eqs~(\ref{e:q=},\ref{e:qpar}).  The fit
parameter $R_=$ reads as $R$-time-like, and this variable measures a
width of the proper-time distribution $H_*$.  The fit parameter
$R_{\parallel}$ reads as $R$-parallel, it measures an invariant length
parallel to the direction of the expansion.  The fit parameter
$R_{\perp}$ reads as $R$-perpendicular or $R$-perp.  For cylindrically
symmetric sources, $R_{\perp}$ measures a transversal rms radius of
the particle emitting source.

The BL radius parameters characterize the lengths of
homogeneity~\cite{sinyukov} in a longitudinally boost-invariant
manner.  The lengths of homogeneity are generally smaller than the
momentum-integrated, total extension of the source, they measure a
region in space and time, where particle pairs with a given mean
momentum ${\bf K}$ are emitted from.

The following Edgeworth expansion can be utilized to characterize
non-Gaussian multidimensional Bose--Einstein correlation functions, in
a longitudinally boost-invariant manner:
\bea
        C_2({\bf k}_1,{\bf k}_2) & = & 1 + \lambda_E
        \exp( - Q_=^2 R_=^2 - Q_{||}^2 R_{||}^2 - Q_{\perp}^2 R_{\perp}^2
                ) \times \nonumber\\
        && \left[ 1 + \frac{\kappa_{3=}}{3!} H_3(\sqrt{2} Q_{=}R_{=} )
        + \frac{\kappa_{4=}}{4!} H_4(\sqrt{2} Q_{=}R_{=} ) + ... \right]
                        \times\nonumber \\
        && \left[ 1 + \frac{\kappa_{3||}}{3!} H_3(\sqrt{2} Q_{||}R_{||} )
        + \frac{\kappa_{4||}}{4!} H_4(\sqrt{2} Q_{||}R_{||} ) + ... \right]
                        \times\nonumber \\
        && \left[ 1 + \frac{\kappa_{3\perp}}{3!} 
                        H_3(\sqrt{2} Q_{\perp}R_{\perp} )
        + \frac{\kappa_{4\perp}}{4!} 
        H_4(\sqrt{2} Q_{\perp}R_{\perp} ) + ... \right] .
        \label{e:bl-edg}
\eea
This yields {\it $5$ free scale parameters} for cylindrically
symmetric, longitudinally expanding sources, and {\it three series of
  shape parameters}.  The scale parameters are $\lambda_E$, $R_=$,
$R_{\parallel}$, $R_{\perp}$ and $\etab$, that characterize the
effective source at a given mean momentum, by giving the vertical
scale of the correlations, the invariant temporal, longitudinal and
transverse extensions of the source and its invariant direction, which
is the space-time rapidity of the effective source in the LCMS frame
(the frame where $k_{1,z} + k_{2,z} = 0$, ~\cite{lcms}).  The three
series of shape parameters are $\kappa_{3=}$,\break\vfill\pagebreak\noindent 
$\kappa_{4=}$, ... ,
$\kappa_{3||}$, $\kappa_{4||}$, ... , $\kappa_{3\perp}$,
$\kappa_{4\perp}$, ... .  Each of these parameters may depend on the
mean momentum $\bK$.

A multi-dimensional Laguerre or a mixed Edgeworth--Laguerre expansion
can be introduced in a similar manner by replacing the Edgeworth
expansion in Eq.~(\ref{e:bl-edg}) by a Laguerre one in any of the
principal directions.

In Eqs~(\ref{e:bl-edg},\ref{e:gaussblperp}), the spatial information
about the source distribution in $(r_x,r_y)$ was combined to a single
perp radius parameter $R_{\perp}$.  In a more general Gaussian form,
suitable for studying rings of fire and opacity effects, the \BL
~invariant BECF can be denoted as
\be
        C_2({\bf k}_1,{\bf k}_2) = 1 + \lambda_* \exp\left( - R_=^2 Q_=^2 -
                         R^2_{\parallel}  Q^2_{\parallel} -
                        R_{..}^2 Q_{..}^2 - R_:^2 Q_:^2 \right).
                \label{e:bl-fring}
\ee
The 6 fit parameters are $\lambda_*$, $R_=$, $R_{\parallel}$,
$R_{..}$, $R_:$ and $\etab$, all are in principle functions of $({\bf
  K},y_0,\Delta\eta)$.  Note that this equation is identical to Eq.\ 
(44) of Ref.\ \cite{3d}, rewritten into the new, symbolic notation of
the Lorentz-invariant directional decomposition.

The above equation may be relevant for a study of expanding shells, or
rings of fire, as discussed first in Ref.\ \cite{3d}.  We shall argue,
based on a simultaneous analysis of particle spectra and correlations,
and on recently found exact solutions of non-relativistic fireball
hydrodynamics~\cite{sol} that an expanding, spherical shell of fire is
formed protons in 30 AMeV $^{40}$Ar + $^{197}$Au reactions, and that a
two-dimensional, expanding ring of fire is formed in the transverse
plane in NA22 h + p reactions at CERN SPS.  The experimental
signatures for the formation of these patterns will be discussed in
Section~\ref{s:na22}.
        
Opacity effects, as suggested recently by H.
Heiselberg~\cite{henning}, also require the distinction between
$R_{..}$ and $R_{:}$.  The lack of transparency in the source may
result in an effective source function, that looks like a crescent in
the side-out reference frame~\cite{henning}.  When integrated over the
direction of the mean momentum, the effective source looks like a ring
of fire in the $(r_x,r_y)$ frame.

The price of the invariant decomposition of the basic building blocks
of any order Bose--Einstein correlation functions in the BL
parameterization is that the correlation functions cannot be directly
binned in the BL variables, as these can determined after the
parameter $\etab$ is fitted to the data --- so the correlation function
has to be binned first in some directly measurable relative momentum
components, e.g.\ the (side, out, long) relative momenta in the LCMS
frame, as discussed in the next subsection. After fitting $\etab$ in
an arbitrary frame, the BECF can be re-binned into the BL form.

\subsection{Gaussian parameterizations of BE Correlations}

We briefly summarize here the Bertsch--Pratt and the Yano--Koonin
parameterization of the Bose--Einstein correlation functions, to point
out some of their advantages as well as drawbacks and to form a basis
for comparison.

\subsubsection{The Bertsch--Pratt parameterization}

The Bertsch--Pratt (BP) parameterization of Bose--Einstein correlation
functions is one of the oldest, widely used multi-dimensional
decomposition, called also as the side--out--longitudinal
decomposition~\cite{bertsch,pratt-bp}.
 
This directional decomposition was devised to extract the contribution
of a long duration of particle emission from an evaporating
Quark--Gluon Plasma, as expected in the mixture of a hadronic and a QGP
phase if the re-hadronization phase transition is a strong first order
transition.
        
The BP parameterization in a compact form reads as
\bea
    C_2({\bf k}_1,{\bf k}_2)   & = & 
        1 +  \lambda \exp\left[ - \rs Q_{s}^2 - \ro Q_{o}^2 
        - \rl Q_l^2 - 2 \rol Q_l Q_{o} \right].
\eea
Here index $o$ stands for {\it out} (and not the temporal direction),
$s$ for {\it side} and $l$ for {\it longitudinal}.  The
out--longitudinal cross-term was introduced by Chapman, Scotto and
Heinz in Refs \cite{uli_s,uli_l} --- this term is non-vanishing for
axially symmetric systems, if the source is not fully boost-invariant,
or if the measurement is made not at mid-rapidity.  In a more detailed
form, the mean momentum dependence of the various components is shown
as
\bea
    C_2({\bf k}_1,{\bf k}_2)
                & = & 
        1 +  
        \lambda({\bf K}) \exp\left[ - \rs({\bf K}) Q_{s}^2({\bf K})
        - \ro({\bf K}) Q_{o}^2({\bf K}) \right. \nonumber  \\
        \null & \null & \hspace{2cm} \left. 
        - \rl({\bf K}) Q_l^2 - 2 \rol({\bf K}) Q_l Q_{o}({\bf K}) \right]\, ,
\eea
where the mean and the relative momenta are defined as
\bea
        {\bf K} & = & 0.5 ({\bf k}_1 + {\bf k}_2 ) ,\\
        {\bf \Delta k} & = & {\bf k}_1 - {\bf k}_2 , \\
        Q_l & = & k_{z,1} - k_{z,2} ,\\
        Q_{o} & = & Q_{o}({\bf K}) 
                \, = \, {\bf \Delta k} \cdot {\bf K} / | {\bf K} | , \\ 
        Q_{s} & = & Q_{s}({\bf K}) \, = \,
                |{\bf \Delta k } \times {\bf K} | / | {\bf K} | . 
\eea
It is emphasized that the BP radius parameters are also measuring
lengths of homogeneity~\cite{sinyukov}.  Not only the radius
parameters but also the decomposition of the relative momentum to the
side and the out components depends on the (direction of) mean
momentum ${\bf K}$.

In an arbitrary frame, Gaussian radius parameters can be defined, and
sometimes they are also referred to as BP radii, when the spatial
components of the relative momentum vector are taken as independent
variables.  The BP radii reflect space-time
variances~\cite{uli_s,uli_l} of the {\it core}~\cite{sk} of the
particle emission, {\it if} a Gaussian approximation to the core is
warranted:
\bea
        C_2({\bf k}_1,{\bf k}_2) & = & 1 + \lambda_*({\bf K}) \,
        \exp\left(- R^2_{i,j}({\bf K}) 
                {\bf \Delta k}_i {\bf \Delta k}_j \right), \\
        \lambda_*({\bf K}) & = & [ N_{\bf c}({\bf K}) / N({\bf K}) ]^2, \\
        R_{i,j}^2({\bf K}) & = &
                 \langle \underline{x}_i \underline{x}_j  \rangle_{\bf c}
                 - \langle \underline{x}_i\rangle_{\bf c}
                   \langle \underline{x}_j \rangle_{\bf c}, \\
        \underline{x}_i & = & x_i - \beta_i t , \\
        \langle f(x,{\bf k}) \rangle_{\bf c} & = & 
                \int d^4 x f(x,{\bf k})  S_{\bf c}(x,{\bf k})  /
                \int d^4 x  S_{\bf c}(x,{\bf k}),
\eea
where $S_{\bf c}(x,{\bf k})$ is the emission function that
characterizes the central core and subscripts $i$ or $j$ stand for
$x$, $y$ or $z$, i.e.\ any of the spatial directions in the frame of
the analysis.  This method is frequently called as
``model-independent" formulation, because the applied Gaussian
approximation is independent of the functional form of the emission
function $S_(x,{\bf k})$~\cite{uli_urs}.  In the literature, this
result is often over-stated, it is claimed that such a
Taylor expansion would provide a general ``proof" that
multi-dimensional Bose--Einstein correlation functions must be
Gaussians.  Although the ``proof" is indeed not depending on the exact
shape of $S(x,K)$, it relies on a second order Taylor expansion of the
shape of the correlation function around its exact value at $Q =0$.
At this point not only the derivatives of the correlation function are
unmeasurable, but the very value of the correlation function $C_2(0)$
is unmeasurable as well, see Figs \ref{f:chalo_mig} and
\ref{f:halo_f1} for graphical illustration.  For exponential or for
power-law type correlations, the building block $\tilde S_c(q,K)$ of
the correlation function is not analytic at $Q=0$, so a Taylor
expansion cannot be applied in their case. For the oscillatory type of
correlation functions, the Gaussian provides a good approximation in
the experimentally unresolvable low $Q$ domain, but it misses the
structure of oscillations at large values of $Q$, which appear because
$S(x,K)$ has more than one maxima, like a source distribution of a
binary star. Thus, the exact shapes of multi-dimensional BECF-s {\it
  cannot be determined a priori} and in case of non-Gaussian
correlators one has to evaluate more (but still not fully)
model-independent relationships, for example
Eqs~(\ref{e:gwig},\ref{e:fpmix},\ref{e:blamp}), which are valid for
broader than Gaussian classes of correlation functions.
        
Note that the tails of the emission function are typically dominated
by the halo of long-lived resonances $S_h(x,{\bf k})$ and even a small
admixture of e.g.\ $\eta$ and $\eta'$ mesons increases drastically the
space-time variances of particle production, and makes the
interpretation of the BP radii in terms of space-time variances of the
total emission function $S = S_c + S_h$ unreliable both qualitatively
and quantitatively, as pointed out already in Ref.\ \cite{sk}.

In the Longitudinal Center of Mass System (LCMS, Ref.\ \cite{lcms}),
the BP radii have a particularly simple form~\cite{lcms}, if the
coupling between the $r_x$ and the $t$ coordinates is also negligible,
$\langle \tilde r_x \tilde t \rangle = \langle \tilde r_x \rangle
\langle \tilde t \rangle$:
\bea
        R_{s}^2({\bf K}) & = &
                 \langle \tilde{r}_y^2 \rangle_{\bf c}\,, \\
        R_{o}^2({\bf K}) & = &
                 \langle \tilde{r}_x^2 \rangle_{\bf c} + 
                 \beta_t^2 \langle \tilde{t}^2 \rangle_{\bf c}\,, \\
        R_{l}^2({\bf K}) & = &
                 \langle \tilde{r}_z^2 \rangle_{\bf c}\,, \\
        R_{ol}^2({\bf K}) & = &
                 \langle \tilde{r}_z 
                (\tilde{r}_x - \beta_t \tilde{t}) \rangle_{\bf c} \,,
\eea
where $\tilde x = x - \langle x \rangle $.  Although this method
cannot be applied to characterize non-Gaussian correlation functions,
the the above form has a number of advantages:\break\vfill\pagebreak\noindent 
it is straightforward
to obtain and it is easy to implement for a numerical evaluation of
the BP radii of Gaussian correlation functions~\cite{uli_urs}.
 
In the LCMS frame, information on the duration of the particle
emission couples {\it only} to the out direction.  This is one of the
advantages of the LCMS frame.  Using the BP, the time distribution
enters the out radius component as well as the out-long cross-term.
Other possible cross-terms were shown to vanish for cylindrically
symmetric sources~\cite{uli_s,uli_l}.

For completeness, we give the relationship between the invariant BL
radii and the BP radii measured in the LCMS, if the BL forms are given
in the Gaussian approximation of Eq.~(\ref{e:bl-fring}):
\bea
        R_{s}^2
                & = & R_{..}^2, \label{e:bp-bl-s} \\
        R_{o}^2
                & = & R_{:}^2 + 
                 \beta_t^2 
                [\cosh^2(\etab) R_{=}^2   + 
                \sinh^2(\etab) R_\parallel^2 ], \\
        R_{ol}^2
                & = & -  \beta_t \sinh(\etab) 
                \cosh(\etab) (R_=^2 + R_\parallel^2 ), \\
        R_{l}^2 & = &
                \cosh^2(\etab) R_\parallel^2   + 
                \sinh^2(\etab) R_=^2, \label{e:bp-bl-l}
\eea
where the dependence of the fit parameters on the value of the mean
momentum, ${\bf K}$ is suppressed.  The advantage of the BP
parameterization is that there are no kinematic constraints between
the side, out and long components of the relative momenta, hence the
BP radii are not too difficult to determine experimentally.  A
drawback is that the BP radii are not invariant, they depend on the
frame where they are evaluated. The BP radii transform as a
well-defined mixture of the invariant temporal, longitudinal and
transverse BL radii, given e.g.\ in Ref.\ \cite{3d}.
         
\subsubsection{The Yano--Koonin--Podgoretskii parameterization}

A covariant parameterization of two-particle correlations has been
worked out for non-expanding sources by Yano, Koonin and Podgoretskii
(YKP) \cite{ykp1,ykp2}.  This parameterization was recently applied
to expanding sources by the Regensburg group~\cite{ykpr1,ykpr2}, by
allowing the YKP radius and velocity parameters be momentum dependent:
\bea 
    C_2({\bf k}_1,{\bf k}_2)
        & = & 
        1 + \exp\left[ - R_{\perp}^2({\bf K}) q_{\perp}^2 - 
                    R_{\parallel}^2({\bf K})  (q_z^2 - q_0^2 )
        \right. \nonumber\\ 
        \null & \null  &  \hspace{2cm}
        \left. - \left(R_0^2({\bf K}) + R_{\parallel}^2({\bf K}) \right)
        \left( q \cdot U({\bf K}) \right)^2 \right] ,
\eea    
where the fit parameter $U({\bf K})$ is interpreted~\cite{ykpr1,ykpr2}
as a {\it four-velocity} of a fluid-element~\cite{seyboth-cf98}.
(Note that in YKP index $0$ refers to the time-like components).
This generalized YKP parameterization was introduced to create a
diagonal Gaussian form in the ``rest frame of a fluid-element''.

This form has an advantage as compared to the BP parameterization: the
three extracted YKP radius parameters, $R_{\perp}$, $R_{\parallel}$
and $R_0$ are invariant, independent of the frame where the analysis
is performed, while $U^{\mu}$ transforms as a four-vector.  The price
one has to pay for this advantage is that the kinematic region may
become rather small in the $q_0$, $q_l$, $q_{\perp}$ space, where the
parameters are to be fitted, as follows from the inequalities $Q^2 = -
q\cdot q \ge 0$ and $q^2_0 \ge 0$:
\be
        0 \le  q_0^2  \le q_z^2 +        q_{\perp}^2,
\ee
and the narrowing of the regions in $q^2_0 - q_z^2$ with decreasing
$q_\perp$ makes the experimental determination of the YKP parameters
difficult, especially when the analysis is performed far from the LCMS
rapidities [or more precisely from the frame where $U^{\mu} =
(1,0,0,0)$\,].

Theoretical problems with the YKP parameterization are explained as
follows.  {\it a)} The YKP radii contain components proportional to
${1 /\beta_t}$, which lead to divergent terms for particles with
very low $p_t$ \cite{ykpr1,ykpr2}.  {\it b)} The YKP fit parameters
are not even defined for all Gaussian sources~\cite{ykpr1,ykpr2}.
Especially, for opaque sources, for expanding shells, or for rings of
fire with $\langle \tilde r_x^2\rangle < \langle \tilde r_y^2 \rangle$
the algebraic relations defining the YKP ``velocity" parameter become
ill-defined and result in imaginary values of the YKP
``velocity",~\cite{ykpr1,ykpr2}.  {\it c)} The YKP ``flow velocity"
$U^{\mu}({\bf K})$ is defined in terms of space-time variances at
fixed mean momentum of the particle pairs \cite{ykpr1,ykpr2},
corresponding to a weighted average of particle {\it coordinates}.  In
contrast, the local flow velocity $u^{\mu}(x)$ is defined as a local
average of particle {\it momenta}.  Hence, in general $U^\mu(\bK) \ne
u^\mu(x)$, and the interpretation of the YKP parameter $U^\mu({\bf
  K})$ as a local {\it flow} velocity of a fluid does not correspond
to the principles of kinetic theory.


\section{Hydrodynamical Parameterization \`a la Buda--Lund (BL-H)}
\label{s:bl-h}

The Buda--Lund hydro parameterization (BL-H) was invented in the same
paper as the BL parameterization of the Bose--Einstein correlation
functions~\cite{3d}, but in principle the general BL forms of the
correlation function do not depend on the hydrodynamical ansatz
(BL-H). The BL form of the correlation function can be evaluated for
any, non-thermalized expanding sources, e.g.\  for the Lund string
model also.

The BL-H assumes, that the core emission function is characterized
with a locally thermalized, volume-emitting source:
\bea
        S_{c}(x,{\bf k}) \, d^4 x &  = & {\dst  g \ov (2 \pi)^3} \,
        \, { k^\mu d^4\Sigma_\mu(x) \ov
        \exp\l({\dst  u^{\mu}(x)k_{\mu} \ov  T(x)} -
        {\dst \mu(x) \ov  T(x)}\r) + s}.
        \label{e:s} 
\eea
The degeneracy factor is denoted by $g$, the four-velocity field is
denoted by $u^{\mu}(x)$, the temperature field is denoted by $T(x)$,
the chemical potential distribution by $\mu(x)$ and $s = 0$, $-1$ or
$1$ for Boltzmann, Bose--Einstein or Fermi--Dirac statistics.  The
particle flux over the freeze-out layers is given by a generalized
Cooper--Frye factor, assuming that the freeze-out hypersurface depends
parametrically on the freeze-out time $\tau$ and that the probability
to freeze-out at a certain value is proportional to $H(\tau)$,
\bea
         k^\mu d^4\Sigma_\mu(x) & = &
         m_t \cosh[\eta - y]  
        H(\tau) d\tau \, \taub d\eta \, dr_x \, dr_y.
\eea
The four-velocity $u^\mu(x)$ of the expanding matter is assumed to be
a scaling longitudinal Bjorken flow appended with a linear transverse
flow, characterized by its mean value $\langle u_t\rangle$, see
Refs \cite{3d,uli_l,ster-beier}:
\bea
        u^{\mu}(x) & = & \l( \cosh[\eta] \cosh[\eta_t],
        \, \sinh[\eta_t]  \frac{r_x}{r_t},
        \, \sinh[\eta_t]  \frac{r_y}{r_t},
        \, \sinh[\eta] \cosh[\eta_t] \r), \nonumber \\
        \sinh[\eta_t]   & = & \ave{u_t} r_t / R_G,
\eea
with $ r_t = [r_x^2 + r_y^2]^{1/2}$.  Such a flow profile, with a
time-dependent radius parameter $R_G$, was recently shown to be an
exact solution of the equations of relativistic hydrodynamics of a
perfect fluid at a vanishing speed of sound, Ref.\ \cite{biro}.

Instead of applying an exact hydrodynamical solution with evaporation
terms, the BL-H characterizes the local temperature, flow and chemical
potential distributions of a cylindrically symmetric, finite
hydrodynamically expanding system with the means and the variances of
these distributions.  The hydrodynamical variables $1/T(x)$,
$\mu(x)/T(x)$, are parameterized as
\ben
        {\dst \mu(x) \ov T(x) } & = & {\dst \mu_0 \ov T_0} -
        { \dst r_x^2 + r_y^2 \ov 2 R_G^2}
        -{ \dst (\eta - y_0)^2 \ov 2 \Delta \eta^2 }, \label{e:mu} \\
        {\dst 1 \ov T(x)} & =  &
        {\dst 1 \ov T_0 } \,\,
        \left( 1 + \Big\langle {\Delta T \over T}\Big\rangle_r\, 
                {\dst  r_t^2 \ov 2 R_G^2} \right) \,
        \left( 1 + \Big\langle {\Delta T \over T}\Big\rangle_t  
                \, {\dst (\tau - \overline{\tau})^2 \ov 2 \Delta\tau^2  } \right),
\enn
the temporal distribution of particle evaporation $H(\tau)$ is assumed
to have the form of
\be
        H(\tau) = \frac{1}{(2 \pi \Delta\tau^2)^{3/2}}
        \exp\left[-\frac{(\tau - \taub)^2} {2  \Delta \tau^2} \right],
\ee
and it is assumed that the widths of the particle emitting sources,
e.g.\ $R_G$ and $\Delta\eta$ do not change significantly during the
course of the emission of the observable particles.  The parameters
$\langle {\Delta T / T}\rangle_r$ and $\langle {\Delta T /
  T}\rangle_t$ control the transversal and the temporal changes of
the local temperature profile, see Refs \cite{3d-cf98,3d-qm,3d} for
further details. This formulation of the BL hydro source includes a
competition between the transversal flow and the transverse
temperature gradient, in an analytically tractable form.  In the
analytic evaluation of this model, it is assumed that the transverse
flow is non-relativistic at the point of maximum
emissivity~\cite{uli_l}, the temperature gradients were introduced
following the suggestion of Akkelin and Sinyukov~\cite{akkelin}.

Note that the shape of the profile function in $\eta$ is assumed to
be a Gaussian in Eq.~(\ref{e:mu}) in the spirit of introducing only
means and variances.  However, in Ref.~\cite{1d} a formula was given,
that allows the {\it reconstruction} of this part of the emission
function from the measured double-differential invariant momentum
distribution in a general manner, {\it for arbitrary sources} with
scaling longitudinal expansions.

\subsection{Correlations and spectra for the BL-Hydro} 

Using the binary source formulation, reviewed in the next section, the
invariant single particle spectrum is obtained as
\bea 
        N_1({\bf k}) & = & 
                {\dst g \ov (2 \pi)^3} \, \overline{E} \, \overline{V} \, 
        \overline{C} 
        \, {
        1 \ov   
        \exp\l({\dst  u^{\mu}(\overline{x})k_{\mu} 
        \ov  T(\overline{x})} -
        {\dst \mu(\overline{x}) \ov  T(\overline{x})}\r) + s}.
        \label{e:bl-n1-h}
\eea
The two-particle Bose--Einstein correlation function was evaluated in
the binary source formalism in Ref.\ \cite{3d-cf98}:
\ben
        C_2({\bf k}_1,{\bf k}_2)
        & = &
        1 +
         \lambda_* \, \Omega(Q_{\parallel}) \,\,\,
                \exp\left(- Q_{\parallel}^2 \overline{R}_{\parallel}^2
                 - Q_{=}^2 \overline{R}_{=}^2
                 - Q_{\perp}^2 \overline{R}_{\perp}^2 \right), 
                \label{e:blosc}
\eea
where the pre-factor $\Omega(Q_\parallel)$ induces oscillations within
the Gaussian envelope as a function of $Q_\parallel$. This oscillating
pre-factor satisfies $0 \le \Omega(Q_\parallel) \le 1$ and $\Omega(0)
= 1$. This factor is given as
\bea
        \Omega(Q_{\parallel}) & = &   \cos^2( Q_{\parallel} 
                        \overline{R}_{\parallel} \,
                               { \Delta\etab} )
                + \sin^2(Q_{\parallel} \overline{R}_{\parallel} \,
                        { \Delta\etab})
                  \tanh^2( {\etab}).
                \label{e:bl-omega}
\eea
The invariant BL decomposition of the relative momentum is utilized to
present the correlation function in the simplest possible form.
Although the shape of the BECF is non-Gaussian, because the factor
$\Omega(Q_{\parallel})$ results in oscillations of the correlator, the
result is still explicitly boost-invariant.  Although the source is
assumed to be cylindrically symmetric, we have 6 free fit parameters
in this BL form of the correlation function: $\lambda_*$, $R_=$,
$R_{\parallel}$, $R_\perp$, $\etab$ and ${\Delta\etab}$.  The latter
controls the period of the oscillations in the correlation function,
which in turn carries information on the separation of the effective
binary sources. This emphasizes the importance of the oscillating
factor in the BL Bose--Einstein correlation function.

The parameters of the spectrum and the correlation function are the
same, defined as follows.  In the above equations, $\overline{a}$
means a momentum-dependent average of the quantity $a$.  The average
value of the space-time four-vector $\overline{x}$ is parameterized by
$(\overline{\tau},\overline{\eta},\overline{r_{x}},\overline{r_{y}})$,
denoting longitudinal proper-time, space-time rapidity and transverse
directions. These values are obtained in terms of the BL-H parameters
in a linearized solution of the saddle-point equations as 
\ben
\overline{\tau} & = & \tau_0, \\
\overline{\eta} & = & {(y_0 - y) /\left[ 1 + \Delta\eta^2 m_t / T_0
  \right]},
                \label{e:etabar}\\
        \overline{r_{x}} & = & \langle u_t \rangle  R_G
                {\dst p_t \ov T_0 +
                \overline{E} \left( \langle u_t\rangle +
                        \langle {\dst \Delta T / T}\rangle_r
                        \right) },       \\
        \overline{r_{y}} & =  0.
\enn
In Eq.~(\ref{e:bl-n1-h}), $\overline{E}$ stands for an average energy,
$\overline{V}$ for an average volume of the effective source of
particles with a given momentum $k$ and $\overline C$ for a correction
factor, each defined in the LCMS frame:
\bea
                \overline{E} & =&
                m_t \cosh(\overline{\eta}),
                \label{e:inve}  \\
        {\overline{V}} & = &
                ( 2 \pi )^{\frac{3}{2}}\, \overline{R}_{\parallel} \,\,
                                    \overline{R}_{\perp}^2 \,
                \, {\dst {\Delta\taub} \,
                \ov {\Delta\tau} },
                        \label{e:invv}
                \\
        \overline{C} \, & = &
                \exp\left( {{\Delta \etab}^2 / 2} \right) /
                \sqrt{\lambda_*}. 
\enn
The average invariant volume ${\overline{V}}$ is given as a
time-averaged product of the transverse area $\overline{R}_{\perp}$
and the invariant longitudinal source size $\overline{R}_{\parallel}$,
given as
\ben
        \overline{R}_{\perp}^2 & = & R^2_{..} \, = \, R^2_{:} \, = \,
                {R_G^2 /\left[ 
                1 + \left( \langle u_t \rangle^2
                + \langle { {\small \Delta T / T } }\rangle_r \right)
                {\overline{E} / 
                T_0}  \right] }, \label{e:rpp-bar} \\
        \overline{R}_{\parallel}^2 & = &
                \taub^2 \, {\Delta\etab}^2, \\
        {\Delta\etab}^2 & = &
                {\Delta\eta^2 /\left( 
                1 +  \Delta\eta^2 {\overline{E} / T_0} \right) },\\
   	\overline{R}_{=}^2 & = &    {\Delta\taub}^2 
		\, = \,
                {\Delta \tau^2/\left( 
                1 + \langle { {\small \Delta T / T} }\rangle_t
                {\overline{E} / T_0} \right) }. 
                \label{e:tau-bar} 
\enn
This completes the specification of the shape of particle spectrum and
that of the two-particle Bose--Einstein correlation function.  These
results for the spectrum correspond to the equations given in
Ref.\ \cite{3d} although they are expressed here using an improved
notation.
        
In a generalized form, the thermal scales are defined as the
$\overline{E}/T_0 $ $ \rightarrow \infty$ limit of Eqs
(\ref{e:rpp-bar}--\ref{e:tau-bar}), while the geometrical scales
correspond to dominant terms in the $\overline{E}/T_0 \rightarrow 0 $
limit of these equations.  In all directions, including the temporal
one, the length-scales measured by the Bose--Einstein correlation
function are dominated by the smaller of the thermal and the
geometrical length-scales.  As shown in Sections~\ref{s:na22} and
\ref{s:bl-pbpb}, the width of the rapidity distribution and the slope
of the transverse-mass distribution is dominated by the bigger of the
geometrical and the thermal length-scales. This is the analytic
reason, why the geometrical source sizes, the flow and temperature
profiles of the source can only be reconstructed with the help of a
simultaneous analysis of the two-particle Bose--Einstein correlation
functions and the single-particle momentum distribution
~\cite{nr,1d,3d,3d-qm,mpd95}.

If the geometrical contributions to the HBT radii are sufficiently
large as compared to the thermal scales, they cancel from the measured
HBT radius parameters. In this case, even if the geometrical source
distribution for different particles (pions, kaons, protons) were
different, the HBT radii (lengths of homogeneity) approach a scaling
function in the large $\overline{E}/T_0$ limit. Up to the leading
order calculation in the transverse coordinate of the saddle-point,
this model predicts a scaling in terms of $\overline{E}$, which
variable coincides with the transverse mass $m_t$ at mid-rapidity.\break
Phenomenologically, the scaling law can be summarized as $R_i \propto
m_t^{\alpha_i}$, where $i$ indexes the directional dependence, and the
exponent $\alpha_i$ may be slightly rapidity dependent, due to the
difference between $\overline{E}$ and $m_t$, and it may
phenomenologically reflect the effects of finite size corrections as
well.  Note also that such a scaling limiting case is only a
possibility in the BL-H, valid in certain domain of parameter space,
but it is not a necessity.  The analysis of Pb + Pb collisions at 158
AGeV indicates that BL-H describes the data fairly well, but the
longitudinal radius component exhibits different scaling behavior
from the transverse radii, see Section~\ref{s:bl-pbpb} for more
details.

\section{Binary Source Formalism} 

Let us first consider the binary source representation of the BL-H
model.  The two-particle Bose--Einstein correlation function was
evaluated in Ref.\ \cite{3d} only in a Gaussian approximation, without
applying the binary source formulation.  An improved calculation was
recently presented in Ref.\ \cite{3d-cf98}, where the correlation
function was evaluated using in the binary source formulation, and the
corresponding oscillations were found.

Using the exponential form of the $\cosh[\eta -y]$ factor, the BL-H
emission function $S_c(x,{\bf k}) $ can be written as a sum of two
terms:
\bea
        S_c(x,{\bf k}) & = & 0.5\, [ S_+(x,{\bf k}) + S_-(x,{\bf k}) ] ,\\
        S_\pm(x,{\bf k}) & = & \frac{g}{(2 \pi)^3}
         m_t \exp[\pm\eta \mp y]  H_*(\tau) \frac{1}{[f_B(x,{\bf k}) + s]} 
                ,\\
        f_B(x,{\bf k}) & = &\exp\left[\frac{k^{\mu}  u_{\mu}(x) 
                                - \mu(x)}{T(x)} \right] .
\eea
Let us call this splitting as the binary source formulation of the
BL-H parameterization.  The effective emission function components are
both subject to Fourier transformation in the BL approach.  In an
improved saddle-point approximation, the two components $S_+(x,k)$ and
$S_-(x,k)$ can be Fourier-transformed independently, finding the
separate maxima (saddle point) $\xb_+$ and $\xb_-$ of $S_+(x,k)$ and
$S_-(x,k)$, and performing the analytic calculation for the two
components separately.

The oscillations in the correlation function are due to this effective
separation of the pion source to two components, a splitting caused by
the Cooper--Frye flux term.  These oscillations in the intensity
correlation function are similar to the oscillations in the intensity
correlations of photons from binary stars in stellar
astronomy~\cite{hbt-bin}.

Due to the analytically found oscillations, the presented form of the
BECF goes beyond the single Gaussian version of the saddle-point
calculations of Refs \cite{uli_s,uli_l}.  This result goes also beyond
the results obtainable in the YKP or the BP parameterizations. In
principle, the binary-source saddle-point calculation gives more
accurate analytic results than the numerical evaluation of space-time
variances, as the binary-source calculation keeps non-Gaussian
information on the detailed shape of the Bose--Einstein correlation
function.

Note that the oscillations are expected to be small in the BL-H
picture, and the Gaussian remains a good approximation to
Eq.~(\ref{e:blosc}), but with modified radius parameters.

\subsection{The general binary source formalism }

In the previous subsection, we have seen how effective binary sources
appear in the BL-H model in high energy physics. However, binary
sources appear generally: in astrophysics, in form of binary stars, in
particle physics, in form of W$^+$W$^-$ pairs, that separate before they
decay to hadrons.
 
Let us consider first the simplest possible example, to see how the
binary sources result in oscillations in the Bose--Einstein or
Fermi--Dirac correlation function. Suppose a source distribution $s(x
- x_+)$ describes e.g.\ a Gaussian source, centered on $x_+$.  Consider
a binary system, where the emission happens from $s_+= s(x-x_+)$ with
fraction $f_+$, or from a displaced source, $s_-= s(x-x_-)$, centered
on $x_-$, with a fraction $f_-$.  For such a binary source, the
amplitude of the emission is
\be
                \rho(x) = f_+ s(x - x_+) + f_- s(x - x_-),
\ee
and the normalization requires
\be
                f_+ + f_- = 1\,.
\ee
The two-particle Bose--Einstein or Fermi--Dirac correlation function
is
\bea
                C(q) & = & 1 \pm | \tilde\rho(q)|^2 =
                1 \pm  \Omega(q) |\tilde s(q)|^2,
\eea
where $+$ is for bosons, and $-$ for fermions.  The oscillating
pre-factor $\Omega(q)$ satisfies $0 \le \Omega(q) \le 1$ and
$\Omega(0) = 1$. This factor is given as 
\bea 
\Omega(q) & = & \left[
  (f_+^2 + f_-^2) + 2 f_+ f_- \cos[q(x_+ - x_-)] \right]\,. 
\eea 
The strength of the oscillations is controlled by the relative
strength of emission from the displaced sources and the period of the
oscillations can be used to learn about the distance of the emitters.
In the limit of one emitter ($f_+ = 1$ and $f_- = 0$, or vice versa),
the oscillations disappear.

The oscillating part of the correlation function in high energy
physics is expected to be much smaller, than that of binary stars in
stellar astronomy.  In particle physics, the effective separation
between the sources can be estimated from the uncertainty relation to
be $x_\pm = |x_+ - x_-| \approx 2 \hbar/M_{\rm W} \approx 0.005$ fm.
Although this is much smaller, the effective size of the pion source,
1 fm, one has to keep in mind that the back-to-back momenta of the
W$^+$W$^-$ pairs can be large, as compared to the pion mass.  Due to
this boost, pions with similar momentum may be emitted from different
W-s with a separation which is already comparable to the 1 fm
hadronization scale, and the resulting oscillations may become
observable.

In stellar astronomy, the separation between the binary stars is
typically much larger than the diameter of the stars, hence the
oscillations are well measurable. In principle, similar oscillations
may provide a tool to measure the separation of the W$^+$ from W$^-$
in 4-jet events at LEP2.  The scale of separation of W$^+$W$^-$ pairs is
a key observable to estimate in a quantum-mechanically correct manner
the influence of the Bose--Einstein correlations on the reconstruction
of the W mass.

In heavy ion physics, oscillations are seen in the long-range part of
the p + p Fermi--Dirac correlation function~\cite{na49-pp}, with a
half-period of $Q_h = 30$ MeV.  This implies a separation of $x_{\pm}
= \pi \hbar/Q_h \approx 20$ fm, which can be attributed to
interference between the the two peaks of the NA49 proton $dn/dy$
distribution \cite{na49-p}, separated by $\Delta y = 2.5$.  As for
the protons we have $m \gg T_0 = 140$ MeV, we can identify this
rapidity difference with the space-time rapidity difference between
the two peaks of the rapidity distribution.  The longitudinal scale of
the separation is then given by $x_{\pm} = 2 \taub
\sinh(\Delta\eta_p/2) $, which can be used to estimate the mean
freeze-out time of protons, $\taub = \pi \hbar /[2 Q_h
\sinh(\Delta\eta_p/2)] \approx 6.4 $ fm/c, in a good agreement with
the average value of $\taub = 5.9 \pm 0.6$ as extracted from the
simultaneous analysis of the single-particle spectra and HBT radii in
NA44, NA49 and WA98 experiments in the Buda--Lund picture, as
summarized in Section~\ref{s:bl-pbpb}.


\section{Particle Correlations and Spectra at 30 -- 160 AMeV} 
\label{s:chic}

\def\bbox#1{{\bf #1} }

There are important qualitative differences between relativistic heavy
ion collisions at CERN SPS and those at non-relativistic energies from
the point of view of particle sources.  Low and intermediate energy
reactions may create a very long-lived, evaporative source, with
characteristic lifetimes of a few 100 fm/c, in contrast to the
relatively short-lived systems of lifetimes of the order of 10 fm/c at
CERN SPS.  During such long evaporation times, cooling of the source
is unavoidable and has to be included into the model.  Furthermore, in
the non-relativistic heavy ion collisions mostly protons and neutrons
are emitted and they have much stronger final-state interactions than
the pions dominating the final state at ultra-relativistic energies,
see Refs \cite{boal,bauer,ardouin} for recent reviews.

The evolution of the particle emission in a heavy-ion collision at
intermediate energies may roughly be described as: production of
pre-equilibrium particles; expansion and possible freeze-out of a
compound source; possible evaporation from an excited residue of the
source.  Note though that this separation is not very distinct and
there is an overlap between the different stages.  The importance of
the various stages above also depends on the beam energy and the
impact parameter of the collision.  See the review paper of
Ref.\ \cite{ardouin} for greater details.

Sophisticated microscopical transport descriptions \cite{aichelin},
such as the BUU\break (Boltzmann--Uehling--Uhlenbeck) and the QMD (Quantum
Molecular Dynamics) models are well-known and believed to provide a
reasonable picture of proton emission in central heavy ion collisions
from a few tenths up to hundreds of MeV per nucleon.  However, the BUU
model predicts too large correlations and underpredicts the number of
protons emitted with low energies, for the reaction $^{36}$Ar +
$^{45}$Sc at $E = 120$ and 160 MeV/nucleon, see Ref.\ \cite{upe}.  This
indicates that the simultaneous description of two-particle
correlations and single-particle spectra is a rather difficult task.
For energies below a few tens of MeV per nucleon, where long-lived
evaporative particle emission is expected to dominate, the measured
two-proton\break\vfill\pagebreak\noindent correlation functions were found to be consistent with
compound-nucleus model predictions \cite{cmpn}.

A simultaneous analysis of proton and neutron single particle spectra
and two-particle correlation was presented in Ref.\ \cite{nrt}.  This
model calculation described the second stage above and, for long
emission times, also part of the third stage.  In Ref.~\cite{nrt}, the
competition among particle evaporation, temperature gradient and flow
was investigated in a phenomenological manner, based on a simultaneous
analysis of quantum statistical correlations and momentum
distributions for a non-relativistic, spherically symmetric,
three-dimensionally expanding, finite source.  The model used can be
considered as a non-relativistic, spherically symmetric version of the
BL-H hydro parameterization~\cite{nrt}.

The non-relativistic kinetic energy is denoted by $E_k(\bbox{k}) =
\bbox{k}^{\, 2} / (2\, m)$.  The following result is obtained for the
effective source size $R_*$:
\begin{eqnarray}
        { R_*^2(\bbox{k}) }
& = &
        \frac{ R_G^2 }{ 1 + 
        \left[ \ave{\Delta T /T}_r E_k(\bbox{k}) 
                + m \ave{u_t}^2\right]/ T_0  }\,. 
\label{e:reff-joh}
\end{eqnarray}
The analytic results for the momentum distribution and the quantum
statistical correlation function are given in the Boltzmann
approximation as
\begin{eqnarray}
    N_1(\bbox{k})
& = &
       \frac{ g }{ (2 \pi)^3 }   
        E_k(\bbox{k}) V_*({\bf k}) 
    \exp\left[
      - \frac{( \bbox{k} - m \bbox{u} (\bbox{r}_s(\bbox{k}\,)) )^2}
             {2 m T(\bbox{r}_s(\bbox{k}\,))}
      + \frac{ \mu(\bbox{r}_s(\bbox{k}))}{ T(\bbox{r}_s(\bbox{k}))  }
        \right]\ , 
\label{e:imd} \nonumber\\
        && \\
        V_*({\bf k}) & = & 
        \left[ 2 \pi R_*^2(\bbox{k}\,)\right]^{3/2} \,, \\
        C(K,  \Delta k\,)
& = &
        1 \pm \exp( - R_*^2( \bbox{K})
        \bbox{\Delta k}^2 - \Delta t^2 \Delta E^2)\ .
\end{eqnarray}
The effects of final-state Coulomb and Yukawa interactions on the
two-particle relative wave-functions are neglected in these analytic
expressions.  When comparing to data, the final-state interactions
were taken into account, see Ref.\ \cite{nrt} for further details.

These general results for the correlation function indicate structural
similarity between the non-relativistic flows in low/intermediate
energy heavy ion collisions~\cite{nr,nrt,sol} and the transverse flow
effects in relativistic high energy heavy ion and elementary particle
induced reactions \cite{3d-qm,mpd95,3d}.  The radius parameters of
the correlation function and the slopes of the single-particle spectra
are momentum dependent both for the non-relativistic versions of the
model, presented in Refs \cite{nr,nrt,sol} and for the model-class
with scaling relativistic longitudinal flows, discussed in
Refs \cite{3d,3d-qm,mpd95,3d-cf98}.

Such a momentum-dependent effective source size has been seen in the
proton--proton correlation functions in the
$^{27}${Al}\,($^{14}${N},\,pp) reactions at $E = 75$ MeV/nucleon
\cite{bauer}: the larger the momentum of the protons the smaller the
effective source size \cite{bauer}, in qualitative agreement with
Eq.~(\ref{e:reff-joh}).

\begin{table}[ht]
\caption{Parameter values obtained from fitting hydro parameters to 
n and p spectra and correlation functions, as measured by the 
CHIC Collaboration in 30~AMeV $^{40}$Ar + $^{197}$Au reactions}
\label{tab_1}
\vspace*{-6pt}
\begin{center}
\begin{tabular}{lcccc}
\hline\\[-10pt]
\null           &   $R_G$ (fm)                 &   $T_0$ (MeV)
           &   $\ave{\Delta T/T}_r $      &   $\ave{u}_t $ \\[2pt] 
\hline\\[-10pt]
Neutrons   &   4.0                        &   3.0
           &   0.0{\phantom 0}                        &   0.018 \\
Protons    &   4.0                        &   5.0
           &   0.16                       &   0.036 \\
\hline
\end{tabular}
\end{center}
\end{table}


This model was applied in Ref.\ \cite{nrt} to the reaction $^{40}$Ar +
$^{197}$Au at 30 MeV/ nucleon.  With the parameter set presented in
Table~\ref{tab_1}, we have obtained a simultaneous description of the
n and p single particle spectra as well as the nn and pp
correlation functions as given by Refs \cite{bo,Cronq,Ghetti}.  See
Ref.\ \cite{nrt} for further details and discussions.

The main effects of the temperature gradient are that it introduces
{\it i)} a mo\-men\-tum-dependent effective temperature which is
decreasing for increasing momentum, resulting in a suppression at high
momentum as compared to the Boltzmann distribution; {\it ii)} a
momentum-dependent effective source size which decreases with
increasing total momentum.  Agreement with the experimental data is
obtained only if the time of duration of the particle emission was
rather long, $\langle t\rangle \approx 520$ fm with a variance of
$\approx 320$ fm/c.

The obtained parameter set reflects a moderately large system
(Gaussian radius parameter $R_G$ = 4.0 fm) at a moderate temperature
($T_0(n)$ = 3 MeV and $T_0(p)$ = 5 MeV) and small flow. The neutrons
and the protons seem to have different local temperature
distributions: the neutron temperature distribution is homogeneous,
while the temperature of the proton source decreases to $T_s(p) = 4.3
$ MeV at the Gaussian radius, a difference that could be attributed to
the difference between their Coulomb interactions~\cite{nrt}.  An
agreement between the model and the data was obtained only if some
amount of flow was included~\cite{nrt}.

After the completion of the data analysis, a new family of exact
solutions of fireball hydrodynamics was found in Ref.\ \cite{sol}, which
features scaling radial Hubble flow, and an initial inhomogeneous,
arbitrary temperature profile.  The competition of the temperature
gradients and flow effects were shown to lead to the formation of
spherical shells of fire in this class of exact hydrodynamical
solutions~\cite{sol}, if the temperature gradient was stronger than
the flow, $\langle \Delta T / T\rangle_r > m \langle
u_t\rangle^2/T_0$.  This is the case found from the analysis of proton
spectra and correlations in Ref.\ \cite{nrt}, while the neutron data do
not satisfy this condition. Assuming the validity of non-relativistic
hydrodynamics to characterize this reaction, one finds that {a slowly
  expanding, spherical shell of fire is formed by the protons}, while
the {neutrons remain in a central, slightly colder and even slower
  expanding, normal fireball} in 30 AMeV $^{40}$Ar + $^{197}$Au heavy
ion reactions.

\section{Description of h + p Correlations and Spectra at CERN SPS} 
\label{s:na22}

\def\vs{\vskip}
\def\\{\hfill\break}
\def\ipp{\"{\char'20}}
\def\ran{\rangle}
\def\lan{\langle}
\def\hf{\hfill}
\def\ds{\displaystyle}

\def\ifmath#1{\relax\ifmmode #1\else $#1$\fi}%
\def\ra{{\mathrm{a}}}
\def\rb{{\mathrm{b}}}
\def\rc{{\mathrm{c}}}
\def\rd{{\mathrm{d}}}
\def\re{{\mathrm{e}}}
\def\rf{{\mathrm{f}}}
\def\rg{{\mathrm{g}}}
\def\rh{{\mathrm{h}}}
\def\rd{\ifmath{{\mathrm{d}}}}
\def\rG{\ifmath{{\mathrm{G}}}}
\def\rK{\ifmath{{\mathrm{K}}}}
\def\rL{\ifmath{{\mathrm{L}}}}
\def\ro{\ifmath{{\mathrm{o}}}}
\def\rp{\ifmath{{\mathrm{p}}}}
\def\rt{\ifmath{{\mathrm{t}}}}
\def\rT{\ifmath{{\mathrm{T}}}}
\def\rs{\ifmath{{\mathrm{s}}}}
\def\rS{\ifmath{{\mathrm{S}}}}
\def\all{\ifmath{{\mathrm{all}}}}
\def\ev{\ifmath{{\mathrm{ev}}}}
\def\eff{\ifmath{{\mathrm{eff}}}}
\def\rms{\ifmath{{\mathrm{rms}}}}
\def\side{\ifmath{{\mathrm{side}}}}
\def\out{\ifmath{{\mathrm{out}}}}
\def\vec#1{{\mbox{\bf #1}}}

The invariant spectra of $\pi^-$ mesons produced in $(\pi^+/\rK^+)\rp$
interactions at 250 GeV/c are analysed in this section in the
framework of the BL-H model of three-dimensionally expanding
cylindrically symmetric finite systems, following the lines of
Ref.\ \cite{na22}.  The EHS/NA22 Collaboration has been the first to
perform a detailed and combined analysis of single-particle spectra
and two-particle Bose--Einstein correlations in high energy
physics~\cite{na22}.  NA22 reported a detailed study of
multi-dimensional Bose--Einstein correlations, by determining the
side, out and the longitudinal radius components at two different
values of the mean transversal momenta in $(\pi^+/\rK^+)\rp$ at CERN
SPS energies~\cite{na22-hbt}.  It turned out, however, that the
experimental two-particle correlation data were equally well described
by a static Kopylov--Podgoretskii parameterization as well as by the
predictions of hydrodynamical parameterizations for longitudinally
expanding, finite systems.  In Refs \cite{3d,3d-qm} we have shown,
that the combined analysis of two-particle correlations and
single-particle spectra may result in a dramatic enhancement of the
selective power of data analysis.

The double-differential invariant momentum distribution of
Eq.\ (\ref{e:bl-n1-h}) can be substantially simplified for
one-dimensional slices
\cite{3d,dkiang}.

{\it i)} At fixed $m_t$, the rapidity distribution reduces to 
\bea
  N_1({\bf k}) & = & C_m \exp\left[ -\frac{(y-y_0)^2}{2\Delta y^2}\right],
  \label{e:na22.6}\\
        \Delta y^2 & = & \Delta\eta^2 + T_0/m_t\,,
  \label{e:na22.2}
\eea
where $C_m$ is an $m_t$-dependent normalization coefficient and $y_0$
is defined above. The width parameter $\Delta y^2$ extracted for
different $m_t$-slices is predicted to depend linearly on $1/m_t$,
with slope $T_0$ and intercept $\Delta \eta^2$ .  Observe, that this
width is dominated by the bigger of the geometrical scale ($\Delta
\eta$) and the thermal scale $T_0/m_t$.

\begin{figure}[th]
\vspace*{-12pt}
\centerline{\epsfig{figure=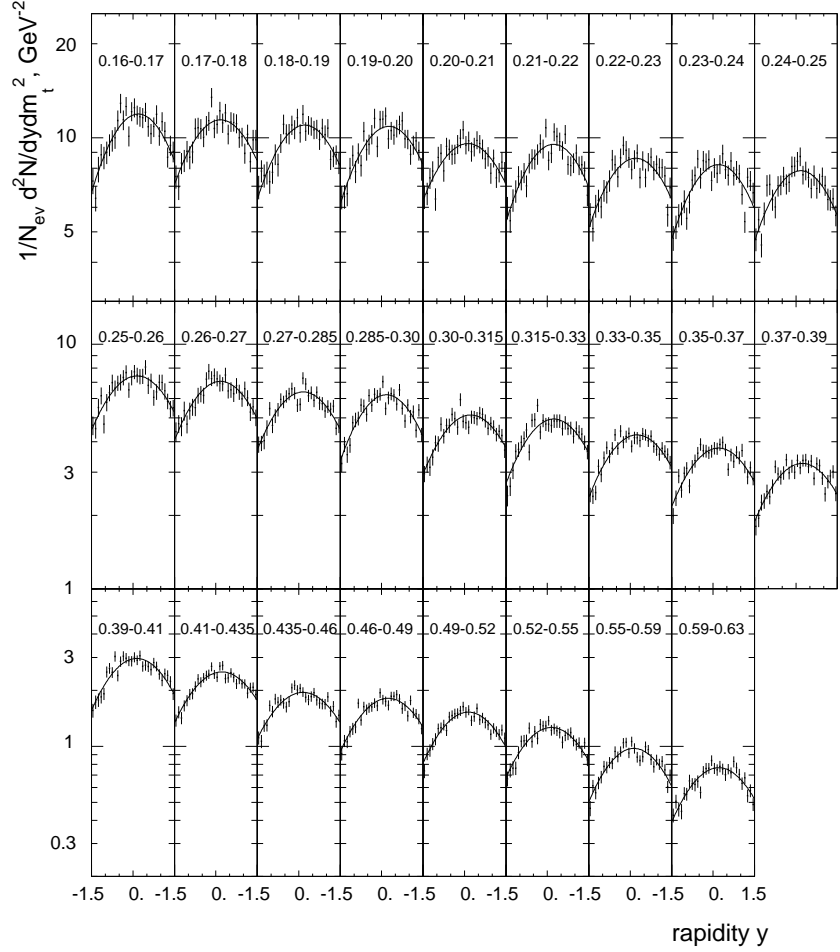,width=12.0cm}}
\vspace*{-1.4cm}
\caption{The rapidity distributions of centrally produced pions ($|y|<1.5$)
  for different $m_t$-slices given. The curves are the fit results
  obtained analytically using the BL-H parameterization.}
\label{f:na22_1}
\end{figure}

\begin{figure}[th]
\vspace*{-12pt}
\centerline{\epsfig{figure=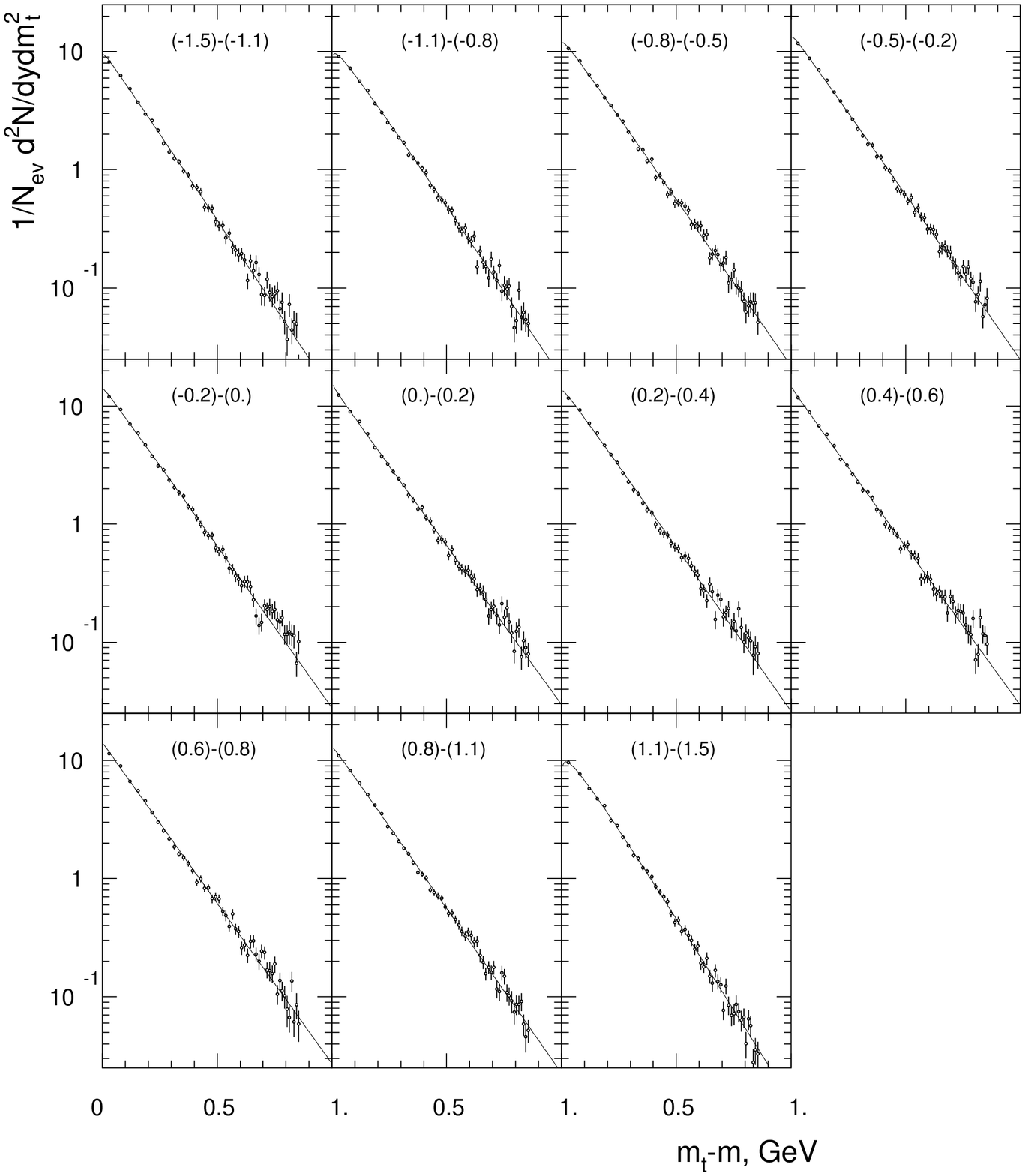,width=12.0cm}}
\vspace*{-1.4cm}
\caption{The $m_t$ distributions of centrally produced pions ($|y|<1.5$)
  for different $y$-slices given. The curves are the fit results
  obtained analytically using the BL-H parameterization.}
\label{f:na22_3}
\end{figure}
Note that for static fireballs or spherically expanding shells
(\ref{e:na22.6}) and (\ref{e:na22.2}) are satisfied with $\Delta\eta =
0$ \cite{dkiang}. Hence the experimental determination of the $1/m_t$
dependence of the $\Delta y$ parameter can be utilized to distinguish
between longitudinally expanding finite systems versus static
fireballs or spherically expanding shells.

{\it ii)} At fixed $y$, the $m_t^2$-distribution reduces to
\be
N_1({\bf k}) = C_y m_t^{\alpha} \exp\left(-\frac{m_t}{T_{\eff}}\right)\,,
  \label{e:na22.7}
\ee
where $C_y$ is a $y$-dependent normalization coefficient and $\alpha$ 
is related to the effective dimensions of inhomogeneity in the source
as $\alpha = 1 - d_{\eff}/2$~\cite{3d}. 
The $y$-dependent ``effective temperature" $T_{\eff}(y)$  reads as~\cite{3d}
\be
T_{\eff}(y) = \frac{T_*}{1+a(y-y_0)^2} ,
  \label{e:na22.8}
\ee
where $T_*$ is the maximum of $T_{\eff}(y)$ achieved at $y=y_0$, and
parameter $a$ can be expressed with the help of the other fit parameters,
see Refs \cite{3d,na22}.

The slope parameter at mid-rapidity, $T_*$ is also determined by an
interplay of the central temperature $T_0$ the flow effects modeled
by $\langle u_t\rangle^2 $ and the temperature difference between the
surface and the center, as characterized by $\langle {\Delta T /
  T}\rangle_r$ \cite{3d,ster-beier}.  Eq.~(66) of Ref.\ \cite{3d} can
be rewritten as
\be
        T_* = T_0 + m \langle u_t\rangle^2 
                \frac{T_0}
                {T_0 + m \langle {\Delta T \over T}\rangle_r}.
        \label{e:tstar-m}
\ee
The approximations of Eqs~(\ref{e:na22.6}) and (\ref{e:na22.7})
explicitly predict a specific narrowing of the rapidity and 
transverse-mass spectra with increasing $m_t$ and $y$, respectively (cf.\
(\ref{e:na22.2}) and (\ref{e:na22.8})). The character of these
variations is expected \cite{dkiang} to be different for the various
scenarios of hadron matter evolution.  These features of the spectra
were found to be in agreement with the NA22 data~\cite{na22}, and were
utilized to reconstruct the particle source of h + p reactions in the
$(t,r_z)$ plane.

\begin{table}[ht]
\caption{Fit results to NA22 h+p data at CERN SPS with a 
Buda--Lund hydro parameterization  for $|y|<1.5$}       
\label{t:na22_1}
\vspace*{-6pt}
\begin{center}                                                
\begin{tabular}{ccccccc}
\hline\\[-10pt]
    $\alpha$  & ${\Delta \eta}$ & $T_0$ (GeV) &  
  $\langle u_t\rangle$ &  $\langle {\Delta T}/{T}\rangle_r $  
   & $\chi^2$/NDF     \\[2pt]
\hline\\[-10pt]
  $0.26 \pm 0.02$ & $1.36 \pm 0.02$ &  $0.140 \pm 0.003$ 
&  $0.20 \pm 0.07$ & $0.71 \pm 0.14$ & 642/683 \\
\hline
\end{tabular}
\end{center}
\end{table}

\subsection{Combination with two-particle correlations}

As already mentioned in the introduction, more comprehensive
information on geometrical and dynamical properties of the hadron
matter evolution are expected from a combined consideration of
two-particle correlations and single-particle inclusive spectra
~\cite{nr,1d,3d,3d-qm,mpd95,nix,akkelin}.

At mid-rapidity, $y=y_0$ and in the LCMS where $k_{1,z} = - k_{2,z}$,
the effective BP radii can be approximately expressed form the BL-H
parameterization as \cite{3d}:
\bea
R_l^2 & = &\taub^2\Delta \etab^2 ,
  \label{e:na22.12}\\
R_o^2 & = &\overline{R}^2_\perp+\beta_t^2\Delta \taub^2  ,
        \label{e:na22.13} \\
R_s^2 & = &\overline{R}^2_\perp
        \label{e:na22.14}
\eea        
with
\bea
\frac{1}{\Delta \etab^2} & = &\frac{1}{\Delta \eta^2}+\frac{M_t}{T_0} \,,
  \label{e:na22.15}\\
\overline{R}^2_\perp & = &
\frac{R_\rG^2}{1+\frac{M_t}{T_0}(\langle u_t\rangle^2+\langle 
\frac{\Delta T}{T}\rangle_r)}, 
  \label{e:na22.16}
\eea
where parameters $\Delta \eta^2,T_0,\langle u_t\rangle$ and $\langle
{\Delta T}/{T}\rangle_r$ are defined and estimated from the
invariant spectra; $R_\rG$ is related to the transverse geometrical
rms radius of the source as $R_\rG(\rms)=\sqrt{2} R_\rG$; $\taub$ is
the mean freeze-out (hadronization) time; $\Delta \taub$ is related to
the duration time $\Delta \taub$ of pion emission and to the temporal
inhomogeneity of the local temperature, as the relation $\Delta \tau
\geq \Delta \taub$ holds; the variable $\beta_t$ is the transverse
velocity of the pion pair.

The effective longitudinal radius $R_l$, extracted for two different
mass ranges, $M_t=0.26\pm 0.05$ and $0.45\pm 0.09$ GeV/c$^2$ are found
to be $R_l=0.93\pm 0.04$ and $0.70\pm 0.09$ fm, respectively.  This
dependence on $M_t$ matches well the predicted one. Using Eq.
(\ref{e:na22.14}) with $T_0=140\pm 3$ MeV and $\Delta \eta^2 =1.85\pm
0.04$ (Table~\ref{t:na22_1}), one finds that the values of $\taub$
extracted for the two different $M_t$ regions are similar to each
other: $\taub =1.44\pm 0.12$ and $1.36\pm 0.23$ fm/c.  The averaged
value of the mean freeze-out time is $\taub =1.4\pm 0.1$ fm/c.

The width of the (longitudinal) space-time rapidity distribution of
the pion source was found to be $\Delta \eta = 1.36 \pm 0.02$.  Since
this value of $\Delta\eta$ is significantly bigger than 0, the static
fireballs or the spherically expanding shells fail to reproduce the
NA22 single-particle spectra~\cite{na22}, although each of these
models was able to describe the NA22 two-particle correlation data in
Ref.\ \cite{na22-hbt}.

The transverse-plane radii $R_o$ and $R_s$ were reported in Ref.\ 
\cite{na22-hbt} for the whole $M_t$ range are: $R_o=0.91\pm 0.08$ fm
and $R_s=0.54\pm 0.07$ fm.  Substituting in (\ref{e:na22.12}) and
(\ref{e:na22.13}), one obtains (at $\beta_t=0.484$\,c
\cite{na22-hbt}): $\Delta \taub=1.3\pm 0.3$ fm/c. The mean
duration time of pion emission can be estimated as $\Delta \tau \geq
\Delta \taub =1.3\pm 0.3$ fm/c.  A possible interpretation of
$\Delta \tau \approx \taub$ might be that the radiation process occurs
during almost all the hydrodynamical evolution of the hadronic matter
produced in meson--proton collisions.

An estimation for the parameter $R_\rG$ can be obtained from
(\ref{e:na22.13}) and (\ref{e:na22.15}) using the quoted values of
$R_s$, $T_0$, $\langle u_t\rangle$ and $\langle {\Delta T}/{T} \rangle
$ at the mean value of $\langle M_t \rangle =0.31\pm 0.04$~GeV/c
(averaged over the whole $M_t$ range): $R_\rG=0.88\pm 0.13$ fm. The
geometrical rms transverse radius of the hydrodynamical tube,
$R_\rG(\rms)=\sqrt{2}R_\rG=1.2\pm 0.2$ fm, turns out to be larger than
the proton rms transverse radius.

The data favour the pattern according to which the hadron matter
undergoes predominantly longitudinal expansion and non-relativistic
transverse expansion with mean transverse velocity $\langle u_t
\rangle = 0.20\pm0.07$, and is characterized by a large temperature
inhomogeneity in the transverse direction: the extracted freeze-out
temperature at the center of the tube and at the transverse rms radius
are $140\pm3$ MeV and $82\pm7$ MeV, respectively.

\subsection{The space-time distribution of $\pi$ emission}
\begin{figure}[th]
\vspace*{-.4cm}
    \centering\epsfig{figure=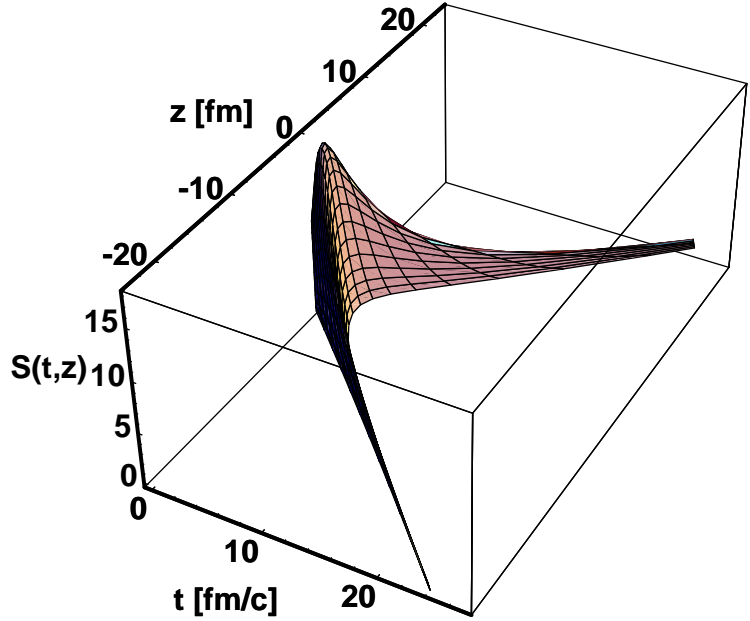,width=10.0cm}
\vspace*{-.8cm}
\caption{The reconstructed $S(t,z)$ emission function in arbitrary units, as a
  function of time $t$ and longitudinal coordinate $z$.  The best fit
  parameters of $\Delta \eta = 1.36$, $y_0 = 0.082$, $\Delta\tau =
  1.3$ fm/c and $\taub = 1.4$ fm/c are used to obtain this plot.
  Note that before we made this reconstruction together with the NA22
  Collaboration, only 1 fm$^2$ area from this extended bumerang shape
  was visible to the intensity interferometry microscope.}
\label{f:na22_5}
\end{figure}

\begin{figure}[th]
\vspace{-2.6cm}
    \centerline{\epsfig{figure=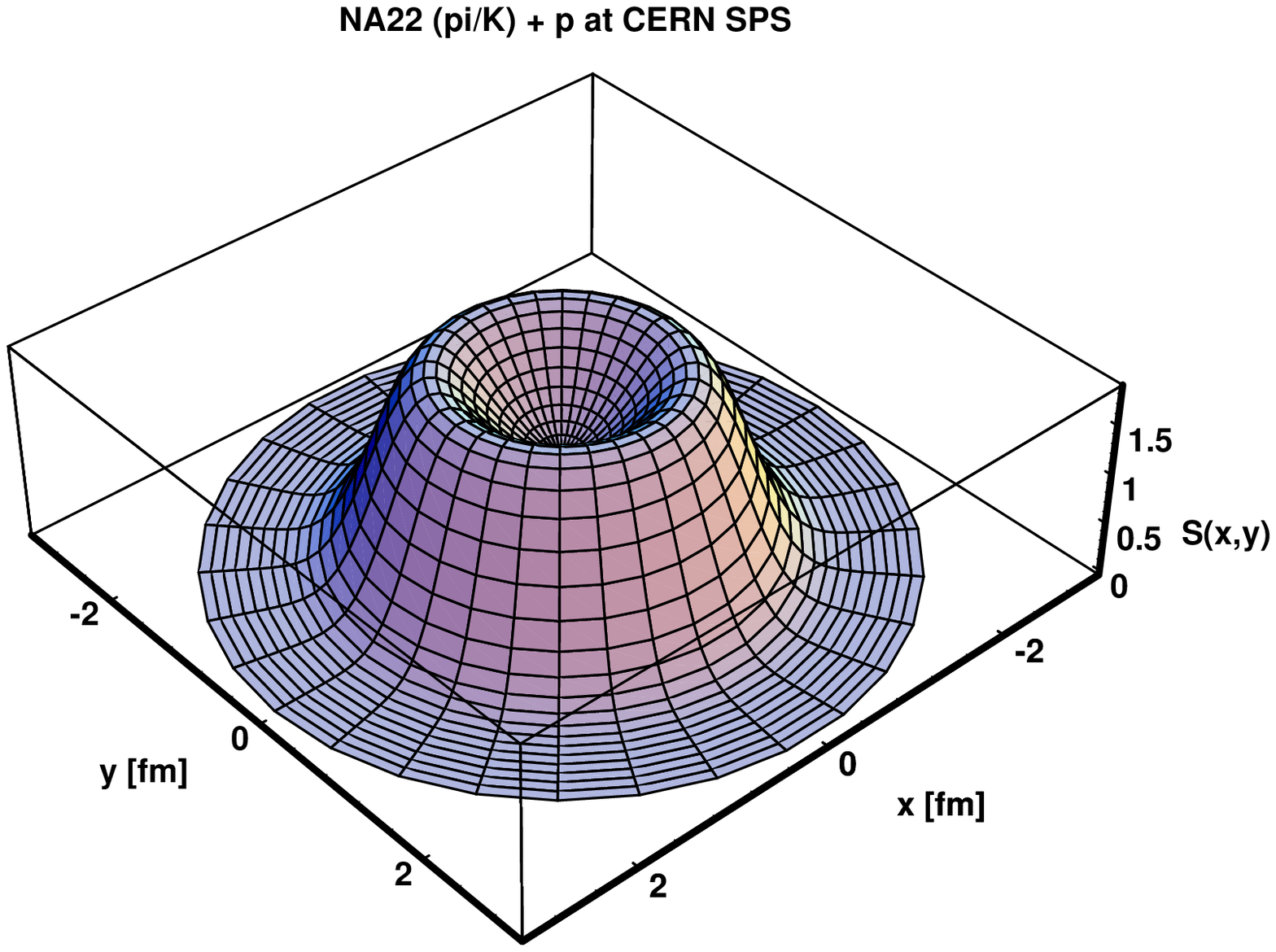,width=11.0cm}}
\vspace{-3.1cm}
\caption{The reconstructed $S(r_x,r_y)$ emission function in 
  arbitrary vertical units, as a function of the transverse
  coordinates $r_x$ and $r_y$.  The shape has been reconstructed
  assuming the validity of a non-relativistic solution of
  hydrodynamics in the transverse direction, and using the values of
  $T_0$, $\langle \Delta T/T\rangle$ and $\langle u_t \rangle$ as
  obtained from the fits to the single-particle spectra. The root mean
  square width of the source distribution was obtained from the fits
  to the NA22 Bose--Einstein correlation functions. The momentum
  variables and the longitudinal and temporal variables are integrated
  over.}
\label{f:na22-r}
\end{figure}

A reconstruction of the space-time distribution of pion emission
points is shown in Fig.~\ref{f:na22_5}, expressed as a function of the
cms time variable $t$ and the cms longitudinal coordinate $z\equiv
r_z$.  The momentum-integrated emission function along the $z$ axis,
i.e.\ at ${\vec r}_t = (r_x, r_y) = (0,0)$ is given by
\begin{equation}
 S(t,z) \propto \exp\left(-  {(\tau - \taub)^2\over 2 \Delta \tau^2} 
\right) \exp\left( - {(\eta - y_0)^2   \over 2 \Delta \eta^2} \right).
  \label{e:na22.20}
\end{equation}
It relates the parameters fitted to the NA22 single-particle spectrum
and HBT radii to the particle production in space-time. The
coordinates $(t,z)$ are expressed with the help of the longitudinal
proper-time $\tau$ and space-time rapidity as $\eta$ as $(\tau
\cosh(\eta), \tau \sinh(\eta) )$.

We find a structure looking like a boomerang, i.e.\ particle
production takes place close to the regions of $z=t$ and $z=-t$, with
gradually decreasing probability for ever larger values of space-time
rapidity. Although the mean proper-time for particle production is
$\taub=1.4$ fm/c, and the dispersion of particle production in
space-time rapidity is rather small, $\Delta \eta = 1.35$ fm, we still
see a characteristic long tail of particle emission on both sides of
the light-cone, giving a total of 40 fm maximal longitudinal extension
in $z$ and a maximum of about 20 fm/c duration of particle
production in the time variable $t$.

In the transverse direction, only the rms width of the source can be
directly inferred from the BP radii.  However, the additional
information from the analysis of the transverse momentum distribution
on the values of $\langle u\rangle_t$ and on the values of
$\ave{\Delta T/T}_r$ can be used to reconstruct the details of the
transverse density profile, as an exact, non-relativistic hydro
solution was found in Ref.\ \cite{sol}, given in terms of the
parameters $\langle u\rangle_t$ and $\ave{\Delta T/T}_r$ and using an
ideal gas equation of state.  Assuming the validity of this
non-relativistic solution in the transverse direction, in the
mid-rapidity range, one can reconstruct the detailed shape of the
transverse density\break\vfill\pagebreak\noindent 
profile.  The result looks like a ring of fire in
the $(r_x,r_y)$ plane, see Fig.~\ref{f:na22-r}.  In this hydro
solution, $\ave{\Delta T/T}_r < m \langle u\rangle_t^2 / T_0$
corresponds to self-similar, expanding fireballs, while $\ave{\Delta
  T/T}_r > m \langle u\rangle_t^2 / T_0$ corresponds to self-similar,
expanding shells or rings of fire.

Due to the strong surface cooling and the small amount of the
transverse flow, one finds that the particle emission in the
transverse plane of  h + p reactions at CERN SPS corresponds to a ring
of fire.  This transverse distribution, together with the scaling
longitudinal expansion, creates an elongated, tube-like source in
three dimensions, with the density of particle production being
maximal on the surface of the tube.


\section{Pb + Pb Correlations and Spectra at CERN SPS} 
\label{s:bl-pbpb}

In Ref.\ \cite{ster-qm99}, an analysis similar to that of the NA22
Collaboration has been performed, with improved analytic
approximations, using Fermi--Dirac or Bose--Einstein statistics $(s =
\pm 1)$ in the analytic expressions fitted to single particle spectra.
The spectra were evaluated with the binary source method, the
Bose--Einstein correlation functions were calculated with the
saddle-point method without invoking the binary source picture.  The
analytical formulas for the BECF and IMD, as were used in the fits,
were summarized in their presently most advanced form in
Section~\ref{s:bl-h}, their development was described in
Refs~\cite{3d,3d-qm,mpd95,3d-cf98,ster-beier,na22}.

\begin{figure}[thb]
\vspace*{-1.cm}
    \centerline{\epsfig{figure=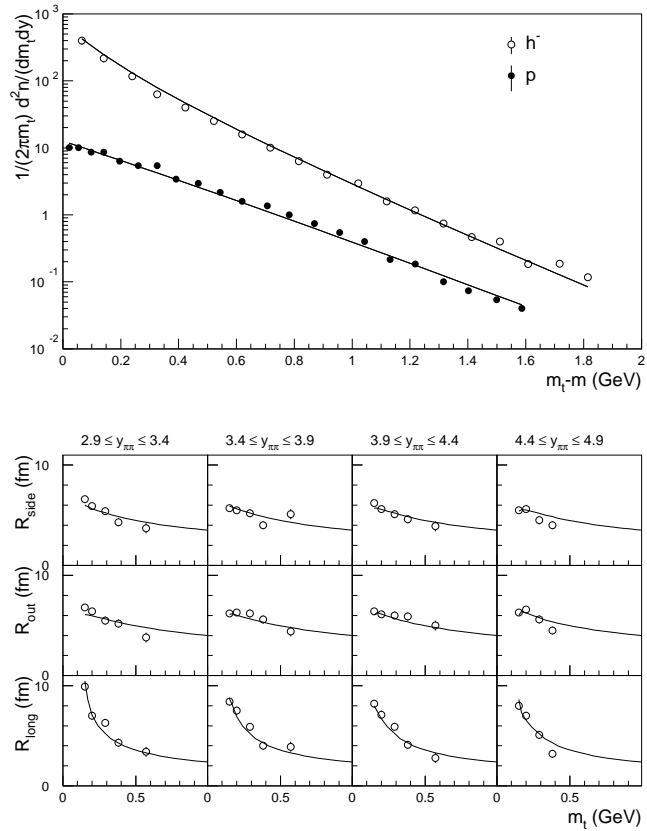,width=9.5cm}}
\vspace*{-1.cm}
\caption{Result of simultaneous fits of the Buda--Lund hydro model
  to particle correlations and spectra in 158 AGeV Pb + Pb reactions
  at CERN SPS (data from the NA49 Collaboration)}
\label{f:xna49}
\end{figure}

\begin{figure}[thb]
\vspace*{-1.cm}
    \centering\epsfig{figure=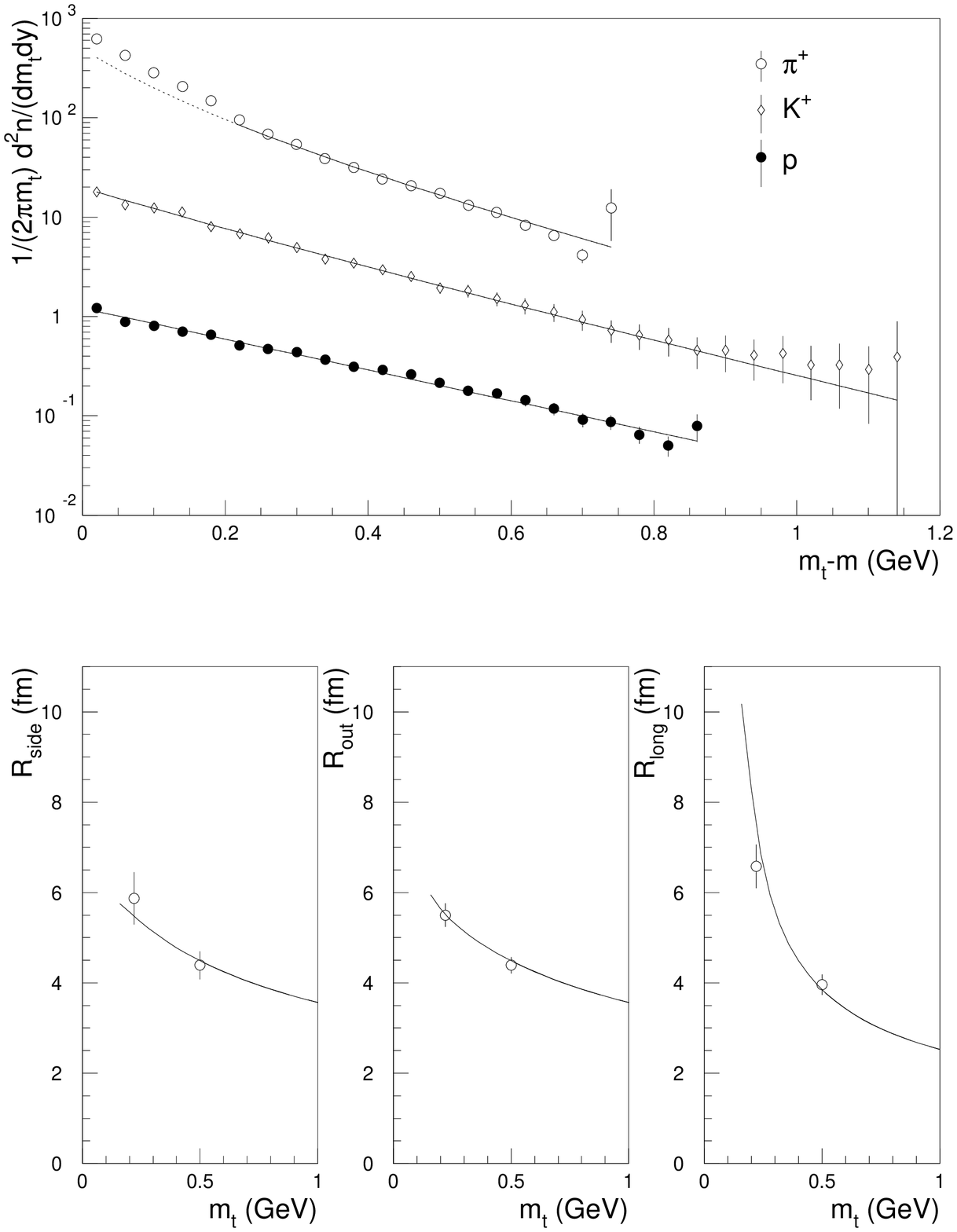,width=9.0cm}
\vspace*{-1.cm}
\caption{Result of simultaneous fits of the Buda--Lund hydro model
  to particle correlations and spectra in 158 AGeV Pb + Pb reactions
  at CERN SPS (data from the NA44 Collaboration)}
\label{f:xna44}
\end{figure}

In case of homogeneous freeze-out temperatures, or particles with
small masses, Eq.~(\ref{e:tstar-m}) implies a linear rise of the slope
with $m$~\cite{3d} as
\be
        T_*(m) = T_0 + m \langle u\rangle_t^2, \qquad
        \mbox{\rm if }\quad \Big\langle {\Delta T \over T}\Big\rangle_r 
\ll T_0/m.
\ee 
For heavy particles, or for large, non-vanishing temperature
gradients, a flattening of the initial linear rise is
obtained~\cite{3d} as
\be
        T_*(m) = T_0 \left[ 1 +  \frac{\langle u\rangle_t^2}
                {\big\langle {\Delta T \over T}\big\rangle_r}\right], \qquad
        \mbox{\rm if }\quad \Big\langle {\Delta T \over T}\Big\rangle_r 
\gg T_0/m .
\ee 
This means that very heavy particles resolve the temperature
inhomogeneities of the source, and they are produced with a
mass-independent effective slope parameter in the BL-H
parameterization, if $T_0/m$ becomes smaller than the temperature
inhomogeneity.  In a general case, the $T_*(m)$ function starts with
an initial linear $m$ dependence, with a slope given by the transverse
flow $\langle u\rangle_t$, then $T_*(m)$ flattenes out to a
mass-independent value if the source has temperature inhomogeneities
in the transverse direction.  Such a behavior was reported by Pb + Pb
heavy ion experiments at CERN SPS~\cite{QM}. The central temperature
is \cite{ster-qm99} $T_0 \approx 140$ MeV, the flattening of the
slopes sets in at about $m = 1400$ MeV~\cite{muller-qm99}, which then
leads to about 10\% temperature inhomogeneity in the transverse
direction of the Pb + Pb source.  This estimate is in a good agreement
with the results of the combined analysis of the single-particle
spectra and the two-particle Bose--Einstein correlation functions, see
Table~5.

\begin{table}[hbt]
\caption{Fit paramaters of Buda--Lund hydro (BL-H) in a simultaneous 
analysis of NA49, NA44 and preliminary WA98  spectra and correlation data}
\label{t:pbpb-fit}
\vspace*{-6pt}
\begin{center}
\begin{tabular}{lcccc}
\hline\\[-10pt]
Parameter & NA49 & NA44 & WA98 & Averaged\\[2pt]
\hline\\[-10pt]
$T_0$ [MeV]      & $134 \pm 3\ \ \ $ & $145 \pm 3\ \ \ $  & $139 \pm 5\ \ \ $  & $139 \pm 6\ \ \ $   \\
$\langle u_t \rangle$ & $0.61 \pm 0.05$ & $0.57 \pm 0.12$ & $0.50 \pm 0.09$ &
$0.55 \pm 0.06$ \\
$R_G$ [fm]       & $7.3 \pm 0.3$ & $6.9 \pm 1.1$ & $6.9 \pm 0.4$ & $7.1
\pm 0.2$  \\
$\tau_0$ [fm/c]  & $6.1 \pm 0.2$  & $6.1 \pm 0.9$ & $5.2 \pm 0.3$  & $5.9 \pm 0.6$  \\
$\Delta\tau$ [fm/c]  & $2.8 \pm 0.4$  & $0.01 \pm 2.2$\ \ \  & $2.0 \pm 1.9$  &
$1.6 \pm 1.5$  \\
$\Delta\eta$     & $2.1 \pm 0.2$  & $2.4 \pm 1.6$ & $1.7 \pm 0.1$  & $2.1
\pm 0.4$  \\
$\langle {\Delta T / T}\rangle_r$ & $0.07 \pm 0.02$ & $0.08 \pm 0.08$ & 
$0.01 \pm 0.02$ & $0.06 \pm 0.05$ \\
$\langle {\Delta T / T}\rangle_t$  & $0.16 \pm 0.05$ & $0.87 \pm 0.72$ & 
$0.74 \pm 0.08$ & $0.59 \pm 0.38$ \\[2pt]
\hline\\[-10pt]
$\chi^2$/{NDF}   & 163/98 = 1.66
                 & 63/71 = 0.89
                 & 115/108 = 1.06
                 & 1.20\\
\hline
\end{tabular}
\end{center}
\vspace{0.5cm}
\end{table}
The NA49, NA44 and WA98 data on single particle spectra of h$^-$,
identified $\pi$, K and p as well as detailed rapidity and $m_t$
dependent HBT radius parameters are found to be consistent with each
other as well as with BL-H.  The BL-H fit results to these data sets
is summarized in Table~5, Ref.\ \cite{ster-qm99}.

\section{Comparison of h + p and Pb + Pb Final States
  at CERN SPS with Heavy Ion Reactions at Low and Intermediate
  Energies}

The final state of central Pb + Pb collisions at CERN SPS corresponds
to a cylindrically symmetric, large ($R_G = 7.1 \pm 0.2$ fm) and
transversally homogenous ($T_0 = 139 \pm 6 $ MeV) fireball, expanding
three-dimensionally with $\langle u_t \rangle = 0.55 \pm 0.06$.  A
large mean freeze-out time, $\taub = 5.9 \pm 0.6$ is found with a
relatively short duration of emission, $\Delta\taub = 1.6 \pm 1.5$ fm,
which is similar to the time-scale of emission in the h + p reaction.
Note that the temporal cooling in Pb + Pb reactions seems to be
stronger than in h + p, which can be expained by the faster,
three-dimensional expansion in the former case, as compared to the
essentially one-dimensional expansion in the case of h + p reactions.
By the time the particle production is over, the surface of Pb + Pb
collisions cools down from 139 MeV to $T_0/( 1 + \langle \Delta T/T_0
\rangle_r ) /( 1 + \langle \Delta T/T_0 \rangle_t ) \approx 83$ MeV.
It is very interesting to note that this value is similar to the
surface temperature of $T_s = 82 \pm 7$ MeV, found in h + p reactions
as a consequence of the transverse temperature inhomogeneities, as
described in Section~\ref{s:na22}, Ref.\ \cite{na22}. Such snowballs
with relatively low values of surface temperature $T_s$ and a possible
hotter core were reported first in 200 AGeV S + Pb reactions in
Ref.\ \cite{3d-qm}.
        
Other hydro parameterizations, as reviewed in Ref.\ \cite{muller-qm99},
frequently neglect the effects of temperature inhomogeneities during
the expansion and particle production stage.  Energy conservation
implies that the temperature cannot be exactly constant when particles
are freezing out in a non-vanishing period of time from a
three-dimensionally expanding source.

The exact solution of non-relativistic, spherically symmetric fireball
hydrodynamics implies~\cite{cspeter,3d} that Gaussian fireballs with
spatially uniform temperature profiles satisfy the collisionless
Boltzmann equation.

Fixing the temperature to a constant in the fits yields an average
freeze-out temperature in the range of $T_f = 110 \pm 30 $
MeV~\cite{muller-qm99,3d-qm,3d-s96,nix}.

\begin{figure}[th]
\vspace*{-2.4cm}
    \centering\epsfig{figure=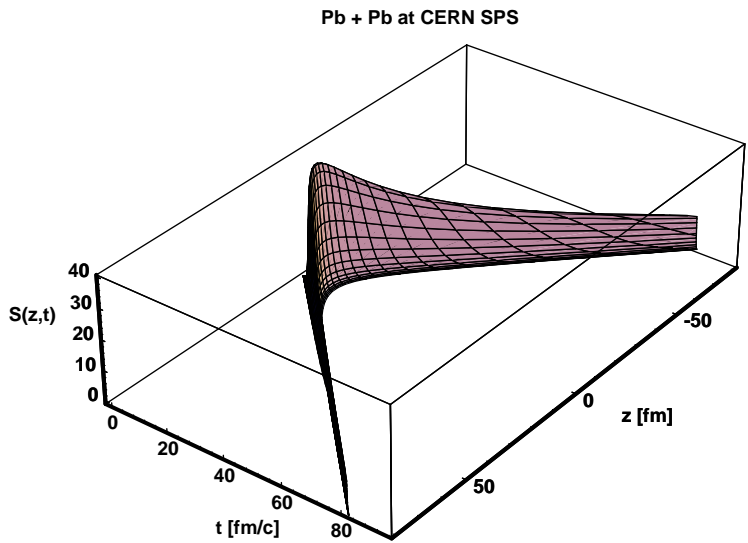,width=10.0cm}
\vspace*{-2.8cm}
\caption{The reconstructed $S(t,z)$ emission function in arbitrary units, 
  as a function of time $t$ and longitudinal coordinate $z$, for 158
  AGeV Pb + Pb reactions}
\label{f:na49-zt}
\end{figure}


\begin{figure}[th]
\vspace{-2.4cm}
    \centering\epsfig{figure=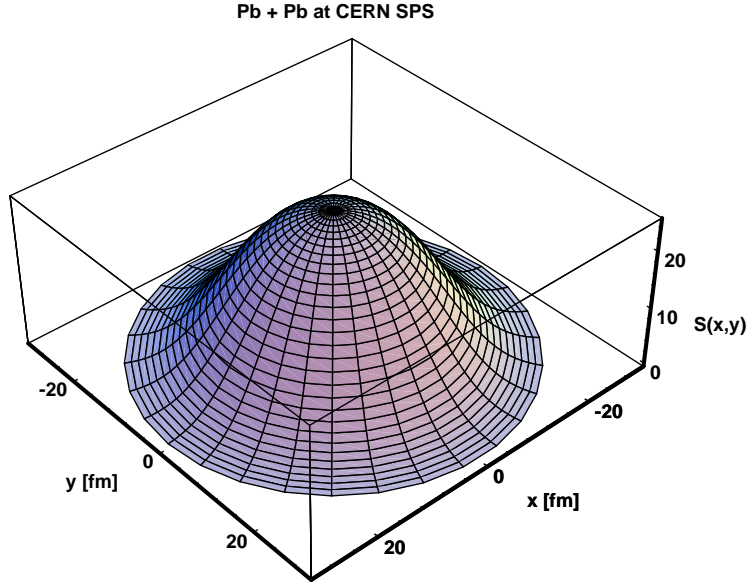,width=10.0cm}
\vspace{-2.8cm}
\caption{The reconstructed $S(r_x,r_y)$ emission function in arbitrary 
  units, as a function of the transverse coordinates $r_x$ and $r_y$}
\label{f:na49-xy}
\end{figure}

Based on the recently found new family of non-relativistic
hydrodynamics~\cite{sol} and on the analysis of h + p single
particle spectra and two-particle Bose--Einstein correlation
function~\cite{na22}, we concluded that the pion emission function
$S(r_x,r_y)$ in h + p reactions corresponds to the formation of a ring
of fire in the transverse plane, because the transverse flow is rather
small and because the sudden drop of the temperature in the transverse
direction leads to large pressure gradients in the center and small
pressure gradients and a density built-up at the expanding radius of
the fire-ring.  We presented arguments for a similar formation of a
spherical shell of fire in the proton distributions at 30 AMeV
$^{40}$Ar + $^{197}$Au reactions.

The formation of shells of fire seems thus to be of a rather generic
nature, related to the initial conditions of self-similar radial
flows.  It is natural to ask the question: can we learn more about
this phenomena in other physical systems?
 
Radial expansion is a well established phenomena in heavy ion
collisions from low energy to high energy reactions.  See Refs
\cite{bauer,ardouin} for recent reviews and for example see
Refs \cite{Herrmann:1996zg,Hong:1998mr,Crochet:1997hz,Kotte:1999gr}
for the evidence of collective flow in central heavy ion collisions
from 100 AMeV to 2 AGeV as measured by the FOPI Collaboration at GSI
SIS.

\begin{figure}[th]
\vspace*{6pt}
    \centerline{\epsfig{figure=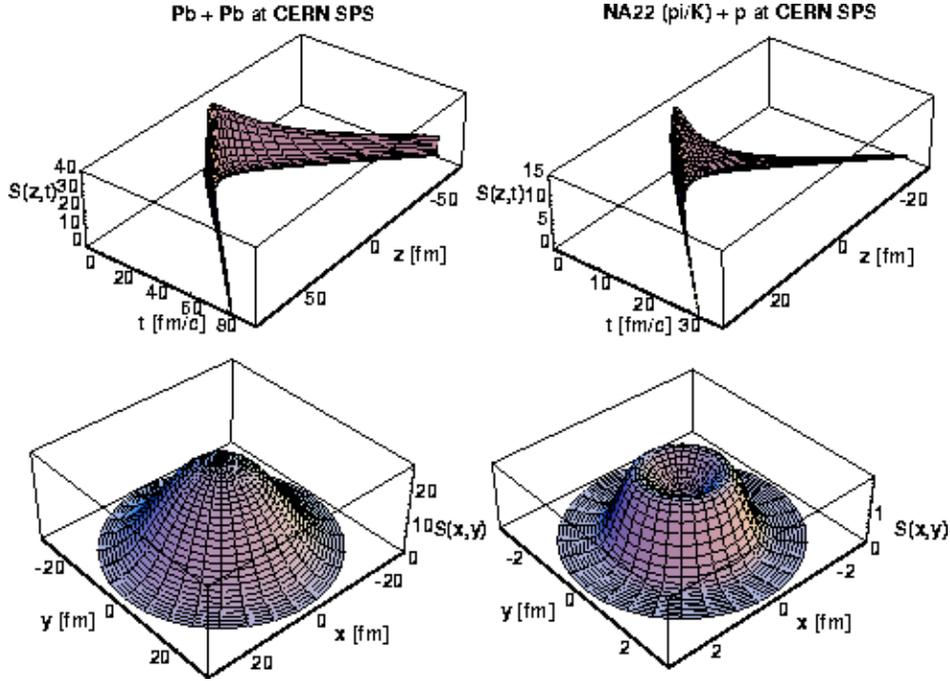,width=12.5cm}}
\caption{Comparison of the reconstructed $S(t,z)$
  and $S(r_x,r_y)$ emission functions for 250 GeV/c h + p reactions
  and for for 158 AGeV Pb + Pb reactions at CERN SPS.  Note the
  different characteristic scales in the transverse and the temporal
  directions, and the different shapes of the transverse density
  distribution.}
\end{figure}


The FOPI Collaboration measured recently the proton--proton correlation
functions at 1.93 AGeV Ni + Ni collisions~\cite{Kotte:1999gr}.  To
interpret their data, they utilized a version of the hydrodynamical
solution, found in Ref.\ \cite{cspeter}.  They assumed a linear flow
profile, a Gaussian density distribution and a constant temperature.
Such a solution of fireball hydrodynamics exists, but it corresponds
to a collisionless Knudsen gas~\cite{cspeter,sol}.  A collisionless
approximation has to break down.  Indeed, only the peak of the FOPI
proton--proton correlation function was reproduced by the collisionless
model, however, the tails had to be excluded from the FOPI analysis.
Perhaps it is worthwhile to search for a possible formation of shells
of fire at the SIS energy domain, by re-analyzing the FOPI
data~\cite{sol}.

\section{Shells of Fire and Planetary Nebulae}

In transport calculations based on the Boltzmann--Uehling--Uhlenbeck
equation, a formation of toroidal density distributions was predicted
for central $^{36}$Ar + $^{45}$Sc collisions at $E = 80$~AMeV in
Ref.\ \cite{upe}, which leads to ring-like configurations for
$S(r_x,r_y)$.

\begin{figure}[thb]
\vspace*{6pt}
    \centerline{\epsfig{figure=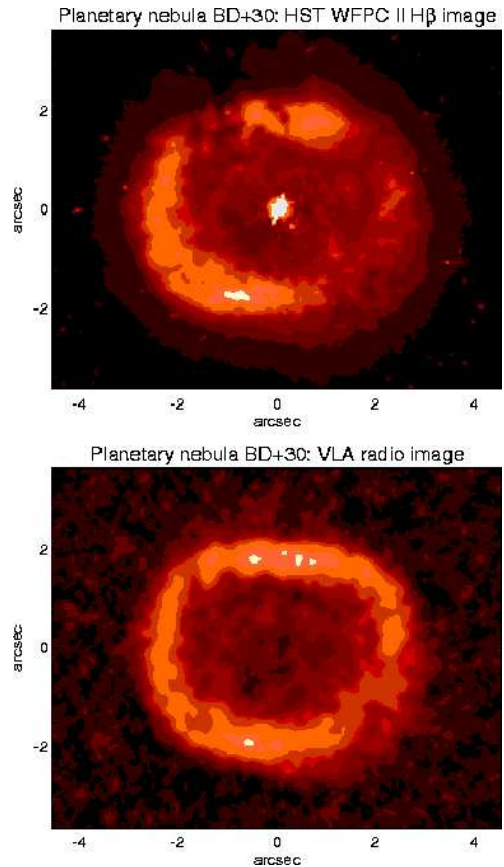,width=6.5cm}}
\caption{Planetary nebula BD+30 imaged by the
  Hubble Space Telescope (top) and by the Very Large Array (VLA)
  radiotelescope in New Mexico (bottom). The latter indicates a
  complete ring of fire, dust blocks some of the visible light on the
  upper image.}
\label{f:bd30}
\end{figure}
        
However, the clearest experimental observation of the development of
expanding shell like structures in the time evolution of exploding
fireballs comes from stellar astronomy.  Stars with initial masses of
less than about eight solar masses end their lives by ejecting
planetary nebulae, stellar remnants turning to white dwarfs.  After
the star has completed its core hydrogen burning, it becomes a red
giant. In the core of the star, helium burns while hydrogen continues
to burn in a thin shell surrounding the core. This hydrogen rich shell
swells to enormous size, and the surface temperature drops to a rather
low value for stars.  A solar wind develops that carries away most of
the hydrogen envelope surrounding the star's central core.  The
envelope material ejected by the star forms an expanding shell of gas
that is known as a planetary nebula.  Planetary nebulae are
illuminated by their central stars and display a variety of often
beautiful structures. Some are spherical or helical, others have
bipolar shapes, and still others are rather irregularly shaped. In a
matter of a few tens of thousands of years, they intermingle with the
interstellar medium and disperse.

The space-time evolution of planetary nebulae is in many aspects
similar to the solution of non-relativistic hydrodynamics given in
Ref.\ \cite{sol}. We argued, that this solution seems to describe also
low and intermediate heavy ion collisions in the 30 -- 80 AMeV energy
domain. A similar hydro solution may also describe the
non-relativistic transverse dynamics at mid-rapidity in hadron +
proton collisions in the CERN SPS energy domain, compare Figs
~\ref{f:na22-r} and \ref{f:bd30}, the latter from Ref.~\cite{bd30}.

In all of these physical systems, expansion competes with the drop of
the pressure gradients, which in turn is induced by the drop of the
temperature on the surface.  If the flow is small enough, the drop of
the temperature on the surface results in a drop of the pressure
gradients on the surface, which implies density pile-up.  On the other
hand, if the flow is strong enough, it blows away the material from
the surface, preventing the formation of shells of fire, and an
ordinary expanding fireball is obtained.

Finally I note that this situation is just a special class of the more
general solutions given in Ref.\ \cite{sol}.  Arbitrary number of
self-similarly expanding, simultaneously existing shells of fire can
be described by the general form of new class of exact solutions of
fireball hydrodynamics~\cite{sol}.


\section{Signal of Partial $U_A(1)$ Symmetry Restoration
from Two-Pion Bose--Einstein Correlations}
\label{s:ua1}

%
%
%
%
\newcommand{\beq}{\begin{equation}}
\newcommand{\eeq}[1]{\label{#1} \end{equation}}

In this section let me summarize Ref.\ \cite{ua1}, where the effective
intercept parameter of the two-pion Bose--Einstein Correlation
function, $\lambda_*$ was shown to carry a sensitive and measurable
signal of partial restoration of the axial $U_A(1)$ symmetry and the
related increase of the $\eta'$ production in ultra-relativistic
nuclear collisions: An increase in the yield of the $\eta'$ meson,
proposed earlier as a signal of partial $U_A(1)$ restoration, was
shown to create a ``hole'' in the low $p_t$ region of $\lambda_*$.

\begin{figure}[htb]
\vspace*{6pt}
\centerline{\psfig{figure=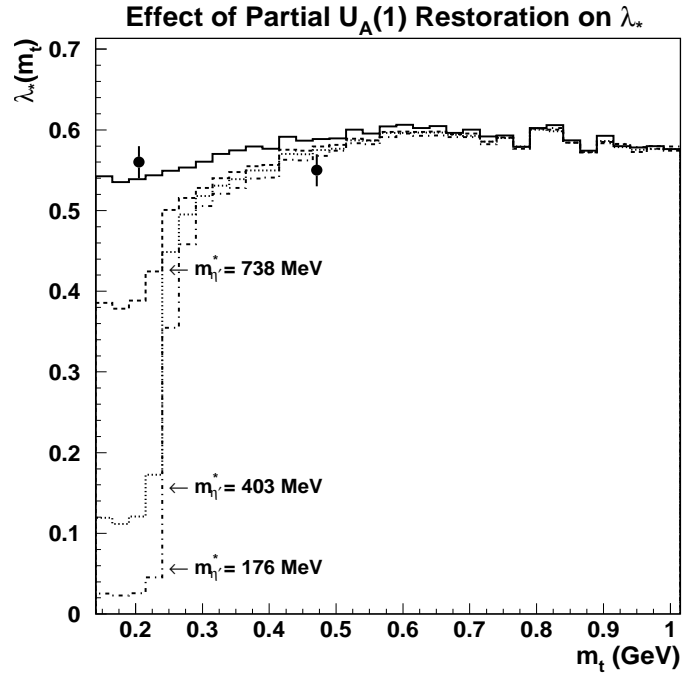,height=3.5in,width=3.5in,angle=0}}
\vspace*{-6pt}
\caption{Using the estimates of pion abundances given by Fritiof,
  the solid line represents $\lambda_*(m_t)$ assuming normal $\eta'$
  abundances while the other lines represent $\lambda_*(m_t)$ with a
  factor of 3 (dashed), 16 (dotted) and 50 (dot--dashed) enhancement of
  $\eta'$ due to partial $U_A(1)$ chiral symmetry restoration and the
  corresponding decrease of the $\eta'$ mass in the hot and dense
  region.  All curves are calculated for $T_0 = 140$ MeV and $\langle
  u_t \rangle = 0.5$.  The datapoints are from 200 AGeV central S + Pb
  reactions at CERN SPS, as measured by the NA44 Collaboration.}
\label{f:lamhole}
\end{figure}

In the chiral limit ($m_u = m_d = m_s = 0$), QCD possesses a $U(3)$
chiral symmetry. When broken spontaneously, $U(3)$ implies the
existence of nine massless Goldstone bosons.  In Nature, there are
only eight light pseudoscalar mesons, a discrepancy which is resolved
by the Adler--Bell--Jackiw $U_A(1)$ anomaly; the ninth would-be
Goldstone boson gets a mass as a consequence of the non-zero density
of topological charges in the QCD vacuum \cite{witten,veneziano}.  In
Refs \cite{kapusta96etap,xnwang}, it is argued that the ninth
(``prodigal'' \cite{kapusta96etap}) Goldstone boson, the $\eta'$,
would be abundantly produced if sufficiently hot and dense hadronic
matter is formed in nucleus--nucleus collisions.  Estimates of
Ref.\ \cite{kapusta96etap} show that the corresponding production cross
section of the $\eta'$ should be enhanced by a factor of 3 up to 50
relative to that for p + p collisions.

If the $\eta'$ mass is decreased, a large fraction of the $\eta'$s
will not be able to leave the hot and dense region through thermal
fluctuation since they need to compensate for the missing mass by
large momentum~\cite{kapusta96etap,xnwang,shuryak}.  These $\eta'$s
will thus be trapped in the hot and dense region until it disappears,
after which their mass becomes normal again; as a consequence, the
$\eta'$-s will have small transverse momenta $p_t$.  Then they decay
to pions via
\beq \eta' \rightarrow \eta + \pi^+ + \pi^- \rightarrow
(\pi^0 + \pi^+ + \pi^-) + \pi^+ + \pi^-.  
\eeq{etap_decay} 
It is important to observe that the $p_t$ of pions produced in this
decay chain is small since many of the $\eta'$ appear at $p_t\simeq 0$
and also since the rest mass of the decay products from the $\eta',
\eta$ decays use up most of the remaining energy.  Based on the
kinematics of the $\eta', \eta$ decay chain to pions, an enhanced
production of $\pi$ mesons was estimated to happen dominantly in the
$p_t \simeq 150$ MeV region, extending to a maximum $p_t \simeq 407$
MeV~\cite{vck}.  In the core/halo picture the $\eta', \eta$ decays
contribute to the halo due to their large decay time
($1/{\Gamma_{\eta', \eta}} \gg 20$ fm/c).  Thus, we expect a hole in
the $0 \leq p_t \leq 150$ MeV region of the effective intercept
parameter, $\lambda_* = [N_{\rm core}({\bf p})/N_{\rm total}({\bf p})]^2$.

To calculate the $\pi^+$ contribution from the halo region, the bosons
($\omega$, $\eta'$, $\eta$ and $ \rK^0_S$) are given both a rapidity
$(-1.0 < y < 1.0)$ and an $m_t$, then are decayed using Jetset 7.4
\cite{JET74}.  The $m_t$ distribution \cite{3d,chalo} of the bosons is
given by
\beq  
N(m_t) = C m_t^{\alpha} e^{-m_t/T_{\rm eff}}, 
\eeq{mt_dist}
where $C$ is a normalization constant, $\alpha = 1 - d/2$
and where \cite{3d,na44-slopes}
\beq
T_{\rm eff} = T_{\rm fo} + m \langle u_t \rangle^2 . 
\eeq{Teff} 
In the above expression, $d=3$ is the dimension of expansion, $T_{\rm fo}
= 140$ MeV is the freeze-out temperature and $\langle u_t \rangle $ is
the average transverse flow velocity.  It should be noted that the
$m_t$ distribution for the core pions is also obtained from
Eq.~(\ref{mt_dist}).  The contributions from the decay products of the
different regions (halo and core) are then added together according to
their respective fractions, allowing for the determination of
$\lambda_*(m_t)$.  The respective fractions of pions are estimated
separately by Fritiof \cite{fritiof_87} and by RQMD \cite{RQMD} as
summarized in Ref.\ \cite{restable}.

\begin{figure}[htb]
\vspace*{6pt}
\centerline{\psfig{figure=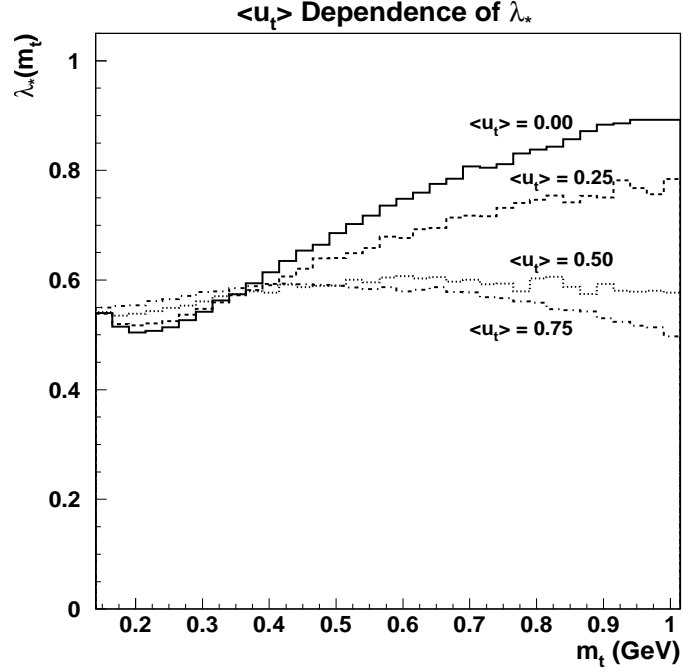,height=3.5in,width=3.5in,angle=0}}
\caption{Using the estimates of pion abundances given by Fritiof,
        $\lambda_*(m_t)$ is calculated using 
        $\langle u_t\rangle = 0.00$ (solid line), 
        $\langle u_t \rangle = 0.25$ (dashed line),
        $\langle u_t \rangle = 0.50$ (dotted line) and
        $\langle u_t\rangle = 0.75$ (dashed--dotted line)}
        \label{f:lamvt}
\end{figure}

Simulating the presence of the hot and dense region involves
increasing the relative abundance of the $\eta '$ and also changing
their $p_t$ spectrum.  The $p_t$ spectrum of the $\eta '$ is obtained
by assuming energy conservation and zero longitudinal motion at the
boundary between the two phases.  This conservation of transverse mass
at the boundary implies
\beq
m^{*2}_{\eta'} + {p_t}^{*2}_{\eta'} = m^{2}_{\eta'} + {p_t}^{2}_{\eta'}, 
\eeq{cons_mt}
where the $*$ denotes the $\eta'$ in the hot dense region.  The $p_t$
distribution then becomes a two-fold distribution.  The first part of
the distribution is from the $\eta'$ which have $p_t^* \leq
[m^{2}_{\eta'} - m^{*2}_{\eta'}]^{1/2}$.  These particles are given a
$p_t = 0$.  The second part of the distribution comes from the rest of
the $\eta'$'s which have big enough $p_t$ to leave the hot and dense
region.  These have the same, flow motivated $p_t$ distribution as the
other produced resonances and are given a $p_t$ according to the $m_t$
distribution
\beq 
N_{\eta'}(m^*_t) = C {m^*_t}^{-0.5} e^{-m^*_t/T'},
\eeq{mt*}
where $C$ is a normalization constant and where $T' = 200$ MeV and
$m^*_{\eta'}$ is the effective temperature and mass, respectively, of
the hot and dense region.  Using the value given above for the
effective temperature and letting $m^*_{\eta'} = 500$ MeV implies an
increase in the production cross section of the $\eta'$ in the hot and
dense region by a factor of 10.

Using three different effective masses for the $\eta'$ in the hot and
dense region, calculations of $\lambda_*(m_t)$ including the hot and
dense regions are compared to those assuming the standard abundances
in Fig.~\ref{f:lamhole}.  A similar $m_t$ dependence but with slightly
higher values of $\lambda_*(m_t)$ is obtained when using RQMD
abundances.  The effective mass of 738 MeV corresponds to an
enhancement of the production cross section of the $\eta'$ by a factor
of 3, while the effective mass of 403 MeV and 140 MeV correspond to
factors of 16 and 50 respectively.  The two data points shown are
taken from NA44 data on central S + Pb reactions at the CERN SPS with
incident beam energy of 200 AGeV \cite{na44-hbt}.  The lowering of the
$\eta'$ mass and the partial chiral restoration result in a hole in
the effective intercept parameter at low $m_t$.  This happens even for
a modest enhancement of a factor of 3 in the $\eta'$ production.
Similar results are obtained when using RQMD abundances.  See
Ref.~\cite{vck} for further details of the simulation.
\vfill\pagebreak

In addition, $\lambda_*(m_t)$ is calculated using Fritiof abundances
with different average flow velocities in Fig.\ \ref{f:lamvt}.  Here it
is shown that $\lambda_*(m_t)$ can also be a measure of the average
collective flow.  In Ref.\ \cite{vck}, an average flow velocity of
$\langle u_t \rangle = 0.50$ resulted in an approximately flat, $m_t$
independent shape for the effective intercept parameter
$\lambda_*(m_t)$ distribution~\cite{vck}.  Calculations using RQMD
abudances result in a similar dependence on $\langle u_t \rangle$, but
with slightly higher values of $\lambda_*(m_t)$.

This analysis of NA44 S + Pb data indicated no visible sign of $U_A(1)$
restoration at SPS energies. In addition, a mean transverse flow of
$\langle u_t\rangle \approx 0.50$ in S + Pb reactions was
deduced~\cite{vck}.  The suggested $\lambda_*$-hole signal of partial
$U_A(1)$ restoration cannot be faked in a conventional thermalized
hadron gas scenario, as it is not possible to create significant
fraction of the $\eta$ and $\eta'$ mesons with $p_t\simeq 0$ in such a
case, Ref.\ \cite{vck}.


\section{Squeezed Correlations and Spectra for \\
Mass-Shifted Bosons}

In this section, let me follow the lines of Refs \cite{ac,acg} to
show that novel back-to-back correlations (BBC) arise for thermal
ensembles of squeezed bosonic states associated with medium-modified
mass shifts.  It was observed in Ref.\ \cite{acg}, that the strength
of the BBC could become unexpectedly large in heavy ion collisions,
and may thus provide an experimentally observable signal of boson
modification in hot and dense matter.
  
Consider, in the rest frame of matter, the following model
Hamiltonian,
\begin{equation}
{H} =  H_0 - \frac{1}{2} \int d^3 {\bf x} d^3 {\bf y} \phi({\bf x})
\delta M^2({\bf x}-{\bf y}) \phi({\bf y}),
\label{ham}
\end{equation}
where $H_0$ is
the asymptotic Hamiltonian,
\begin{equation}
        H_0 = \frac{1}{2} \int d^3 {\bf x} \left(
                \dot{\phi}^2+ |\nabla \phi|^2
                +
                        m_0^2 \phi^2  \right).
\end{equation}
The scalar field $\phi({\bf x})$ in this Hamiltonian, $H$, corresponds
to quasi-particles that propagate with a momentum-dependent
medium-modified effective mass, which is related to the vacuum mass,
$m_0$, via
$$
 m_*^2({|{\bf k}|}) =  m_0^2 - \delta M^2({|{\bf k}|}).
$$
The mass shift is assumed to be limited to long wavelength 
collective modes:
$$
\delta M^2({|{\bf k}|}) \ll m_0^2 \quad\quad 
        \mbox{\rm if}\quad |{\bf k}| > \Lambda_s.
$$
The invariant single-particle and two-particle momentum
distributions are given as:
\begin{eqnarray}
N_1({\bf k}_1) & = & \omega_{{\bf k}_1}{d^3N \over d{\bf k}_1} 
        = \omega_{{\bf k}_1} \langle
 a^\dagger_{{\bf k}_1} a^{\phantom\dagger}_{{\bf k}_1}\rangle , \\
N_2({\bf k}_1,{\bf k}_2) & = & 
\omega_{{\bf k}_1} \omega_{{\bf k}_2} 
        \langle a^\dagger_{{\bf k}_1} a^\dagger_{{\bf k}_2} 
        a^{\phantom\dagger}_{{\bf k}_2}
        a^{\phantom\dagger}_{{\bf k}_1} \rangle ,\\
\langle a^\dagger_{{\bf k}_1} a^\dagger_{{\bf k}_2} 
a^{\phantom\dagger}_{{\bf k}_2} a^{\phantom\dagger}_{{\bf k}_1} \rangle 
& = &  
\langle a^\dagger_{{\bf k}_1} a^{\phantom\dagger}_{{\bf k}_1}\rangle
\langle  a^\dagger_{{\bf k}_2} a^{\phantom\dagger}_{{\bf k}_2} \rangle  + 
        \langle a^\dagger_{{\bf k}_1} a^{\phantom\dagger}_{{\bf k}_2}\rangle
\langle  a^\dagger_{{\bf k}_2} a^{\phantom\dagger}_{{\bf k}_1} \rangle
\nonumber \\
        \null &  &  +
\langle a^\dagger_{{\bf k}_1} a^\dagger_{{\bf k}_2}\rangle
\langle  a^{\phantom\dagger}_{{\bf k}_2}
         a^{\phantom\dagger}_{{\bf k}_1} \rangle,
\label{rand}
\end{eqnarray}
where $a_{\bf k}$ is the annihilation operator for asymptotic quanta
with four-momentum $k^{\mu}\, = \, (\omega_{\bf k},{\bf k})$,
$\omega_{\bf k}=\sqrt{m^2 + {\bf k}^2}$ and the expectation value of
an operator $\hat{O}$ is given by the density matrix $\hat{\rho}$ as
$\langle \hat{O} \rangle = {\rm Tr} \, \hat{\rho}\, \hat{O}$.
Eq.(\ref{rand}) has been derived as a generalization of Wick's theorem
for {\em locally} equilibriated (chaotic) systems in
Ref.~\cite{sinyukov}.

The chaotic and squeezed amplitudes were introduced~\cite{acg} as
\begin{eqnarray}
G_c(1,2) & = & 
\sqrt{\omega_{{\bf k}_1} \omega_{{\bf k}_2} }
  \langle a^{\dagger}_{{\bf k}_1} a^{\phantom\dagger}_{{\bf k}_2}\rangle,\\
G_s(1,2) & = & 
\sqrt{\omega_{{\bf k}_1} \omega_{{\bf k}_2} }
  \langle a^{\phantom\dagger}_{{\bf k}_1}
  a^{\phantom\dagger}_{{\bf k}_2} \rangle .
\label{gs}
\end{eqnarray}
In most situations, the chaotic amplitude, $G_c(1,2) \equiv G(1,2)$
is dominant, and carries the Bose--Einstein correlations,
while the squeezed amplitude, $G_s(1,2)$ vanishes.

\begin{figure}[htb]
\hspace{18pt}\begin{center}
\vspace*{9.2cm}
\includegraphics{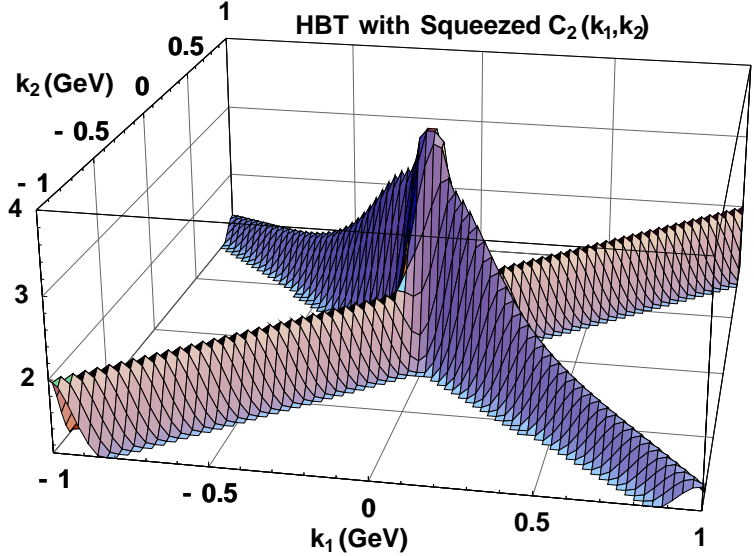}
\end{center}
\vspace{-70pt}
\caption{Illustration of the new back-to-back correlations
  for mass-shifted $\pi^0$ pairs, assuming $T = 140 $ MeV,
  $G_{c/s}(p_1,p_2) \propto \exp[-q_{12}^2 R_G^2/2 ] $, with $R_G = 2$
  fm. The fall of BBC for increasing values of $|{\bf k}|$ is
  controlled here by a momentum-dependent effective mass, $m_{\pi,*} =
  m_{\pi} [ 1 + \exp( - {\bf k}^2 / \Lambda_{s}^2)]$, with $\Lambda_s
  = 325$ MeV in the sudden approximation.}
\label{f:acg}
\end{figure}

\subsection{Mass modification in a homogenous heat bath} 
The terms involving $G_s(1,2)$ become non-negligible when mass shift
becomes non-vanishing, i.e.\ $\delta M^2({\bf k})\ne 0$. Given such a
mass shift, the dispersion relation is modified to $\Omega_{\bf k}^2
=\omega^2_{\bf k}-\delta M^2({\bf k})$, where $\Omega_{\bf k}$ is the
frequency of the in-medium mode with momentum ${\bf k}$.  The
annihilation operator for the in-medium quasi-particle $b_{\bf k}$, and
that of the asymptotic field, $a_{\bf k}$, are related by a Bogolyubov
transformation \cite{ac}:
\begin{equation}
a^{\phantom{\dagger}}_{{\bf k}_1}
        = c^{\phantom{\dagger}}_{{\bf k}_1}
          b^{\phantom{\dagger}}_{{\bf k}_1} + 
                s^{*\phantom{\dagger}}_{-{\bf k}_1} b^\dagger_{-{\bf k}_1}
\equiv C^{\phantom{\dagger}}_1 + S^\dagger_{-1},
\label{asq}
\end{equation}
where
$c_{\bf k}=\cosh[r_{\bf k}]$, 
$s_{\bf k}=\sinh[r_{\bf k}]$  and $r_{\bf k}$
 reads as
\begin{equation}
r_{\bf k}=\frac{1}{2}\log(\omega_{\bf k}/\Omega_{\bf k})\;\; .\label{sqr}
\end{equation}
We introduce the shorthand, $C^{\phantom{\dagger}}_1$ and
$S_{-1}^\dagger$, to simplify later notation. As the Bogolyubov is a
squeezing transformation, let us call $r_{\bf k}$ mode dependent
squeeze parameter.  While it is the {\it $a$-quanta that are
  observed}, it is the {\it $b$-quanta that are thermalized} in
medium~\cite{aw}.  Let us consider the average for a globally
thermalized gas of the $b$-quanta, that is homogenous in volume $V$:
\begin{equation}
\hat{\rho} = {\displaystyle\phantom{|} 1 \over Z} \exp\left(-\frac{1}{T}
        \frac{V}{(2 \pi)^3} \int d^3{\bf k}\, \Omega_{\bf  k}
                \, b^\dagger_{\bf k} b^{\phantom\dagger}_{\bf k}\right).
\end{equation}
When this thermal average is applied,
\begin{eqnarray}
G_c(1,2) & = &
\sqrt{\omega_{{\bf k}_1} \omega_{{\bf k}_2}}
\left[\langle C^\dagger_1C^{\phantom\dagger}_2\rangle + 
\langle
S^{\phantom\dagger}_{-1} S^\dagger_{-2}\rangle\right], \label{gc1}\\
G_s(1,2)&=&
\sqrt{\omega_{{\bf k}_1} \omega_{{\bf k}_2}}
\left[
\langle S^\dagger_{-1} C^{\phantom\dagger}_2
\rangle + \langle C^{\phantom\dagger}_1 S^\dagger_{-2}
\rangle 
\right]
. \label{gs1}
\end{eqnarray}
If this thermal $b$ gas freezes out suddenly at some time at
temperature $T$,
the observed single $a$-particle distribution takes the following form:
\begin{eqnarray}
N_1({\bf k}) & = & \frac{V}{(2 \pi)^3}\, \omega_{\bf k}\, n_1({\bf k}),\\
n_1({\bf k}) & = &
        | c_{\bf k}^{\phantom{\dagger}}|^2 
        n_{\bf k}^{\phantom{\dagger}} 
        + |  s_{-\bf k}|^2 (n_{-\bf k} + 1), \\
n_{\bf k} & =  &\frac{1}{\exp(\Omega_{\bf k}/T) -1} \; .
\end{eqnarray}
This spectrum includes a squeezed vacuum contribution in addition to
the mass-modified thermal spectrum.

In this homogeneous limiting case, the two particle correlation
function is unity except for the parallel (HBT) and antiparallel (BBC)
cases:
\begin{eqnarray}
C_2({\bf k}, {\bf k}) & = & 2, \label{e:c2}\\
C_2({\bf k}, {\bf -k}) & = & 1 +
{\displaystyle\phantom{|} {|c^*_{\bf k}s_{\bf k}^{\phantom{\dagger}}
        n_{\bf k}^{\phantom{\dagger}}+
        c^*_{-\bf k} s_{-\bf k}^{\phantom{\dagger}}
        (n_{-\bf k}^{\phantom{\dagger}}  + 1) |^2 }
\over \displaystyle\phantom{|} {n_1({\bf k}) \, n_1({- \bf k})} }. 
        \label{e:c2b}
\end{eqnarray}
The {\em dynamical} correlation due to the two mode squeezing
associated with mass shifts is therefore {\em back-to-back}, as first
pointed out in Ref.\ \cite{ac}.  The strength of the HBT correlations
remains 2 for identical momenta.

It follows from Eq.~(\ref{e:c2b}) that the intercept of the BBC is
unlimited from above: $ 1 \le C_2({\bf k}, - {\bf k}) < \infty$.  As
$|{\bf k}| \rightarrow \infty$, $C_2({\bf k}, - {\bf k}) \simeq 1 +
1/|s_{-{\bf k}}|^2 \simeq 1 + 1/n_1({\bf k}) \rightarrow \infty$.
Hence, at large values of $|{\bf k}|$, the particle production is {\it
  dominated} by that of back-to-back correlated pairs for any
non-vanishing value of the in-medium mass shifts~\cite{acg}.

\subsection{Suppression by finite duration of emission }

To describe a more gradual freeze-out, the probability distribution
$F(t_i)$ of the decay times $t_i$ is introduced. The sudden
approximation is recovered in the $F(t_i ) = \delta(t_i - t_0)$
limiting case.  The time evolution of the operators is given by
$a_{\bf k}(t) = a_{\bf k}(t_i) \exp[- i \omega_{\bf k} (t - t_i)]$.
This leads to a suppression of BBC as
\be
        C_2({\bf k}, - {\bf k})  =   
                1 + |\tilde F(\omega_{\bf k} + \omega_{- {\bf k}})|^2
        {\displaystyle\phantom{|} {|c^*_{\bf k}s_{\bf k}^{\phantom{\dagger}}
        n_{\bf k}^{\phantom{\dagger}}+
        c^*_{-\bf k} s_{-\bf k}^{\phantom{\dagger}}
        (n_{-\bf k}^{\phantom{\dagger}}  + 1) |^2 }
\over \displaystyle\phantom{|} {n_1({\bf k}) \, n_1({- \bf k})} }. 
\ee
Here $\tilde F(\omega) = \int dt F(t) \exp(-i\omega t)$, so for 
an exponential decay, $F(t) = \Theta(t - t_0)\times$ 
$\Gamma \exp[ -\Gamma (t - t_0)]$ the suppression factor is
\be
        |\tilde F(\omega_{\bf k} + \omega_{- {\bf k}})|^2
        = 1/[1 + (\omega_{\bf k} + \omega_{- {\bf k}})^2/\Gamma^2].
\ee
In the adiabatic limit, $\Gamma \rightarrow 0$, this factor suppresses
completely the BBC, while in the sudden approximation, $\Gamma
\rightarrow \infty$, the full strenght of the BBC is preserved. For a
typical $\delta t = \hbar/\Gamma = 2$ fm/c decay time, and for BBC of
$\phi$ mesons with $m_* = 0.6 - 1.4$~GeV, this suppression factor is
about 0.001, which decreases the BBC of $\phi$ mesons from the scale
of 2000 to 2, the scale of the HBT correlations.  This emphasizes the
enormous strength of the BBC \cite{acg}.

The formalism to evaluate the BBC for {\it locally thermalized,
  expanding sources} was also developed, see Ref.\ \cite{acg} for
greater details.

As the Bogolyubov transformation always mixes particles with
anti-particles, the above considerations hold only for particles that
are their own anti-particles, e.g.\ the $\phi$ meson and $\pi^0$.  The
extension to particle--anti-particle correlations is
straightforward.  Let $+$ label particles, $-$ antiparticles if
antiparticle is different from particle, let $0$ label both particle
and antiparticle if they are identical.  The non-trivial correlations
from mass modification for pairs of $(++)$, $(+-)$ and $(00)$ type
read as follows:
\begin{eqnarray}
C^{++}_2({\bf k}_1,{\bf k}_2)\! & =  & \!
        1 
        + 
        {|G_c(1,2) |^2 \over G_c(1,1) G_c(2,2) }, \\
C^{+-}_2({\bf k}_1,{\bf k}_2)\! & =  & \!
        1 
        + {|G_s(1,2)|^2\over G_c(1,1) G_c(2,2) } , \\
C^{00}_2({\bf k}_1,{\bf k}_2)\! & =  &\!
        1 
        + 
        {|G_c(1,2) |^2 \over G_c(1,1) G_c(2,2) }
        + {|G_s(1,2)|^2\over G_c(1,1) G_c(2,2) } ,  
        \label{e:cfin} 
\end{eqnarray}
where we assume that mass modifications of particles and
anti-particles are the same as happens at vanishing baryon density.

This theory of particle correlations and spectra for bosons with
in-medium mass shifts predicts huge back-to-back correlations of
$\phi^0,\phi^0$ and $\rK^+,\rK^-$ meson pairs~\cite{acg}.  These BBC could
become observable at the STAR and PHENIX heavy ion experiments at RHIC
\cite{lz}, and could be looked for in present CERN SPS experiments.
Further model calculations are required to study the mass-shift
effects on realistic source models.


\section{A Pion-Laser Model and Its Solution}

\def\ket#1{\vert#1\rangle}
\def\bra#1{\langle#1\vert}
\def\brak#1#2{\langle#1\vert #2\rangle}
\def\I#1{\int d^3#1}
\def\bx{{\bf{x}}}
\def\bp{{\bf{p}}}
\def\bk{{\bf{k}}}
\def\bpi{{\bf{\pi}}}
\def\bxi{{\bf{\xi}}}
\def\bq{{\bf{q}}}
\def\br{{\bf{r}}}
\def\axd{\hat{ a}^{\dag} (\bx)}
\def\apd{\hat a^{\dag} (\bp)}
\def\ax{\hat{ a}^{} (\bx)}
\def\ap{\hat  a^{} (\bp)}
\def\psxd{\hat \Psi^{\dag} (\bx)}
\def\psx{\hat \Psi^ (\bx)}
\def\psxde{\hat \Psi^{\dag} (\bx)}
\def\psxe{\hat \Psi^ (\bx)}
\def\pdpx#1{\hat \Psi^{\dag}(\mathbf{x}_{#1},\mathbf{\pi}_{#1})\ket{0}}
\def\be{\begin{equation}}
\def\ee{\end{equation}}
\def\bea{\begin{eqnarray}}
\def\eea{\end{eqnarray}}
\def\dst{\displaystyle\phantom{|}}
\def\ov{\over\dst}
\def\bak{{\bf K}}
\def\dek{{\bf \Delta k}}
 
In high energy heavy ion collisions hundreds of bosons are created in
the present CERN SPS reactions when Pb + Pb reactions are measured
at 160 AGeV laboratory bombarding energy. At the RHIC accelerator,
thousands of pions could be produced in a unit rapidity
interval~\cite{QM}.  If the number of pions in a unit value of
phase-space is large enough these bosons may condense into the same
quantum state and a pion laser could be created~\cite{pratt}.

In this section a consequent quantum mechanical description of
multi-boson systems is reviewed, based on properly normalized
projector operators for overlapping multi-particle wave-packet states
and a model of stimulated emission, following the lines of
Refs \cite{cstjz,jzcst,brood}.  One of the new analytic results is
that multi-boson correlations generate {\it momentum-dependent} radius
and intercept parameters even for {\it static} sources, as well as
induce a special {\it directional dependence} of the correlation
function. This is to be contrasted to the simplistic but very
frequently invoked picture of Eq.~(\ref{e:rhoq}), where sources
without expansion correspond to a correlation function that depends
only on the relative momentum, but not on the mean momentum of the
particle pairs.

A solvable density matrix of a generic quantum mechanical 
system is  
\bea
\hat \rho\! & = &\! \sum_{n=0}^{\infty} \,
        {\dst  {p}_n  \ov {\cal N}{(n)}} \!
        \int \!\! 
        \,\,
         \prod_{i=1}^n d\alpha_i\rho_1(\alpha_i)  \!
        \left(\sum_{\sigma^{(n)}} \prod_{k=1}^n \, 
        \brak{\alpha_k}{\alpha_{\sigma_k}}
        \right)
        \,\ket{\alpha_1,...,\alpha_n} \bra{\alpha_1,...,\alpha_n} \,.
        \nonumber\\
        && \label{e:dtrick}
\eea
Here the index $n$ characterizes sub-systems with particle number
fixed to $n$, the multiplicity distribution is prescribed by the set
of $\left\{ {p}_n\right\}_{n=0}^{\infty}$, normalized as $\sum_{n=
  0}^{\infty} p_n = 1$.  The density matrixes are normalized as
$\mbox{Tr} \, \hat \rho = 1$ and $\mbox{Tr} \, \hat \rho_n =
1$.  The states $\ket{\alpha_1,...,\alpha_n}$ denote properly
normalized $n$-particle wave-packet boson states:
\be
\ket{\ \alpha_1, \ ...\ , \ \alpha_n} = 
 {\left({ \displaystyle{\strut \sum_{\sigma^{(n)} }
        \prod_{i=1}^n \brak{\alpha_i}{\alpha_{\sigma_i}} } } 
        \right)^{- {1\over 2}} } \!\!
\  \alpha^{\dag}_n  \ ... \
\alpha_1^{\dag} \ket{0}.
\label{e:expec2}
\ee
Here $\sigma^{(n)}$ denotes the set of all the permutations of the
indexes $\left\{1, 2, ..., n\right\}$ and the subscript sized
${\sigma_i}$ denotes the index that replaces the index $i$ in a
given permutation from $\sigma^{(n)}$.  The wave-packet creation
operators, $\alpha_i^{\dag}$ create the normalized single-particle
states $\ket{\alpha_i} = \alpha_i^{\dag} \ket{0}$, with
$\brak{\alpha_i}{\alpha_i} = 1$.  The $\alpha_i =
(\xi_i,\pi_i,\sigma_i,t_i)$ stands for a given value of the parameters
of a single-particle wave-packet: the mean coordinate, the mean
momentum, the width of the wave-packet in coordinate space and the
time of the production.  The distribution function $\rho_1(\alpha_i)$
provides the probability distribution for a given value of the
wave-packet parameters.  For simplicity, we assume a static source at
rest, uniform wave-packet widths and simultaneous production,
$\sigma_i = \sigma$ and $t_i = t_0$.  A Gaussian distribution of the
centers of wave-packets is also assumed: in the coordinate space, the
distribution of $\xi_i$ is a characterized with a radius $R$, while in
the momentum-space, the centers of wave-packets $\pi_i$ are assumed to
have a non-relativistic Boltzmann distribution corresponding to a
temperature $T$ and mass $m$.  The coefficient of proportionality,
${\cal N}{(n)}$, can be determined from the normalization condition.

The density matrix given in Eq.~(\ref{e:dtrick}) describes a
quantum-mechanical wave-packet system with induced emission, and the
amount of the induced emission is controlled by the overlap of the $n$
wave-packets~\cite{jzcst}, yielding a weight in the range of $[1,n!]$.
Although it is very difficult numerically to operate with such a
wildly fluctuating weight, the problem of overlapping multi-boson
wave-packets with stimulated emission was reduced in
Refs \cite{cstjz,jzcst} to an already discovered ``ring"-algebra of
permanents for plane-wave outgoing states~\cite{pratt}, with modified
source parameters~\cite{cstjz,jzcst}.
 
Assuming a non-relativistic, non-expanding Gaussian source at rest,
and a Poisson multiplicity distribution $p_n^{(0)}$ in the rare gas
limiting case:
\be
        {p}^{(0)}_n = {n_0 ^n \over n!} \exp(-n_0),
\ee
the ring-algebra was reduced in Ref.\ \cite{pratt} to a set of
recurrences, which reduced the complexity of the problem from the
numerically impossible $n!$ to the numerically easy $n^2$.  These
recurrences were solved analytically in Refs \cite{cstjz,jzcst},
further reducing the complexity of the problem to $n^0$, and yielding
analytic insight to the behavior of the multi-boson symmetrization
effects.

The probability of events with fixed multiplicity $n$, the
single-particle and the two-particle momentum distribution in such
events are given as
\bea
        {p}_n \!\! & = & \!\!
                \omega_{n} \left( \sum_{k=0}^{\infty} \omega_{k} \right)^{-1},
                        \label{e:d.1}\\
        N^{(n)}_1(\bk_1)  \!\! &  = &  \!\!
                \sum_{i=1}^n  {\dst  \omega_{n-i} \ov \omega_{n}}
                 G_i (1,1) , \label{e:d.2} \\
        N^{(n)}_2(\bk_1,\bk_2)  \!\! &  =  &  \!\!
                \sum_{l=2}^n 
                \sum_{m=1}^{l-1} 
                {\dst \omega_{n-l}  \ov \omega_n}
                \left[ G_m(1,1) G_{l-m} (2,2) 
                + G_m(1,2) G_{l-m}(2,1)\right] , 
                 \nonumber \\
        && \label{e:d.3} 
\eea
where $\omega_n = p_n / p_0$.  Averaging over the multiplicity
distribution $p_n$ yields the inclusive spectra as
\bea
        G(1,2) & = & \sum_{n=1}^{\infty} G_n(1,2), \\
        N_1(\bk_1) & = & \sum_{n=1}^{\infty} p_n N^{(n)}_1(\bk_1)
                \, = \,  G(1,1), \\
        N_2(\bk_1,\bk_2) & = & G(1,1) G(2,2) + G(1,2) G(2,1).
\eea

Let us introduce the following auxiliary quantities:
\bea
        \gamma_{\pm} 
                 =  {\dst 1\ov 2} \left( 1 + x \pm \sqrt{1 + 2 x}\, \right),
                        \label{e:gam.s} & \qquad  & 
        x  =  R_{e}^2 \sigma_T^2\,,
                        \label{e:di.x}\\
        \sigma_T^2  =  \sigma^2 + 2 m T\,, 
        \label{e:sigt} 
        & \qquad & 
        R_{e}^2  =  R^2 + {\dst m T \ov \sigma^2 \sigma_T^2}\,,  
        \label{e:reff}
\eea
The {\it general analytical solution} of the model is given through
the generating function of the multiplicity distribution $p_n$ 
\bea
        G(z) & = & \sum_{n = 0}^{\infty} p_n z^n  \label{e:d.g}
        \, = \, \exp\left( \sum_{n=1}^{\infty} C_n (z^n - 1) \right),
        \label{e:gsolu}
\eea
where $C_n$ is introduced as
\be
        C_n \!  = \! {\dst 1 \ov n}\! \int \! d^3 \bk_1 
                \, G_n(1,1) \label{e:3.4} 
          =  {\dst n_0^n \ov n} 
                        \left[ \gamma_+^{n\over 2} - 
                                \gamma_-^{n\over 2} \right]^{-3 }.
                \label{e:c.s} 
\ee
The {\it general analytic solution} for the functions $G_n(1,2)$ is
given as:
\bea
        G_n(1,2)\! & = & \! j_n 
        \exp \left\{ 
        - {b_n \over 2} \left[ 
                \left(\gamma_+^{n\over 2} \bk_1 
                - \gamma_-^{n \over 2} \bk_2\right)^2
                + \left(\gamma_+^{n\over 2} \bk_2 - 
                  \gamma_-^{n \over 2} \bk_1\right)^2 \right]
        \right\},\!\! \\
        j_n \! & = & \! n_0^n\! \left[{ b_n \over \pi}\right]^{3 \over 2}\,, 
        \qquad 
        b_n \,  = \, {\dst 1 \over  \sigma_T^2} 
               {\dst \gamma_+ - \gamma_- \ov \gamma_+^n - \gamma_-^n}\,. 
\eea
The detailed proof that the analytic solution to the multi-particle
wave-packet model is indeed given by the above equations is described
in Ref.\ \cite{jzcst}.

The representation of Eq.~(\ref{e:gsolu}) indicates that the
quantities $C_n$-s are the so called combinants
~\cite{gyul-comb,hegyi-c1,hegyi-c2} of the probability distribution of
$p_n$ and in this case their explicit form is known for any set of
model parameters.  In the generator functional formalism of
multi-particle production, the combinants can be introduced in general
as the integrals of the exclusive correlation
functions~\cite{hegyi-cgen}.  The form of the multiplicity
distribution, given by Eqs~(\ref{e:gsolu},\ref{e:c.s}) does not
correspond to the multiplicity distributions described in standard
textbooks of mathematical statistics, e.g.\ Ref.\ 
\cite{kendallstuart}.  It has the very interesting property, that the
probability distribution simultaneously corresponds to an infinite
convolution of independently distributed clusters of particle
singlets, pairs, triplets and higher-order $n$-tuples, as well as to
an infinite convolution of strongly correlated Bose--Einstein
distribution~\cite{cstjz,jzcst} of particle singlets, pairs, triplets
etc.  As far as I know, this is a new type of physically motivated
discrete distribution in the theory of probability and statistics.

The large $n$ behavior of $p_n $ depends on the ratio of $ n_0 / n_c$,
where the critical value of $n_0$ is $n_c = \gamma_+^{3/2}$,
~\cite{pratt,zhang-clarif,cstjz,jzcst}.  If $n_0 < n_c$, one finds
$\langle n (n - 1)\rangle > \langle n \rangle^2 $, a super-Poissonian
multiplicity distribution, and a chaotic or thermal behavior of the
inclusive correlations, $C_2({\bf k},{\bf k}) = 2$.  If $n_0 \ge n_c$,
the multiplicity distribution, the inclusive spectra and the inclusive
correlations become mathematically undefined, but the exlusive
quantities remain finite for any fixed value of $n$.  To calculate
inclusive observables, a regularization has to be introduced similarly
to the description of Bose--Einstein condensation of massive quanta in
the limit of $\mu \rightarrow m$ in standard statistical mechanics.

{\it Highly condensed limiting case}.  In Refs \cite{cstjz,jzcst} we
have related the divergence for $n_0 \ge n_c$ of the mean multiplicity
$ \langle n \rangle $ to the {\it onset} of a generalized type of
Bose--Einstein condensation of the wave-packets to the wave-packet
state with the smallest energy. Note that the {\it onset} of
Bose--Einstein condensation happens in the limit when $p_n/p_{n+1}
\rightarrow 1$, which happens if $n_0 \rightarrow n_c$ from
below~\cite{jzcst}, and this limiting case formally corresponds to an
``infinite temperature" case~\cite{diosi} --- if the finite slope
parameters of the $N_1^{(n)}({\bf k})$ single-particle distributions
in exclusive events are not taken into account and the concept of the
temperature is inferred only from the number distribution.

In a physical situation, the total number of pions is limited: $n \le
n_{E} = E_{\rm tot}/E_0$, where $E_0$ is the energy of the wave-packet
with the smallest energy (including the mass $m$).  Thus, energy
conservation induces a cut-off in the number of pions, that has to be
taken into account explicitly~\cite{zhang-ener,coh-cre}.  Such a cut
in the multiplicity distribution can be straightforwardly implemented,
as the basic building block, the fully symmetrized $n$-particle
invariant momentum distribution in events with exactly $n$ particles
is always finite for every fixed values of $n$, similarly to the
bosonic enhancement factor $\omega_n$.  At $n_0 = n_c$, the series
$S_n = \sum_{j=0}^n \omega_j$ changes from a convergent to a divergent
one.  After the regularization of the model, by assigning a zero
probability to multiplicities greater that $n_E$, one can show that
for $n_0 > n_c$ a Bose--Einstein condensation develops more and more
with increasing values of $n_0$.
        
{\it Utmost care} is required when evaluating the results published in
the literature regarding the nature of coherence and Bose--Einstein
condensation in the pion-laser model: some papers identify the
``Bose--Einstein condensation" with the ``infinitely hot" $n_0 = n_c$
limiting case. At this point, however, the condensate just appears
with non-zero probability (and one has to introduce the cut
multiplicity distribution to describe it with a $p_{n_E} > 0$), but
the number of quanta in the condensate is rather small, $p_{n_E}
\propto 1/n_E$ at the $n = n_c$ critical point.

The nature of the Bose--Einstein condensation was discussed and
clarified in Ref.\ \cite{coh-cre}, where it was shown that the
condensate will fully develop and dominate the density matrix in the
$R \rightarrow 0$ and $T \rightarrow 0$ simultaneous limiting cases,
confirming the intuitive picture that Bose--Einstein condensation
happens in very cold and very small systems.

In the highly condensed limiting case, the multiplicity distribution
of the produced particles will be sub-Poissonian, a very narrow, cut
power-law distribution that increases with $n$ as
\be
        p_n \propto  \left(\frac{n_0}{n_c}\right)^n \, 
                \Theta(n_{E} - n)\qquad
        \mbox{\rm for} \,\,\, n_0 \gg n_c,
\ee
and vanishes after $n > n_{E}$.  In the limit when the number of
particles in the condensate is very large, the exclusive and the
inclusive correlation functions become unity~\cite{jzcst},
\begin{eqnarray}
        C(\bk_1,\bk_2) & = & C^{(n)}(\bk_1,\bk_2) = 1 \\
        &&\hspace{-2.5cm} \mbox{\rm (the highly condensed limiting case,
        $n_0 \gg n_c$).} \nonumber
\end{eqnarray}
By definition, the above equalities imply optical coherence in the
highly condensed limiting case~\cite{jzcst,coh-cre}.  It is worthwhile
to emphasize, that optical coherence is not to be confused with the
appearance of the coherent states of the annihilation
operator~\cite{coh-cre}.  Instead of being an eigenstate of the
annihilation operator, the fully developed Bose--Einstein condensate
is an eigenstate of the creation operator, with zero eigenvalue.  This
is due the cutoff induced by the conservation of energy: it is not
possible to add one more pion to the condensate if already all the
pions allowed by the constraints are in the condensate.

{\it Rare gas limiting case}.  In contrast, the large source sizes or
large effective temperatures correspond to a rare Boltzmann gas, the $
x \gg 1 $ limiting case.  The general analytical solution of the model
becomes particularly simple in this limiting case.  The leading order
multiplicity distribution can be found from
Eqs~(\ref{e:gsolu},\ref{e:c.s}), corresponding to independently
distributed particles with a small admixture of independently
distributed particle pairs~\cite{cstjz}:
\be
        p_n = {\dst n_0^n \ov n!} \exp(-n_0) \,
                \left[ 1 + {\dst n(n-1) - n_0^2 \ov 2 (2 x)^{3\over 2}
                        } \right].
        \label{e:pn.sol}
\ee
The mean multiplicity, the factorial cumulant moments of the
multiplicity distribution, the inclusive and exclusive momentum
distributions were obtained to leading order terms in $1/x$ in
Ref.\ \cite{cstjz}.  Figure \ref{f:pl1} indicates that {\it the radius
  parameter of the exclusive correlation function becomes mean
  momentum momentum-dependent, even for static sources\/}!  This genuine
multi-particle symmetrization effect is more pronounced for higher
values of the fixed multiplicity $n$, in contrast to the momentum
dependence of $\lambda_{\bak}$ that is independent of
$n$~\cite{cstjz}.
\begin{figure}[htb]
\vspace*{6pt}
\centerline{\epsfig{file=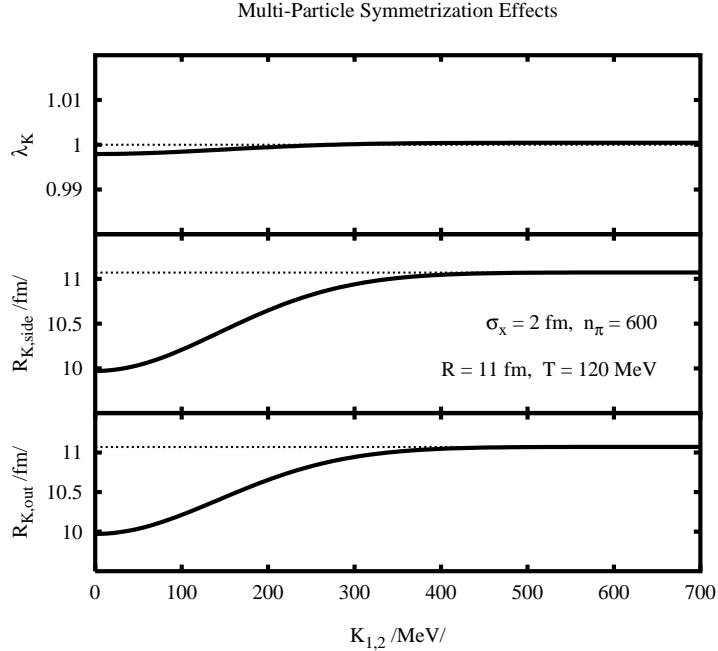,width=10.0cm}}
\vspace*{-12pt}
\caption{Multi-particle symmetrization results at low $\bak$ in a
  momentum-dependent reduction of the intercept parameter
  $\lambda_{\bak}$, the side-wards and the outwards radius parameters,
  $R_{\bak,s}$ and $R_{\bak,o}$ from their static values of 1 and
  $R_{e}$, respectively.  The enhancement of these parameters at high
  momentum is hardly noticeable for large and hot systems.}
\label{f:pl1}
\end{figure}

One finds that multi-boson symmetrization effects lead to the
development of a Bose--Einstein condensate. Before the onset of the
Bose--Einstein condensation, the stimulated emission becomes
significant in the low momentum modes earlier than in the high
momentum modes.  This is the reason, why even the exclusive
correlation functions develop a mean momentum dependent radius
parameter, as well as a direction-dependent radius component and a
mean momentum dependent intercept parameter.


\section{Summary and Outlook}
        
In this review, new kind of similarities were highlighted between
stellar astronomy and intensity interferometry in high energy physics.
The model independent characterization of short-range correlation was
given in terms of expansions in complete orthonormal sets of
polinomials, the core/halo model and the recently found Coulomb
wave-function correction method was reviewed, for Bose--Einstein
$n$-particle correlations.

The invariant Buda--Lund (BL) parameterization of Bose--Einstein
correlation functions was derived in a general form, and compared to
the Bertsch--Pratt and the Yano--Koonin--Podgoretskii parameterization
in the particular Gaussian limiting case.  The Buda--Lund
hydrodynamical parameterization, BL-H was fitted to hadron--proton and
Pb + Pb collisions at CERN SPS energies.  Larger mean freeze-out
proper-times and larger transverse radii were found in the Pb + Pb
reactions.  Although the central values of freeze-out temperatures
were rather similar in both reactions, the transverse temperature
gradient is larger while the transversal flow is smaller in h + p
reactions, than in the Pb + Pb system.  This resulted in different
shapes for the transverse density profiles, that were approximately
reconstructed assuming the applicability of a new family of solutions
to fireball hydrodynamics~\cite{sol}.  Although Pb + Pb reactions were
found to be rather homogenous expanding fireballs, the h + p reactions
were found to be similar to a cold and expanding ring of fire when
viewed in the transverse plane.  The central freeze-out temperature is
about $T_0 = 140 $ MeV in both reactions, the surface temperature
after the emission of particles is over seems to be also similar,
about $T_s = 82$ MeV, the duration of the particle emission is also
about $\Delta \taub \approx 1.5$ fm in both cases.

Inspecting the results of a non-relativistic version of BL-H to
$^{40}$Ar + $^{197}$Au proton and neutron correlations and spectra, an
indirect signal was observed for the formation of a shell of fire,
made of protons, while the neutrons seems to come from an ordinary
fireball.  The hydrodynamics of cooling and expanding shells of low
energy heavy ion reactions was shown to be similar to that of
spherical planetary nebulae, indicating a new connection between
stellar astronomy and particle interferometry in heavy ion physics.

Another similarity between stellar astronomy and high energy physics
was discussed in terms of the interferometry of binary sources: the
binary stars in stellar astronomy create oscillations in the HBT
effect~\cite{hbt-bin} similarly to the oscillations that were shown to
exist in the Buda--Lund type of hydrodynamical parameterization in
heavy ion physics and to the expected oscillations of pion
correlations in particle interferometry in W$^+$W$^-$ decays at LEP2.
The first positive evidence for the existence of such binary sources
in heavy ion physics seems to be the recent measurement of oscillating
proton--proton correlations by the NA49 Collaboration~\cite{na49-pp},
which may be a consequence of the existence of the two maxima in the
proton rapidity distribution and the attractive final-state
interactions of protons, that enhance the large $Q$ part of the pp
intensity correlation function and make these oscillations clearly
visible.
        
The question of non-Gaussian oscillations of three-dimensional
Bose--Einstein correlation functions in heavy ion physics has not yet
been experimentally investigated. I think it is time to start the
experimental search for non-Gaussian structures in multi-dimensional
Bose--Einstein correlation functions in high energy heavy ion and
particle physics. I hope that experiments will decide to publish in
the future not only the (Gaussian) fit parameters of
(multi-dimensional) Bose--Einstein correlation functions, but, most
importantly, the measured {\it data points} and the corresponding {\it
  the error bars}.  It was shown already in Refs
\cite{nr,1d,3d,3d-qm,mpd95}, that the reconstruction of the space-time
picture of the particle emission: the extraction of density, flow and
temperature profiles requires the {\it simultaneous} analysis of the
double-differential {\it single-particle spectra} and the
momentum-dependent multi-dimensional Bose--Einstein {\it correlation
  functions}. These data sets should be made public in as much detail
as possible, including multi-dimensional data and error-bar tables.

In the chiral limit, when the up, down and strange quarks become
massless and the $U_A(1) $ symmetry is fully restored, the mass of the
$\eta'$ meson vanishes in the $U_A(1)$ symmetric, new phase.  The
appearance of such a phase implies that the intercept parameter of the
two-pion correlation function vanishes in the $p_t \le 150$~MeV
region.  In this sense, the transverse mass-dependent intercept
parameter $\lambda_*(m_t)$, was interpreted as an effective order
parameter of partial $U_A(1)$ symmetry restoration~\cite{vck,ckv}.

Bosonic mass shifts in medium were shown to result in unlimitedly
large back-to-back correlations of the observable boson--anti-boson
pairs. Although a finite time suppression factor may reduce the
strength of these correlations substantially, the magnitude of the
back-to-back correlations is estimated to be observably strong for
typical mass shifts and freeze-out time distributions in
ultra-relativistic heavy ion collisions.
       
Multi-boson symmetrization effects were shown to generate {\it
momentum-de\-pend\-ent} radius and intercept parameters even for {\it
static} sources.

The proposed $\lambda_*$-hole signal of the $U_A(1)$ symmetry
restoration, the new kind of back-to-back correlations and optically
coherent, effectively lasing pion sources could be searched for in
future in heavy ion experiments at CERN SPS and at RHIC. The
oscillations in multi-dimensional Bose--Einstein and Fermi--Dirac
correlations could be searched for in e$^+$e$^-$ annihilation
experiments at LEP2, as well as in heavy ion collisions at CERN SPS
and at RHIC. Note that this paper is a review of particle
interferometry before 2000; a substantially shortened version of the
present material has been published in Ref.\ \cite{hbt-database}.

\section*{Acknowledgments} I would like to thank
to my co-authors: M. Asakawa, J. Beier, M. Gyulassy, R.~Hakobyan, S.
Hegyi, J. Helgesson, D. Kiang, W. Kittel, D. Kharzeev, B. L\"orstad,
S. Nickersson, A. Ster, S. Vance and J. Zim\'anyi, and to the NA22
Collaboration, for their various contributions to some of the sections
in this review.  I also would like to thank the Organizers of the
NATO School Nijmegen'99 for creating a pleasent atmosphere and an inspiring
working environment.  I am grateful to Professors Kittel, Hama and
Padula for inspiration and for stimulating working environment.
        
This research was supported by the grants Hungarian OTKA T024094,
T026435, T029158 and T034269, the US--Hungarian Joint Fund MAKA grant
652/1998, NWO--OTKA N025186, OMFB--Ukraine S\&T grant 45014 and FAPESP
98/2249-4 and \mbox{99/09113-3}.  \vfill\pagebreak

\end{document}